\pdfoutput=1

\documentclass[11pt,twoside,a4paper,cmspaper,final,collab]{cms-tdr}
\def\svnVersion{185700}\def\cmsCernNo{2012-348}\def\cmsCernDate{\today}\def\cmsMessage{Submitted to the European Physical Journal C}

\begin{document}\cmsNoteHeader{SUS-12-010}

\hyphenation{had-ron-i-za-tion}
\hyphenation{cal-or-i-me-ter}
\hyphenation{de-vices}
\RCS$Revision: 173957 $
\RCS$HeadURL: svn+ssh://svn.cern.ch/reps/tdr2/papers/SUS-12-010/trunk/SUS-12-010.tex $
\RCS$Id: SUS-12-010.tex 173957 2013-02-26 01:50:45Z richman $
\newcommand\met{\ensuremath{\eslash_\text{T}}\xspace}
\newcommand{\sMET}{\ensuremath{Y_\text{MET}}\xspace}
\newcommand{\cPK}{\ensuremath{\cmsSymbolFace{K}}\xspace}
\newcommand\Lp{\ensuremath{L_{\mathrm{P}}}\xspace}
\newcommand\Ht{\ensuremath{H_{\cmsSymbolFace{T}}}\xspace}
\newcommand\stlep{\ensuremath{S_{\cmsSymbolFace{T}}^{\text{lep}}}\xspace}
\newcommand{\mT}{\ensuremath{M_{\cmsSymbolFace{T}}}\xspace}
\newcommand{\amT}{\ensuremath{M_{\cmsSymbolFace{T}}}\xspace }
\newcommand\zANN{\ensuremath{z_{\mathrm{ANN}}}\xspace}
\ifthenelse{\boolean{cms@external}}{\providecommand{\cmsLeft}{top}}{\providecommand{\cmsLeft}{left}}
\ifthenelse{\boolean{cms@external}}{\providecommand{\cmsRight}{bottom}}{\providecommand{\cmsRight}{right}}
\ifthenelse{\boolean{cms@external}}{\providecommand{\cmsLLeft}{Top}}{\providecommand{\cmsLLeft}{Left}}
\ifthenelse{\boolean{cms@external}}{\providecommand{\cmsRRight}{Bottom}}{\providecommand{\cmsRRight}{Right}}
\newlength\cmsFigWidth
\ifthenelse{\boolean{cms@external}}{\setlength\cmsFigWidth{0.45\textwidth}}{\setlength\cmsFigWidth{0.8\textwidth}}
\cmsNoteHeader{SUS-12-010} 
\title{Search for supersymmetry in pp collisions at $\sqrt{s} =7$\TeV in events with a single lepton, jets, and missing transverse
momentum}%
\ifthenelse{\boolean{cms@external}}{\titlerunning{Search for supersymmetry at $\sqrt{s}=7\TeV$}}{}

\date{\today}

\abstract{Results are reported from a search for new physics processes in
events containing a single isolated high-transverse-momentum lepton (electron or muon), energetic jets, and
large missing transverse momentum. The analysis is based on a 4.98\fbinv sample of proton-proton
collisions at a center-of-mass energy of 7\TeV, obtained with the CMS detector at the LHC.
Three separate background estimation methods, each relying primarily on control
samples in the data, are applied to a range of signal regions, providing
complementary approaches for estimating the background yields.
The observed yields are consistent with the predicted standard model
backgrounds. The results are interpreted in terms of limits on the parameter space for the
constrained minimal supersymmetric extension of the standard model, as well as on cross sections for
simplified models, which provide a generic description of the production and decay
of new particles in specific, topology based final states.}

\hypersetup{%
pdfauthor={SUSY RA4 Team},%
pdftitle={Search for supersymmetry in pp collisions at sqrt(s) = 7 TeV in events with a single lepton, jets, and missing transverse
momentum},%
pdfsubject={CMS},%
pdfkeywords={CMS, physics, supersymmetry, SUSY}}

\maketitle 

\section{Introduction}
\label{sec:Introduction}

This paper reports results from an updated and improved search for new physics processes in
proton-proton collisions at a center-of-mass energy of 7~TeV, focusing on the
signature with a single isolated lepton (electron or muon),
multiple energetic jets, and large missing momentum transverse
to the beam direction ($\ETslash$). The data sample was collected by the Compact Muon Solenoid (CMS)
experiment during 2011 at the Large Hadron Collider (LHC)
and corresponds to an integrated luminosity of 4.98\fbinv,
roughly one hundred times larger than the sample used for our previous
search~\cite{Chatrchyan:2011qs}.

The lepton + jets + $\ETslash$ signature is prominent in models based on
supersymmetry (SUSY)~\cite{Martin:1997ns,Wess:1974tw,Nilles:1983ge,Haber:1984rc,Barbieri:1982eh,Dawson:1983fw}.
In $R$-parity-conserving models~\cite{Farrar:1978xj}, SUSY particles are produced in pairs, and
their decay chains end with the lightest supersymmetric particle (LSP). In some
scenarios, the LSP is a neutralino ($\PSGcz$), a heavy, electrically neutral, weakly
interacting particle with the properties of a dark-matter candidate~\cite{Feng:2010gw}.
The presence of two such LSPs in each SUSY event typically leads to a large missing transverse momentum,
depending on the details of the SUSY mass spectrum. The isolated lepton indicates
a weak decay of a heavy particle, such as a \PW\ boson or a chargino ($\PSGc^\pm$).
Multiple jets can be produced in complex decay chains of SUSY particles.
This signature arises in many SUSY models, including the constrained
minimal supersymmetric extension of the standard model
(CMSSM)~\cite{Kane:1993td,Chamseddine:1982jx}, and in
simplified models~\cite{ArkaniHamed:2007fw,Alwall:2008ag,Alwall:2008va,Alves:2011wf},
which are based on simplified mass spectra and decays of new particles.
Both of these frameworks are used to interpret the results.
Searches in this or similar channels have been reported by
CMS~\cite{Chatrchyan:2011qs,Chatrchyan:2012pc} and ATLAS~\cite{Aad:2011hh,ATLAS:2011ad,Aad:2012ms}.

Searches for SUSY particles are complicated by the presence of standard model (SM) backgrounds that can share
many of the features of signal events. In the single-lepton final state,
backgrounds arise primarily from the
production of \cPqt\cPaqt~and \PW+jets events, with smaller contributions
from \cPZ+jets, single-top quark production, and QCD multijet events. In the
event topology studied here, a large observed value of  $\ETslash$
in a standard model event is usually genuine,
resulting from the production of one or more
high-momentum neutrinos. A smaller contribution to events in the high-$\ETslash$ tail
in this search can arise from
the mismeasurement of jets in high cross section
processes such as QCD multijet events.
To determine the contributions from these backgrounds,
we use methods that are primarily based
on control samples in data, sometimes in conjunction
with specific information from simulated
event samples or from additional measurements that provide constraints
on the background processes.

Three complementary methods are used to analyze the data, providing valuable cross-checks
and probing different signal regions.
The \textit{Lepton Spectrum (LS) method} was used in the CMS
single-lepton~\cite{Chatrchyan:2011qs} and opposite-sign dilepton~\cite{Chatrchyan:2011bz}
SUSY searches performed using the 2010 data sample. It uses the
observed lepton transverse momentum (\pt) spectrum and other control samples
to predict the \ETslash distribution associated with the dominant SM backgrounds.
This method is sensitive to SUSY models in which the \ETslash distribution is decoupled
from the lepton \pt spectrum, as is the case when two undetected LSPs produce a large missing
transverse momentum. The \textit{Lepton-Projection Variable (\Lp)
method} uses the \Lp variable, which was developed for the CMS measurement of the \PW\ polarization
in \PW+jets events~\cite{Wpol-PRL}. This variable, described in
Section~\ref{sec:LPMethod}, is correlated with the helicity angle of the
lepton in the \PW-boson rest frame. Both the \Lp and the LS methods
take advantage of well-understood properties of the \PW\ polarization
in \cPqt\cPaqt\ and \PW+jets events for the background determination.
The methods are complementary in that they rely on significantly different 
approaches to determining the backgrounds, based on different kinematic 
variables and different signal regions.
The \textit{ANN method} uses an artificial neural network discriminant built from several
kinematic quantities. The ANN discriminant is then used
in conjunction with $\ETslash$ to define signal and sideband regions, from which the background yield is
determined. A key variable in the ANN is $M_\mathrm{T}$, an approximate
invariant mass of the system comprising the lepton and the $\ETslash$,
computed with the momentum components
transverse to the beam direction. Background events usually have $M_\mathrm{T}<M(\PW)$,
where $M(\PW)$ is the $\PW$ boson mass,
because the observed $\ETslash$ is associated with the neutrino from $\PW\to\ell\bar\nu$
decay.

This paper is organized as follows. Sections~\ref{sec:CMSDetector} and \ref{sec:EventSamples}
describe the CMS detector and the event samples.
The event preselection requirements that are common to all methods are discussed
in Section~\ref{sec:EventSelection}.
Sections~\ref{sec:LeptonSpectrumMethod}, ~\ref{sec:LPMethod}, and ~\ref{sec:ANN}
describe the LS, \Lp, and ANN methods, respectively, for obtaining SM background estimates from control samples in
data. The observed yields in data are compared with the background estimate obtained
for each method. Systematic uncertainties are described in Section~\ref{sec:SystematicUncertainties}.
Finally, the results, interpretation, and conclusions of the analysis are
presented in Sections~\ref{sec:Results} and \ref{sec:Conclusions}.

\section{The CMS detector}
\label{sec:CMSDetector}

The CMS detector, described in detail in Ref.~\cite{ref:CMS}, is a multipurpose
apparatus designed to study high-\pt  physics processes in proton-proton collisions,
as well as a broad range of phenomena in heavy-ion collisions.
The central element of CMS is a 3.8\unit{T} superconducting solenoid,
13\unit{m} in length and 6\unit{m} in diameter. Within the magnet are (in order
of increasing distance from the beam pipe) high-precision silicon pixel
and silicon strip detectors for charged particle tracking;
a lead-tungstate crystal electromagnetic calorimeter for
measurements of photons, electrons, and the electromagnetic component
of jets; and a hadron calorimeter, constructed from scintillating tiles and brass
absorbers, for jet energy measurements.
Beyond the magnet is the muon system, comprising drift tube, cathode strip,
and resistive-plate detectors interleaved with steel absorbers. Most of the
detector systems are divided into subsystems that cover the central (barrel)
and forward (endcap) regions. The first level of the CMS trigger
consists of custom hardware processors that use information
from the calorimeter and the muon system to select up to 100 kHz of the
most interesting events. These events are then analyzed in the High Level Trigger
(HLT) processor farm, which uses information from all CMS detector systems
to reduce the event rate to about 300 Hz.

In describing the angular distribution of particles and the acceptance of the detector, we
frequently make use of the pseudorapidity, $\eta = -\ln[\tan(\theta/2)]$, where
the polar angle $\theta$ of the particle's momentum vector is measured
with respect to the $z$ axis of the CMS coordinate system. The $z$ axis
points along the direction of the counterclockwise-moving proton beam; the azimuthal
angle $\phi$ is measured in a plane perpendicular to this axis. The separation
between two momentum vectors in $\eta$-$\phi$ space is characterized by the
quantity $\Delta R = \sqrt{(\Delta\eta)^2+ (\Delta\phi)^2}$, which is approximately
invariant under Lorentz boosts along the $z$ axis.

\section{Data and simulated event samples}
\label{sec:EventSamples}

The data samples used in the analysis were selected using triggers
based on $\ETslash$, lepton $\pt$, and the transverse momenta ($\pt^j$) of the
observed jets $j$. The overall level of jet activity was measured with
the quantity $\HT^\text{trigger}=\sum_j \pt^j$,
the scalar sum of jet transverse momenta satisfying $\pt^j>40$\GeV.
The missing transverse momentum $\ETslash^\text{trigger}$ was computed in the trigger using
particle-flow algorithms~\cite{ref:PAS-PFT-09-001,ref:PAS-PFT-10-002}.
To maintain an acceptable trigger rate,
the thresholds on $\ETslash^\text{trigger}$, lepton $\pt$, and $\HT^\text{trigger}$,
were raised as the LHC luminosity increased over the course of the data collection period.
The highest thresholds applied in the muon trigger selection
were $\ETslash^\text{trigger}>50$\GeV, muon $\pt>15$\GeV,
and $\HT^\text{trigger}>300$\GeV. For electron triggers, the highest
thresholds applied
were $\ETslash^\text{trigger}>50$\GeV, electron $\pt>15$\GeV
and $\HT^\text{trigger}> 250$\GeV; a
loose electron isolation requirement
was also applied to help control the rate.
The offline analysis requirements for both muon and electron events
are more restrictive than those used in the trigger.

The analysis procedures are designed using simulated event samples.
Except for certain scans of the SUSY parameter space
discussed later, the detector simulation is performed using the
\GEANTfour\ package~\cite{GEANT4}. A variety of Monte Carlo (MC) event generators are
used to model the backgrounds. The QCD multijet samples are generated with the
\PYTHIA6.4.22~\cite{pythia} MC generator with tune Z2~\cite{ref:TuneZ2}.
The dominant background, \cPqt\cPaqt, is studied with a sample generated using
\MADGRAPH 5.1.1.0~\cite{madgraph}. The \PW +jets and \cPZ +jets processes are
also simulated with \MADGRAPH. Single-top ($s$-channel,
$t$-channel, and tW) production is simulated with {\POWHEG}~\cite{powheg}. To model
the effect of multiple pp interactions per beam crossing (pileup),
simulated events are generated with a nominal distribution of multiple vertices,
then reweighted to match the distribution of
the number of collision vertices per bunch crossing as measured in data.

Event samples for SUSY benchmark models are generated with \PYTHIA.
As example CMSSM scenarios, we use LM3 and LM6, which are
among the standard benchmarks~\cite{PTDR2} used in CMS. The
CMSSM benchmarks are described by the universal scalar mass parameter
$m_0$, the universal gaugino mass parameter $m_{1/2}$,
the universal trilinear soft-SUSY-breaking parameter $A_0$,
the ratio of the two Higgs-doublet vacuum expectation values $\tan\beta$,
and the sign of the Higgs mixing parameter $\mu$. The
LM3 (LM6) benchmark is described by $m_0=330$\GeV ($85$\GeV), $m_{1/2}=240$\GeV ($400$\GeV),
$A_0=0$\GeV ($0$\GeV), $\tan\beta=20$ (10), and $\mu>0\ (0)$.
For LM3, the masses of
the gluino and squarks are very similar (${\approx}600$\GeV), except for
$m(\widetilde{\mathrm{t}})\approx 440$\GeV, while the mass of the LSP
is $m(\PSGczDo) = 94$\GeV. The LM6 spectrum is heavier,
with $m(\PSg)\approx 930$\GeV, $m(\PSQ)\approx 800$\GeV,
$m(\widetilde{\mathrm{t}})\approx 650$\GeV, and $m(\PSGczDo)\approx 160$\GeV.
The next-to-leading-order (NLO) cross sections for these models
are approximately 4.8\unit{pb} (LM3), and 0.4\unit{pb} (LM6).

The ANN method uses the LM0 model~\cite{PTDR2} to train the neural
network. Because of its large cross section (54.9\unit{pb} at NLO),
LM0 has already been excluded~\cite{Chatrchyan:2011qs}, but its kinematic distributions
still provide a reasonably generic description of SUSY behavior with respect
to the variables used in the neural network. The parameters for LM0 are
$m_0=200$\GeV, $m_{1/2}=160$\GeV, $A_0=-400$\GeV, $\tan\beta=10$, and
$\mu>0$.

The results are interpreted in two ways: (i) as constraints on CMSSM parameter space
and (ii) as constraints on cross sections for event topologies described
in the framework of simplified models. In both cases, a large number of simulated
event samples are required to scan over the relevant space of model parameters. For this reason, the
scans are performed with the CMS fast simulation package~\cite{Abdullin:2011zz},
which reduces the time associated with the detector simulation.

Both the LS and $L_\mathrm{P}$ background determination methods rely on knowledge of the
\PW-boson polarization in \PW+jets and in \cPqt\cPaqt\ events. The polarization
effects are well modeled in simulated event samples, which are used in
conjunction with control samples in data.
The angular distribution of the (positively) charged lepton in the \PWp\
rest frame can be written as:
\ifthenelse{\boolean{cms@external}}{
\begin{multline}
\frac{\rd N}{\rd\cos\theta^*_{\ell}} = f_{+1}\,\frac{3}{8}(1+\cos\theta^*_{\ell})^2 +\\
f_{-1}\,\frac{3}{8}(1-\cos\theta^*_{\ell})^2+f_{0}\,\frac{3}{4}\sin^2\theta^*_{\ell},
\label{Eq:Wpolarization}
\end{multline}
}
{
\begin{equation}
\frac{\rd N}{\rd\cos\theta^*_{\ell}} = f_{+1}\,\frac{3}{8}(1+\cos\theta^*_{\ell})^2 +
f_{-1}\,\frac{3}{8}(1-\cos\theta^*_{\ell})^2+f_{0}\,\frac{3}{4}\sin^2\theta^*_{\ell},
\label{Eq:Wpolarization}
\end{equation}
}
where $f_{+1}$, $f_{-1}$, and $f_{0}$ denote the polarization fractions associated with
the \PW-boson helicities $+1$, $-1$, and 0, respectively. The angle $\theta^*_{\ell}$
is the polar angle of the charged lepton in the \PWp\  rest frame, measured
with respect to a $z$ axis that is aligned with the momentum direction of the \PWp\
in the top-quark rest frame.
The polarization fractions thus determine the angular distribution of
the lepton in the \PW\  rest frame and, together with the Lorentz boosts,
control the \pt distributions of the lepton and the neutrino in the laboratory
frame.

The \PW\  polarization fractions in top-quark decays have been calculated~\cite{ref:Czarnecki}
with QCD corrections to next-to-next-to-leading order (NNLO), and the polarization is predominantly longitudinal.
For \cPqt $\to$ \cPqb \PWp\ these fractions are
$f_{0}=0.687\pm0.005$, $f_{-1}=0.311\pm0.005$,
and $f_{+1}=0.0017\pm0.0001$.
These precise calculations reduce the uncertainties
associated with the \PW\  polarization in \cPqt\cPaqt\ events to a low level.
The theoretical values are consistent with measurements from ATLAS~\cite{Aad:2012ky},
which obtained $f_{0}=0.67\pm 0.03 \pm 0.06$, $f_{-1}=0.32\pm 0.02\pm 0.03$, and
$f_{+1}=0.01\pm 0.01\pm 0.04$, expressed for the \PWp\ polarizations.

The \PW\  polarization in \PW+jets events exhibits a more complex behavior than
that in \cPqt\cPaqt\ production. Both CMS~\cite{Wpol-PRL}
and ATLAS~\cite{ATLAS:2012au} have reported measurements of these
effects, which are consistent with
\ALPGEN~\cite{Mangano:2002ea} and \MADGRAPH~\cite{madgraph} simulations predicting that
the \PWp\  and \PWm\ bosons are both
predominantly left-handed in \PW +jets events at high \pt .
An NLO QCD calculation~\cite{ref:Blackhat} has demonstrated that
the predicted polarization fractions are stable with respect to QCD
corrections. As discussed in later sections, this detailed knowledge
of the \PW-boson polarization provides key information for
measuring the SM backgrounds using control samples in data.

\section{Event preselection}
\label{sec:EventSelection}

Table~\ref{tab:Preselection} summarizes the main variables
and requirements used in the event preselection, which is
designed to be simple and robust. Except where noted, a
common set of preselection requirements is used by each of the
three analysis methods. Events are required to have at least one good
reconstructed primary vertex, at least three
jets (\Lp method and ANN method) or four jets (LS method),
and exactly one isolated muon or exactly one isolated electron.
These basic requirements select an event sample that is dominated by genuine, single-lepton
events from SM processes.

The primary vertex must satisfy a set of quality requirements,
including $|z_{\rm PV}|<24$ cm and $\rho_{\rm PV}<2$ cm, where $z_{\rm PV}$ and $\rho_{\rm PV}$ are
the longitudinal and transverse distances of the primary vertex
with respect to the nominal interaction point in the CMS detector.

Jets are reconstructed offline using the anti-$k_{\mathrm T}$ clustering
algorithm~\cite{ref:antikt} with a distance parameter of 0.5. The
particle four-vectors reconstructed by the CMS particle-flow algorithm~\cite{ref:PAS-PFT-09-001,ref:PAS-PFT-10-002},
are used as inputs to the jet clustering algorithm. The particle-flow algorithm
combines information from all CMS sub-detectors to provide a complete list of
long-lived particles in the event. Corrections based on simulation are applied to the jet energies
to establish a uniform response across the detector and a first
approximation to the absolute energy scale~\cite{JES}.
Additional jet energy corrections are applied to the data using measurements of energy
balance in dijet and photon + jet control samples in data. These additional corrections take into
account residual differences between the jet energy scale in data and simulation.
The effect of pileup was significant during much of the data-taking period.
Extra energy clustered into jets due to pileup is taken into account with an event-by-event
correction to the jet momentum four-vectors. Jet candidates are required to satisfy quality
criteria that suppress noise and spurious energy deposits in the calorimeters.
The performance of jet reconstruction and the corrections are described in
Ref.~\cite{JES}. In this analysis, reconstructed jets are required to satisfy $\pt>40$\GeV and $|\eta|< 2.4$.
The $\ETslash$ vector is defined as the negative of the vector sum of the transverse momenta of all
the particles reconstructed and identified by the particle-flow algorithm.

\begin{table}[tb!]
\begin{center}
\topcaption{Main preselection requirements. The term lepton designates either an electron or a muon.
Definitions of the quantities and further details are given in the text.}
\label{tab:Preselection}
\begin{tabular}{ll}
\hline
Quantity                &     Requirement \\
\hline
Primary vertex position &     $\rho_{\mathrm{PV}}<2$ cm, $|z_{\mathrm{PV}}|<24$ cm \\
Jet \pt threshold       &     $>40$\GeV    \\
Jet $\eta$ range        &     $|\eta| <2.4$ \\
Number of jets          &     $\ge 3$ ($\Lp$ and ANN methods), \\
                        &     $\ge 4$ (LS method)\\
Lepton \pt threshold    &     $>20$\GeV    \\
Muon $\eta$ range       &     $|\eta|<2.1$ \\
Muon isolation (relative) & $<0.10$ \\
Electron $\eta$ range         & $|\eta|<1.442$, $1.56<|\eta|<2.4$ \\
Electron isolation (relative) & $<0.07$ (barrel)\\
                              & $<0.06$ (endcaps) \\
Lepton \pt thresh. for veto & $>15$\GeV \\
\hline
\end{tabular}
\end{center}
\end{table}

In the muon channel, the preselection requires
a single muon candidate~\cite{Chatrchyan:2012xi} satisfying
$\pt(\mu)>20$\GeV and $|\eta|<2.1$.
Several requirements are imposed on the elements that form the muon candidate.
The reconstructed track must satisfy quality criteria related to the number of hits
in the pixel, strip, and muon detectors, and it must have
an impact parameter $d_0$ in the transverse plane with respect to the beam spot
satisfying $|d_0|<0.02$\unit{cm} and an impact parameter $d_z$ with respect to the
primary vertex along the $z$ direction satisfying
$|d_z|<1.0$~cm.

To suppress background in which the muon
originates from a semileptonic decay of a hadron containing a bottom or charm quark, we
require that the muon candidate be spatially isolated from other energy in the event. A
cone of size $\Delta R=0.3$ is constructed around the initial muon momentum
direction in $\eta$-$\phi$ space. The muon combined isolation variable,
$I^\text{comb}=\sum_{\Delta R<0.3}(\et + \pt)$, is defined as the sum
of the transverse energy \et (as measured in the electromagnetic and hadron
calorimeters) and the transverse
momentum \pt  (as measured in the silicon tracker) of all reconstructed objects
within this cone, excluding the muon.
This quantity is used to compute
the combined isolation relative to the muon transverse momentum,
$I^\text{comb}_\text{rel}=I^\text{comb}/\pt(\mu),$
which is required to satisfy
$I^\text{comb}_\text{rel}<0.1$.

Electron candidates~\cite{ref:PAS-EGM-10-004} are reconstructed by matching
energy clusters in the ECAL with tracks in the silicon tracking system.
Candidates must satisfy $\pt>20$\GeV and $|\eta|<2.4$, excluding the barrel-endcap transition
region ($1.442<|\eta|<1.56$). Quality and photon-conversion rejection requirements
are also imposed. The relative isolation variable, defined in a manner similar to that in
the muon channel, must satisfy $I^\text{comb}_\text{rel}<0.07$ in the barrel region and $I^\text{comb}_\text{rel}<0.06$ in the
endcaps. The requirements on $d_0$ and $d_z$ are the same as those used in the muon channel.

The preselection requirements have a large effect on the sample composition.
The lepton isolation requirement is
critical for the rejection of QCD multijet processes, which have very large
cross sections but are reduced to a low level by the isolation
and the other preselection requirements.
While many lepton candidates are produced in the semileptonic
decays of hadrons containing \cPqb\  or \cPqc\  quarks, from $\pi$ and \cPK\ decays in flight, and
from misidentification of hadrons, the vast majority
of these candidates are either within or near hadronic jets.
The background from  \PW +jets events (primarily from
$\PW\to\Pe\nu$ or $\PW\to\mu\nu$, but also $\PW\to\tau\nu$) is initially also very large.
This contribution is heavily suppressed by the three- or four-jet requirement. Depending on
the particular signal region, either \cPqt\cPaqt\ or \PW+jets production
emerges as the largest contribution to the background in the sample of events with
moderate to large $\ETslash$.

Events with a second isolated-lepton candidate satisfying the criteria listed in
Table~\ref{tab:Preselection}
are vetoed. This requirement not only
suppresses SM background, but also minimizes the statistical overlap between the
event sample used in this search and those used in multilepton searches.
However, \cPqt\cPaqt\ events with dileptons
can still be present, and this contribution must be determined, particularly because
the presence of two neutrinos in the decay chains can result in large values of \ETslash. The background involving
\PW~$\to \tau\nu$ decays, both from \cPqt\cPaqt\ events and from direct \PW\  production, must
also be determined. To help suppress the dilepton
background, the requirements on the veto leptons are somewhat looser than those on the
signal lepton. For both muons and electrons,
the \pt threshold is $\pt>15$\GeV, the
isolation requirement is $I^\text{comb}_\text{rel}<0.15$, and the impact parameter requirement is
$|d_0|<$0.1\unit{cm} (the $d_z$ requirement is kept the same as for the signal lepton). In addition, some of the quality requirements
for both the muon and electron are loosened.

Further event selection requirements are used in the individual
background estimation methods described in Sections~\ref{sec:LeptonSpectrumMethod},
\ref{sec:LPMethod}, and \ref{sec:ANN}.
The methods use the quantity \HT, which
is defined as the scalar sum of the transverse momenta of particle-flow jets $j$ with
$\pt^j>40$\GeV and $|\eta^j|<2.4$,
\begin{eqnarray}
\HT = \sum_{j}\pt^{j}.
\label{eq:HTdef}
\end{eqnarray}

The three background determination methods presented in the following three sections
use different approaches to estimating the SM backgrounds using control samples
in data.
In Section~\ref{sec:Results}, we compare the results of the
different methods and make some observations about their features.

\section{Lepton Spectrum method}
\label{sec:LeptonSpectrumMethod}

\subsection{Overview of the Lepton Spectrum method}
\label{ssec:LeptonSpectrumOverview}

This section describes
the Lepton Spectrum (LS) method, which is named for
the technique used to determine the dominant background source:
genuine, single-lepton processes.
Such processes account for about 75\%
of the total SM background in the signal regions and arise
primarily from \cPqt\cPaqt, single-top, and \PW+jets events.
Their contribution to the $\ETslash$ distribution is
estimated by exploiting the fact that, when the
lepton is produced in \PW-boson decay, the $\ETslash$
distribution is fundamentally related
to the lepton \pt spectrum, unlike the $\ETslash$ for many
SUSY models. A more detailed description
of the Lepton Spectrum method is given in the
references~\cite{Chatrchyan:2011qs,ref:Pavlunin}.

Non-single-lepton backgrounds are also determined
using control samples in the data. Such events
arise mainly from
(i) \cPqt\cPaqt\ dilepton events, in which
zero, one, or both of the leptons is a $\tau$ and
(ii) \cPqt\cPaqt\ and \PW+jets events with a single
$\tau\to(\mu,\Pe)$ decay. Background from QCD multijet events
is expected from simulation to be very small. However, the uncertainties
in such simulations are difficult to quantify, because
the QCD multijet background in the phase space relevant to this analysis
arises from extreme tails of processes
with very large cross sections. We therefore use control
samples in data to measure the QCD multijet background.
Simulated event samples are used for the determination
of the \cPZ+jets contribution, which is estimated with
sufficient precision to be below one event for most of
the signal regions.

The signal regions are defined with three thresholds in $\HT$ ($\HT\ge 500$\GeV,
$\HT\ge 750$\GeV, and $\HT\ge 1000$\GeV) and four bins in $\ETslash$
($250\le\ETslash<350$\GeV, $350\le\ETslash<450$\GeV, $450\le \ETslash< 550$\GeV,
and $\ETslash>550$\GeV).

\subsection{Estimation of single-lepton backgrounds}
\label{ssec:SingleLeptonBackgroundsLSMethod}

The physical foundation of the Lepton Spectrum method is that,
when the lepton and neutrino are produced together in two-body
\PW\  decay  (either in \cPqt\cPaqt\ or in \PW+jets events), the
lepton \pt spectrum is directly related to the $\ETslash$ spectrum.
The lepton and the neutrino share a common Lorentz boost from the
\PW~rest frame to the laboratory frame.
As a consequence, the lepton spectrum reflects the \pt distribution of
the \PW, regardless of whether the lepton was produced in a top-quark decay or in
a \PW+jets event. With suitable corrections, discussed below,
the lepton \pt spectrum can therefore be used to predict the $\ETslash$
spectrum for SM single-lepton backgrounds.

\begin{figure}[tbp!]
\begin{center}
\includegraphics[width=0.48\textwidth]{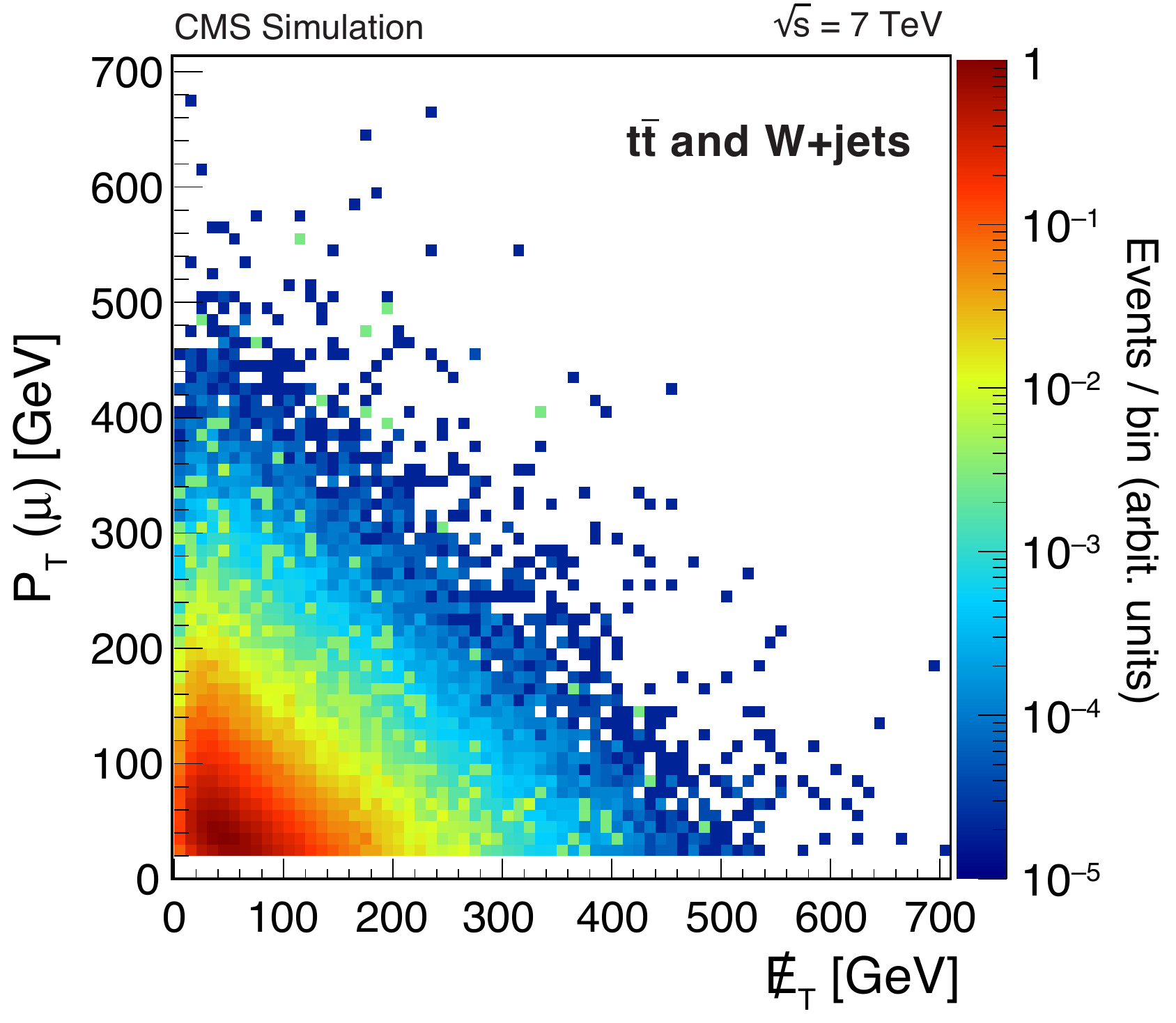}
\includegraphics[width=0.48\textwidth]{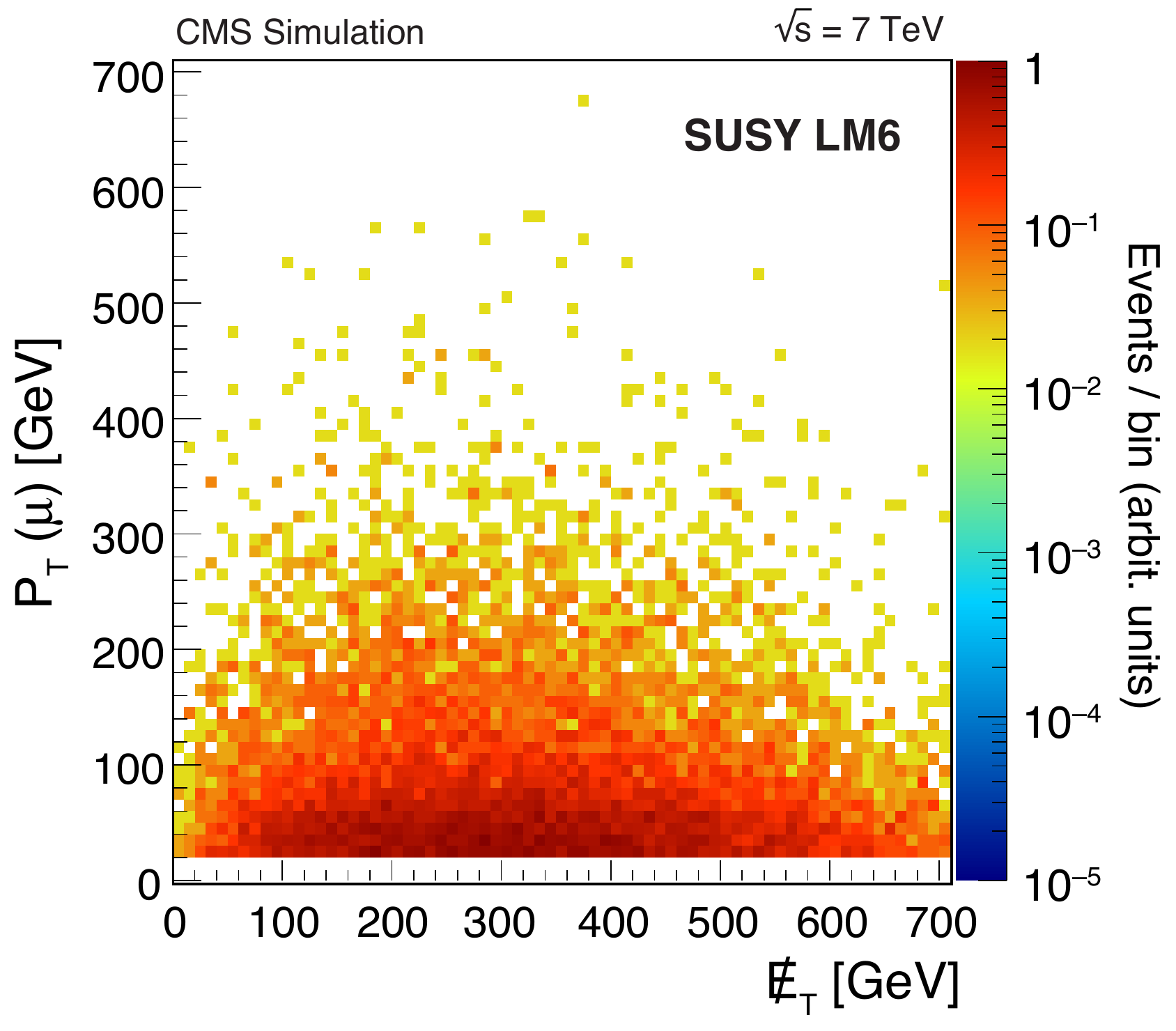}
\end{center}
\caption{Distributions of muon \pt  vs.~$\ETslash$ in the $\mu$ channel for
simulated \cPqt\cPaqt\ and \PW+jets events (\cmsLeft) and for the LM6 SUSY benchmark model (\cmsRight).
}\label{fig:2D_pTvsMET}
\end{figure}

The $\ETslash$ distribution in many SUSY models is dominated by
the presence of two LSPs. In contrast to the SM backgrounds,
the $\ETslash$ and lepton \pt distributions in SUSY processes are therefore nearly
decoupled. The $\ETslash$ distribution for such models
extends to far higher values than the lepton spectrum. Figure~\ref{fig:2D_pTvsMET}
shows the relationship between the lepton-\pt  and $\ETslash$ distributions
in the laboratory frame for two simulated event samples:
(i) the predicted SM mixture of \cPqt\cPaqt\ and \PW+jets events and (ii)
the SUSY LM6 benchmark model. When taken from data, the upper-left region in
Fig.~\ref{fig:2D_pTvsMET} (\cmsLeft) provides the key control sample
of high-\pt leptons from SM processes. This region typically has very little
contamination from SUSY events, which populate the high-$\ETslash$ region but
have relatively low lepton \pt values.

The lepton $\pt$ spectrum is measured with a muon control sample
defined by the preselection criteria and the $\HT$ requirements.
Unlike the signal region, no $\ETslash$ requirement is applied, because even
a modest one ($\ETslash>25$\GeV) would bias the high end of the lepton \pt
spectrum, which is critical for making the background prediction.
Only muon events are used as a control sample, because the QCD multijet background is
significant in the low-$\ETslash$ region of the electron sample.
The number of events that are common to both the control sample and the signal region
is small. For example, the overlap as measured in simulated
\cPqt\cPaqt\ events is 3.6\% for $\HT\ge750$\GeV, $\ETslash\ge250$\GeV,
and $\pt\ge250$\GeV. Because no \ETslash requirement is placed on the
muon control sample, a small amount of QCD background
remains and must be measured and subtracted. The scaling from the
muon to the electron samples is obtained by fitting their ratio in
the data over the range $60 \le \ETslash \le 250$\GeV, with systematic uncertainties
evaluated by varying the fit range. The resulting correction factor is
$N(\Pe)/N(\mu)=0.88\pm 0.03\pm0.03$, where the uncertainties are statistical and
systematic, respectively.

To use the lepton spectrum to predict the $\ETslash$ spectrum in single-lepton
SM background processes, three main issues must be understood:
(i) the effect of the \PW-boson polarization in both \cPqt\cPaqt\ and \PW+jets events,
(ii) the effect of the applied lepton \pt threshold, and
(iii) the difference between the experimental resolutions on the
measurements of lepton \pt  and $\ETslash$.

The status of theoretical and experimental knowledge
of \PW-boson polarization in \cPqt\cPaqt\ and in \PW+jets events
is discussed in Section~\ref{sec:EventSamples}. The helicity zero
polarization state results in a forward-backward symmetric
angular distribution of the lepton and the neutrino in the
\PW\ rest frame (with respect to the \PW\ momentum direction),
leading to identical lepton and neutrino spectra in the laboratory
frame. In contrast, the helicity $\pm 1$ states result
in angular asymmetries that lead to somewhat different lepton and neutrino
\pt spectra in the laboratory frame. These effects are taken
into account by applying correction factors obtained
from simulation to the measured lepton spectrum,
with uncertainties as described in Section~\ref{sec:SystematicUncertainties}.

The second key issue in the Lepton Spectrum method is
the effect of the threshold ($\pt>20$\GeV) applied to the leptons in both the
signal and control samples.
Because of the anticorrelation between the lepton \pt and the $\ETslash$
arising from non-zero \PW-boson helicity states, the threshold
requirement removes SM background events in the high-$\ETslash$ signal region
but not the events in the control sample with high-\pt muons that
are used to predict the high tail of the $\ETslash$ spectrum.
For the \cPqt\cPaqt\ background, this effect
partially compensates for the bias from the \PW\  polarization. For \PW +jets events,
in contrast, the polarization effects for \PWp\  and \PWm\ approximately cancel, but the
lepton \pt  threshold shifts the predicted yield upward. Correction
factors from simulation are used to account for these effects
(as well as for polarization effects), which are well defined and understood.

Finally, the resolution on the reconstructed $\ETslash$ is poorer than that for the lepton \pt ,
so the $\ETslash$ spectrum is somewhat broadened with respect to the
prediction from the lepton spectrum.
We measure $\ETslash$ resolution functions
in the data using QCD multijet events obtained with a set of single-jet
triggers spanning the range from $\ET\ge 30$\GeV to $\ET\ge 370$\GeV. These
resolution functions, or templates,
quantify the \ETslash resolution as a function of the number of jets and the \HT
of the event. These templates are used
to smear the measured lepton momenta. Because the
templates are taken from data, they include not only the intrinsic
detector resolutions, but also acceptance effects.
The overall effect of the smearing is modest, changing the background prediction by
5--15\%, depending on the $\ETslash$ threshold applied.

The raw background predictions for the single-lepton background
are corrected to account for the effects described above, as well as
for the small contamination of the single-lepton control sample
arising from dilepton and single-$\tau$ events
with high-\pt  leptons. These backgrounds are measured separately,
as described below. The overall correction factor is defined such that the
single-lepton prediction in a given signal region in simulation matches
the yield from single-lepton processes.

The predicted single-lepton background yield varies from about 150 events
for the signal region with $250\le\ETslash<350$\GeV and $\HT\ge500$\GeV to about 3 events
for the region with $\ETslash\ge 550$\GeV and $\HT\ge 1000$\GeV. These
predictions, as well as the expectations from simulation, are presented in
Tables~\ref{tab:table-ht500},
\ref{tab:table-ht750}, and \ref{tab:table-ht1000} and discussed in
more detail in Section~\ref{ssec:ResultsLSMethod}.

\begin{table*}[tbp!]
\begin{center}
\topcaption{Event yields for the Lepton Spectrum method for $\HT\ge 500$\GeV. The upper part of the table
gives the background predictions that are based on simulated (MC) event samples and the yield for the SUSY signal
points LM3 and LM6. The lower part gives
the backgrounds predicted using control samples in the data (data-driven prediction).
The actual yield observed in data is given at the bottom, with
the separate muon and electron yields given in parentheses $(N_{\mu},N_{\rm e})$ after the total yield.
The uncertainties on the background predictions are statistical and systematic.
The MC yields are not used in setting limits and are included only for reference.
The uncertainties on the MC yields are statistical only.}\label{tab:table-ht500}
{\small
\begin{tabular}{lcccc}
\hline
  \ETslash range [\GeVns{}] & [250, 350) & [350, 450) &  [450, 550) & $\ge$550 \\
\hline
\multicolumn{5}{c}{MC yields}            \\[1ex]

 1 $\ell$             & 146.7 $\pm$ 2.1 & 34.8 $\pm$ 1.1 & 8.5 $\pm$ 0.6  & 2.9 $\pm$ 0.3  \\
 Dilepton             & 19.9 $\pm$ 0.5  & 3.8 $\pm$ 0.2  & 0.7 $\pm$ 0.1  & 0.3 $\pm$ 0.1  \\
 1 $\tau$             & 30.6 $\pm$ 0.9  & 7.9 $\pm$ 0.5  & 2.1 $\pm$ 0.3 & 0.8 $\pm$ 0.2 \\
 Z+jets               & 1.3 $\pm$ 0.8   & $<$ 0.1  & $<$ 0.1  & $<$ 0.1  \\[1ex]

 Total (MC)          & 198.6 $\pm$ 2.5 & 46.5 $\pm$ 1.2 & 11.3 $\pm$ 0.6 & 4.0 $\pm$ 0.4 \\
 SUSY LM3 (MC)       & 266.3 $\pm$ 3.7   & 91.0 $\pm$ 2.2 & 23.3 $\pm$ 1.1  & 9.9 $\pm$ 0.7 \\
 SUSY LM6 (MC)       & $23.4\pm0.3$    & $20.0\pm0.3$   & $13.4\pm0.2$   & $10.8\pm0.2$ \\
\hline
\multicolumn{5}{c}{Data-driven prediction} \\[1ex]

 1 $\ell$ & 109 $\pm$ 13 $\pm$ 18 & 32.0 $\pm$ 7.5 $\pm$ 5.8 & 3.9 $\pm$ 2.7 $\pm$ 1.2 & 3.1 $\pm$ 2.3 $\pm$ 1.0 \\
 Dilepton & 15.8  $\pm$ 1.9 $\pm$ 1.8 & 3.0 $\pm$ 0.9 $\pm$ 0.5 & 0.5 $\pm$ 0.3 $\pm$ 0.2 & 0.1 $\pm$ 0.2 $\pm$ 0.2 \\
 1 $\tau$ & 33.0  $\pm$ 1.8 $\pm$ 1.7 & 8.9 $\pm$ 1.0 $\pm$ 0.5 & 2.1 $\pm$ 0.5 $\pm$ 0.2 & 1.1 $\pm$ 0.3 $\pm$ 0.2 \\
 QCD      & 0.0   $\pm$ 1.0 $\pm$ 1.2 & 0.0 $\pm$ 1.0 $\pm$ 1.2 & 0.0 $\pm$ 1.0 $\pm$ 1.2 & 0.0 $\pm$ 1.0 $\pm$ 1.2 \\
 Z+jets (MC)   & 1.3   $\pm$ 0.8 $\pm$ 1.3 & $<$ 0.1 & $<$ 0.1 & $<$ 0.1 \\[1ex]

 Total (predicted) & 159 $\pm$ 14 $\pm$ 18 & 44.0 $\pm$ 7.7 $\pm$ 6.0 & 6.6 $\pm$ 2.9 $\pm$ 1.7 & 4.3 $\pm$ 2.6 $\pm$ 1.6 \\

Data (observed) & 163~(84, 79) & {46}~(21, 25) & {9}~(8, 1) & {2}~(1, 1)\\
\hline
\end{tabular}
}
\end{center}
\end{table*}

\begin{table*}[tbp!]
\begin{center}
\topcaption{Event yields for the Lepton Spectrum method for $\HT\ge 750$\GeV.
Further details are given in the Table~\ref{tab:table-ht500} caption.}
\label{tab:table-ht750}
{\small
\begin{tabular}{lcccc}
\hline
\ETslash range [\GeVns{}] & [250, 350) & [350, 450) &  [450, 550) & $\ge$550 \\
\hline
\multicolumn{5}{c}{MC yield} \\[1ex]

 1 $\ell$ & 47.3 $\pm$ 1.2  & 14.9 $\pm$ 0.7  & 5.4 $\pm$ 0.4  & 2.7 $\pm$ 0.3  \\
 Dilepton & 8.2 $\pm$ 0.4  & 2.3 $\pm$ 0.2  & 0.6 $\pm$ 0.1  & 0.3 $\pm$ 0.1  \\
 1 $\tau$ & 9.2 $\pm$ 0.5  & 3.0 $\pm$ 0.3  & 1.2 $\pm$ 0.2  & 0.7 $\pm$ 0.2  \\
 Z+jets & 0.7 $\pm$ 0.6  & $<$ 0.1  & $<$ 0.1  & $<$ 0.1  \\[1ex]

Total (MC) & 65.4 $\pm$ 1.5  & 20.2 $\pm$ 0.8 & 7.2 $\pm$ 0.5  & 3.6 $\pm$ 0.4  \\
SUSY LM3 (MC) & 114.6 $\pm$ 2.5  & 47.1 $\pm$ 1.6  & 16.1 $\pm$ 0.9  & 8.6 $\pm$ 0.7  \\
SUSY LM6 (MC) & $14.9\pm0.3$  & $13.8\pm0.2$  & $10.3\pm0.2$ & $9.8\pm0.2$ \\
\hline
\multicolumn{5}{c}{Data-driven prediction}\\[1ex]

 1 $\ell$ & 41.7 $\pm$ 8.7 $\pm$ 5.4  & 11.7 $\pm$ 5.0 $\pm$ 1.9  & 2.6 $\pm$ 2.3 $\pm$ 0.6  & 3.1 $\pm$ 2.4 $\pm$ 0.8  \\
 Dilepton & 5.9 $\pm$ 1.1 $\pm$ 0.7  & 1.3 $\pm$ 0.5 $\pm$ 0.2  & 0.5 $\pm$ 0.2 $\pm$ 0.1  & 0.1 $\pm$ 0.1 $\pm$ 0.3  \\
 1 $\tau$ & 9.6 $\pm$ 0.9 $\pm$ 0.6  & 3.1 $\pm$ 0.6 $\pm$ 0.3  & 1.1 $\pm$ 0.3 $\pm$ 0.2  & 0.8 $\pm$ 0.2 $\pm$ 0.1  \\
 QCD & 0.0 $\pm$ 0.2 $\pm$ 0.4  & 0.0 $\pm$ 0.2 $\pm$ 0.4  & 0.0 $\pm$ 0.2 $\pm$ 0.4  & 0.0 $\pm$ 0.2 $\pm$ 0.4  \\
 Z+jets (MC) & 0.7 $\pm$ 0.6 $\pm$ 0.7  & $<$ 0.1  & $<$ 0.1  & $<$ 0.1  \\[1ex]

 Total (predicted) & 57.9 $\pm$ 8.9 $\pm$ 5.6  & 16.2 $\pm$ 5.0 $\pm$ 2.0  & 4.2 $\pm$ 2.4 $\pm$ 0.8  & 4.0 $\pm$ 2.4 $\pm$ 1.0  \\

Data (observed) & 48~(27, 21) & 16~(7, 9) & 5~(4, 1) & 2~(1, 1)\\
\hline
\end{tabular}
}
\end{center}
\end{table*}

\begin{table*}[tbp!]
\begin{center}
\topcaption{Event yields for the Lepton Spectrum method for $\HT>1000$\GeV.
Further details are given in the Table~\ref{tab:table-ht500} caption.}
\label{tab:table-ht1000}
{\small
\begin{tabular}{lcccc}
\hline
\ETslash range [\GeVns{}] & [250, 350) & [350, 450) &  [450, 550) & $\ge$550 \\
\hline
\multicolumn{5}{c}{MC yield}   \\[1ex]

 1 $\ell$ & 13.4 $\pm$ 0.6  & 4.8 $\pm$ 0.4  & 2.1 $\pm$ 0.3  & 1.3 $\pm$ 0.2  \\
 Dilepton & 2.7 $\pm$ 0.2  & 1.0 $\pm$ 0.1  & 0.3 $\pm$ 0.1  & 0.2 $\pm$ 0.1  \\
 1 $\tau$ & 2.1 $\pm$ 0.2  & 0.7 $\pm$ 0.1  & 0.5 $\pm$ 0.1  & 0.4 $\pm$ 0.1  \\
 Z+jets & 0.5 $\pm$ 0.5  & $<$ 0.1  & $<$ 0.1  & $<$ 0.1  \\[1ex]

 Total (MC)    & 18.8 $\pm$ 0.9  & 6.4 $\pm$ 0.5  & 2.9 $\pm$ 0.3  & 1.9 $\pm$ 0.2  \\
 SUSY LM3 (MC) & 38.1 $\pm$ 1.4  & 18.3 $\pm$ 1.0   & 7.0 $\pm$ 0.6  & 5.5 $\pm$ 0.5 \\
 SUSY LM6 (MC) & $7.0\pm 0.2$    & $6.0 \pm0.2$   & $4.6\pm0.1$    & $5.2\pm 0.2$ \\
\hline
\multicolumn{5}{c}{Data-driven prediction}   \\[1ex]

 1 $\ell$ & 11.7 $\pm$ 4.6 $\pm$ 1.8 & 5.5 $\pm$ 3.6 $\pm$ 1.0 & 2.0 $\pm$ 2.2 $\pm$ 0.6 & 3.1 $\pm$ 2.3 $\pm$ 1.0 \\
 Dilepton & 1.2 $\pm$ 0.6 $\pm$ 0.1 & 0.4 $\pm$ 0.4 $\pm$ 0.1 & 0.2 $\pm$ 0.2 $\pm$ 0.1 & 0.1 $\pm$ 0.2 $\pm$ 0.2 \\
 1 $\tau$ & 3.0 $\pm$ 0.5 $\pm$ 0.5 & 0.9 $\pm$ 0.3 $\pm$ 0.2 & 0.4 $\pm$ 0.2 $\pm$ 0.2 & 0.8 $\pm$ 0.2 $\pm$ 0.2 \\
 QCD & 0.0 $\pm$ 0.1 $\pm$ 0.1 & 0.0 $\pm$ 0.1 $\pm$ 0.1 & 0.0 $\pm$ 0.1 $\pm$ 0.1 & 0.0 $\pm$ 0.1 $\pm$ 0.1 \\
 Z+jets (MC) & 0.5 $\pm$ 0.5 $\pm$ 0.5 & $<$ 0.1 & $<$ 0.1 & $<$ 0.1  \\[1ex]

 Total (predicted) & 16.4 $\pm$ 4.7 $\pm$ 1.9 & 6.8 $\pm$ 3.6 $\pm$ 1.0 & 2.6 $\pm$ 2.2 $\pm$ 0.6 & 4.0 $\pm$ 2.4 $\pm$ 1.0 \\

Data (observed) & 14~(7, 7) & 4~(1, 3) & 0~(0, 0) & 2~(1, 1) \\
\hline
\end{tabular}
}
\end{center}
\end{table*}

\subsection{Estimation of non-single-lepton backgrounds}
\label{ssec:NonSingleLeptonBackgroundsLSMethod}

The non-single-lepton backgrounds include dilepton events in several categories,
events with $\PW\to\tau\nu$ followed by $\tau\to\ell$ decays
(in either \cPqt\cPaqt\ or \PW +jets events), and QCD multijet processes. These subdominant backgrounds
are estimated using control samples in data, in conjunction with information from simulation.
The contribution from Drell-Yan and \cPZ+jets is very small and is
estimated directly from simulation.

Dilepton background events (including the $\tau$ as one of the leptons)
contain at least two neutrinos, so these events can be important in the
tails of the $\ETslash$ distributions. These backgrounds
are divided into the following categories:
(i) $2\ell$ events with one lost or ignored lepton ($\ell=\Pe, \mu$),
(ii) $\ell+\tau$ events with $\tau\to\text{hadrons}$, and
(iii) $\ell+\tau$ events with $\tau\to\text{lepton}$.
A lost lepton is one that is either not reconstructed or is out of the
detector acceptance. An ignored lepton is one that is reconstructed but
fails either the lepton-identification requirements (including isolation) or the \pt  threshold requirement.

To estimate the background from dilepton events with lost or ignored leptons,
we compute the ratio of the combined yield of dilepton events in the $\Pe\Pe$, $\Pe\mu$, and $\mu\mu$ channels in data
to the corresponding combined yield in simulated event samples. This ratio,
which is $0.91\pm0.07$ for $\HT\ge500$\GeV, $0.93\pm0.15$ for $\HT\ge750$\GeV, and
$0.87\pm0.37$ for $\HT\ge1000$\GeV,
is used to rescale
the $\ETslash$ distribution of dilepton events that appear in the signal
region in simulation. (Events within 20\GeV of the nominal \cPZ\ mass
are excluded in the $\Pep\Pem$ and $\mu^+\mu^-$ channels.)
This approach is used
because the dilepton control sample in data is small, and using it to obtain the
shapes of $\ETslash$ distributions would result in
large statistical uncertainties.
For all $\ETslash$ bins above 250\GeV, the predicted
yield from this background contribution is less than
6 events, and for all $\ETslash$ bins above 350\GeV, the yield is at or below 1 event.
The $\ETslash$ distribution associated with the reconstructed dilepton events in data
is well described by the simulation.

Dilepton events can also involve $\tau$ decays, either $\tau\to$ hadrons or $\tau\to\ell$.
The $\ETslash$ distributions in the dilepton events in data, when suitably modified
to reflect the presence of a leptonic or hadronic $\tau$ decay,
provide an accurate description of the shape of the $\ETslash$
distribution of these backgrounds. Thus, to estimate the shape from the
$\tau\to$ hadrons background, we effectively replace a lepton in a reconstructed
dilepton event with a hadronic $\tau$ jet. Both hadronic and leptonic $\tau$
response functions are used, providing a probability distribution for a
$\tau$ to produce a jet or a lepton with a given fraction $\pt(\text{jet})/\pt(\tau)$
or $\pt({\ell})/\pt(\tau)$.
These response functions, obtained from simulation, are computed in bins of $\pt(\tau)$.
This procedure can change the total
number of jets above threshold in the event, as well as other properties
such as $\HT$ and $\ETslash$, which are recalculated.
Simulated event samples are used to determine,
for each of these processes $i$, the ratio $r_i=N_\text{feed}^i/N_\text{control}$
of the number of events observed in the single-lepton channel to the number of events
in the control sample, as a function of $\ETslash$. This procedure effectively
normalizes all such contributions to the control samples in data.
For all $\ETslash$ bins above 250\GeV, the number of dilepton events with a
$\tau\to\text{hadrons}$ decay is predicted to be about 7 events or less and
is much smaller in the higher $\ETslash$ bins. The number of dilepton events
with a $\tau\to\ell$ decay is predicted to be less than 3 events for all
$\ETslash$ bins above 250\GeV and is much smaller in the higher $\ETslash$ bins.

Estimates for the $\tau\to\ell$ single-lepton backgrounds from
\cPqt\cPaqt\ and \PW +jets processes are based on a procedure similar
to that used for the dilepton backgrounds, but in this case
the single-lepton sample itself is used as the control sample. The $\ETslash$
distribution obtained by applying the $\tau\to\ell$ response function
to the data is rescaled by a ratio from simulation that gives the
yield of $\tau\to\ell$ background events divided by the yield of events
in the single-lepton control sample, as a function of $\ETslash$.
The number of background events from the
single $\tau\to\ell$ contribution falls from 33 for $\HT\ge500$\GeV
and $250\le\ETslash<350$\GeV to 1.1 event for $\HT\ge500$\GeV and
$\ETslash\ge 550$\GeV.

\begin{figure}[tbp!]
 \begin{center}
\includegraphics[width=0.45\textwidth]{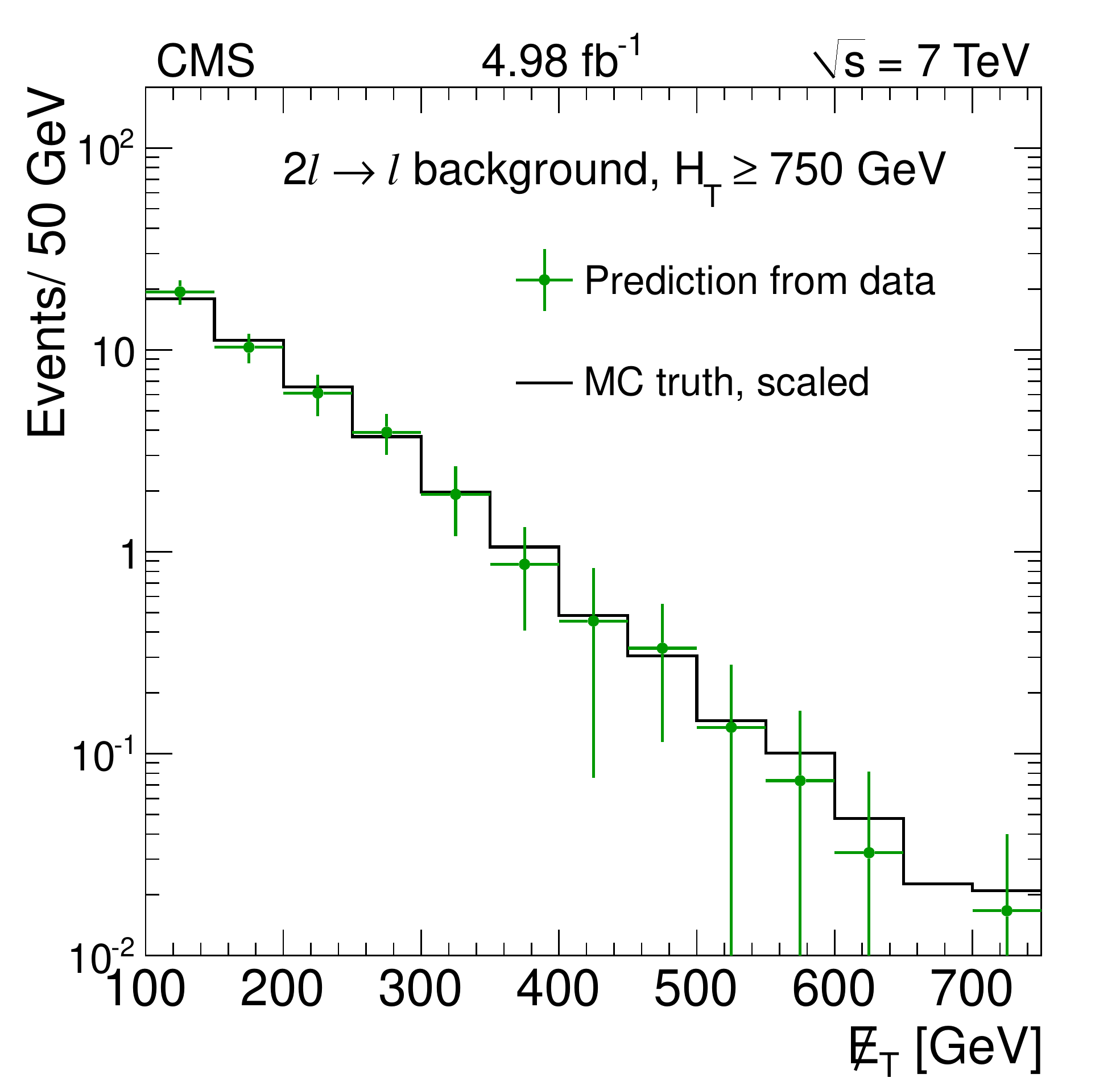}
\includegraphics[width=0.45\textwidth]{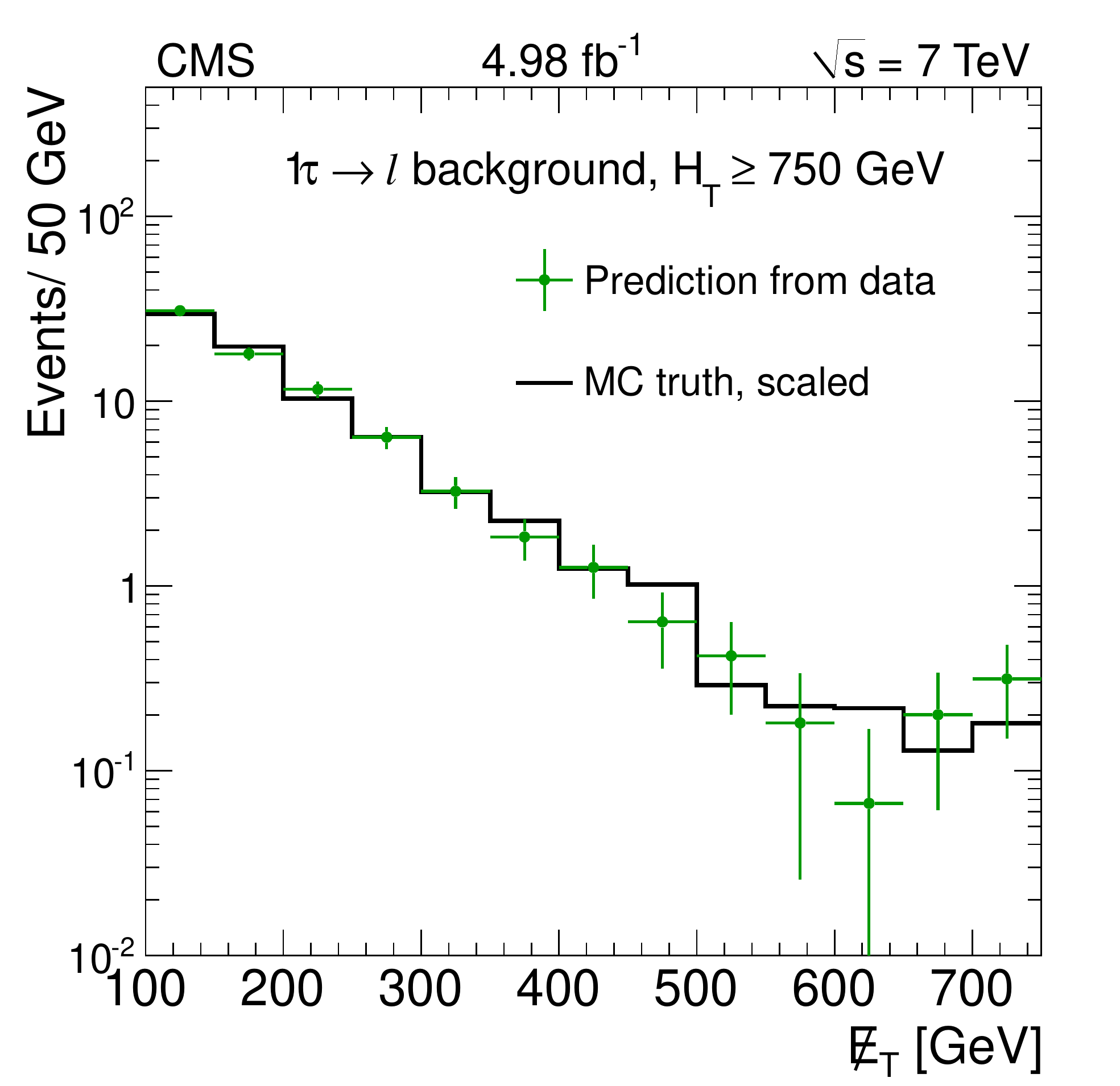}
 \end{center}
 \caption{Predictions for dilepton and $\tau\to\ell$ backgrounds after requiring $\HT\ge 750$\GeV:
control samples in data (green points with error bars) vs.~MC predictions (black solid histogram) for
(\cmsLeft) dilepton background and
(\cmsRight) $\tau\to\ell$ \,  background. The MC prediction has been scaled to the integral of the data prediction.
}
\label{fig:DileptonAndTauPredictionsData}
\end{figure}

The background predictions in data are shown in Fig.~\ref{fig:DileptonAndTauPredictionsData},
where the expectation based on simulation is shown for comparison.
The total predicted dilepton plus single $\tau\to\ell$ background
yield ranges from about 50 events for $\HT\ge 500$\GeV and $250\le\ETslash<350$\GeV
to about 1 event for $\HT\ge 1$\TeV and $\ETslash\ge 550$\GeV.
All of these predictions, as well as the expecations from simulation, are presented in
Tables~\ref{tab:table-ht500},
\ref{tab:table-ht750}, and \ref{tab:table-ht1000}, which are
discussed in more detail in Section~\ref{ssec:ResultsLSMethod}.

Background from QCD multijet events is suppressed to a level well below 1 event
in nearly all signal regions, as shown in Tables~\ref{tab:table-ht500},
\ref{tab:table-ht750}, and \ref{tab:table-ht1000}. The
QCD multijet background is determined by first defining a control
sample with small missing transverse momentum ($\ETslash<50$\GeV) and
with a lepton impact parameter relative to the beam spot $|d_0|>0.02$ cm. These
requirements select a sample with little contamination from other SM processes
such as $\cPqt\cPaqt$ and $\PW$+jets processes. Using this control sample,
we measure the shape of the distribution in the combined relative isolation variable, $I_\text{rel}^\text{comb}$
(see Section~\ref{sec:EventSelection}).
The shape of this distribution has very little correlation to $\ETslash$ or to the lepton
impact parameter ($d_0$), and so
can be applied in the high-$\ETslash$ signal regions. For each signal region in the
data, we determine the background at low values of $I_\text{rel}^\text{comb}$
by first scaling the measured QCD multijet background shape in the relative isolation variable
to the high-$I_\text{rel}^\text{comb}$
sideband of the signal region. The shape is then used to extrapolate the yield to the low-
$I_\text{rel}^\text{comb}$ signal region. In the high-$\ETslash$ signal regions,
some non-QCD SM background can be present at high $I_\text{rel}^\text{comb}$,
where the QCD background shape is normalized. We therefore subtract the
estimated background from \cPqt\cPaqt, \PW+jets, and \cPZ+jets from this
region. These yields are taken from simulation, with systematic uncertainties
determined from a comparison with a control region in the data.

\subsection{Results from the Lepton Spectrum method}
\label{ssec:ResultsLSMethod}

Tables~\ref{tab:table-ht500}, \ref{tab:table-ht750}, and \ref{tab:table-ht1000}
compare the background yields predicted
from the control samples in data with the yields obtained directly from simulation
for $\HT\ge 500$\GeV, $\HT\ge 750$\GeV, and $\HT\ge 1000$\GeV, respectively.
We observe that the single-lepton background is the dominant contribution in
all regions. The various sources of
uncertainties associated with these background determinations are
discussed in Section~\ref{sec:Results}.
Finally, the yields observed in the signal
regions in the data, which are listed at the bottom of each table, are
consistent with the total background predictions based on the control samples. Thus,
we observe no evidence for any excess of events in the data above the SM
contributions.

\begin{figure*}[p]
\centering
\includegraphics[width=0.40\textwidth]{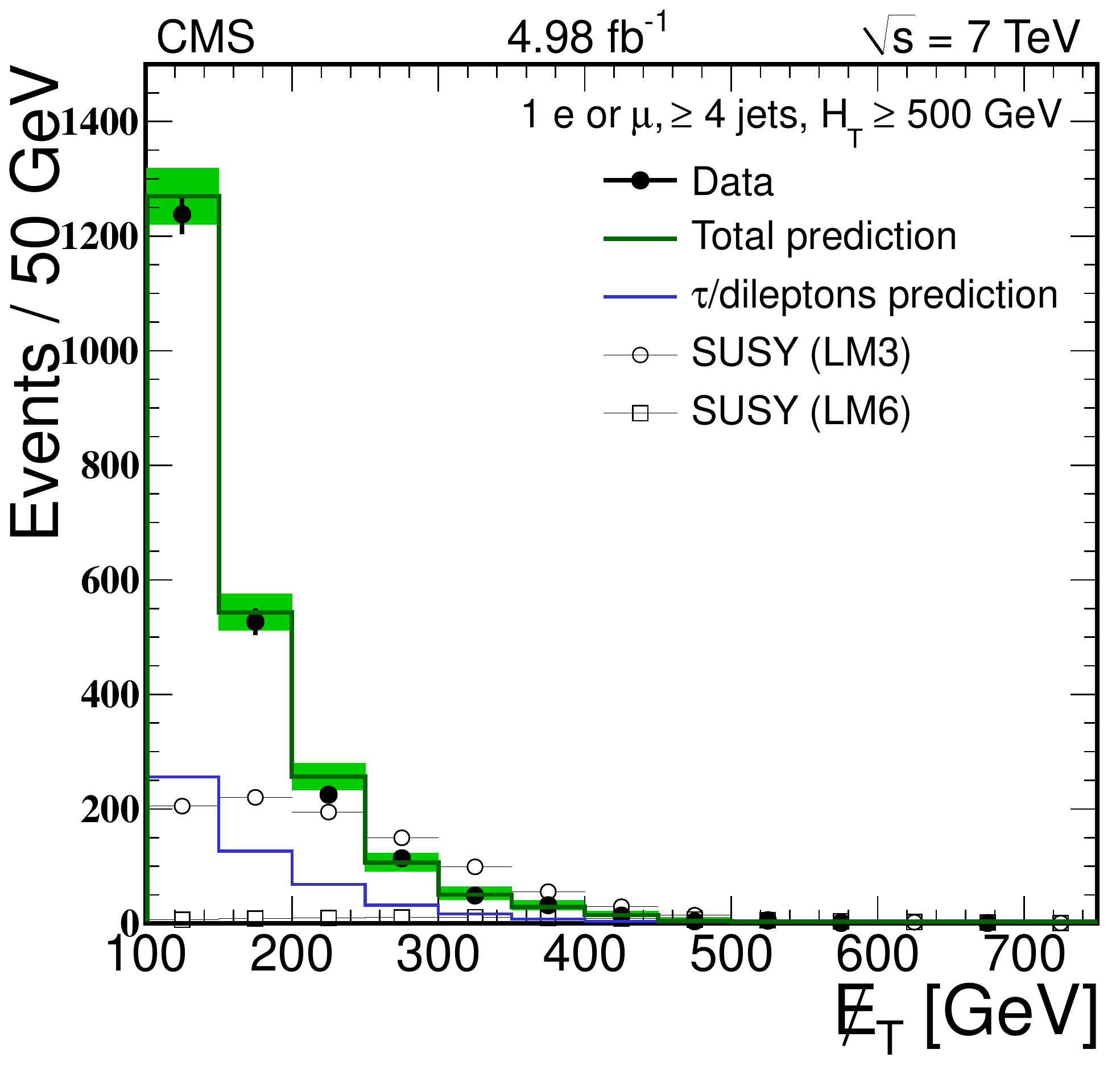}
\includegraphics[width=0.40\textwidth]{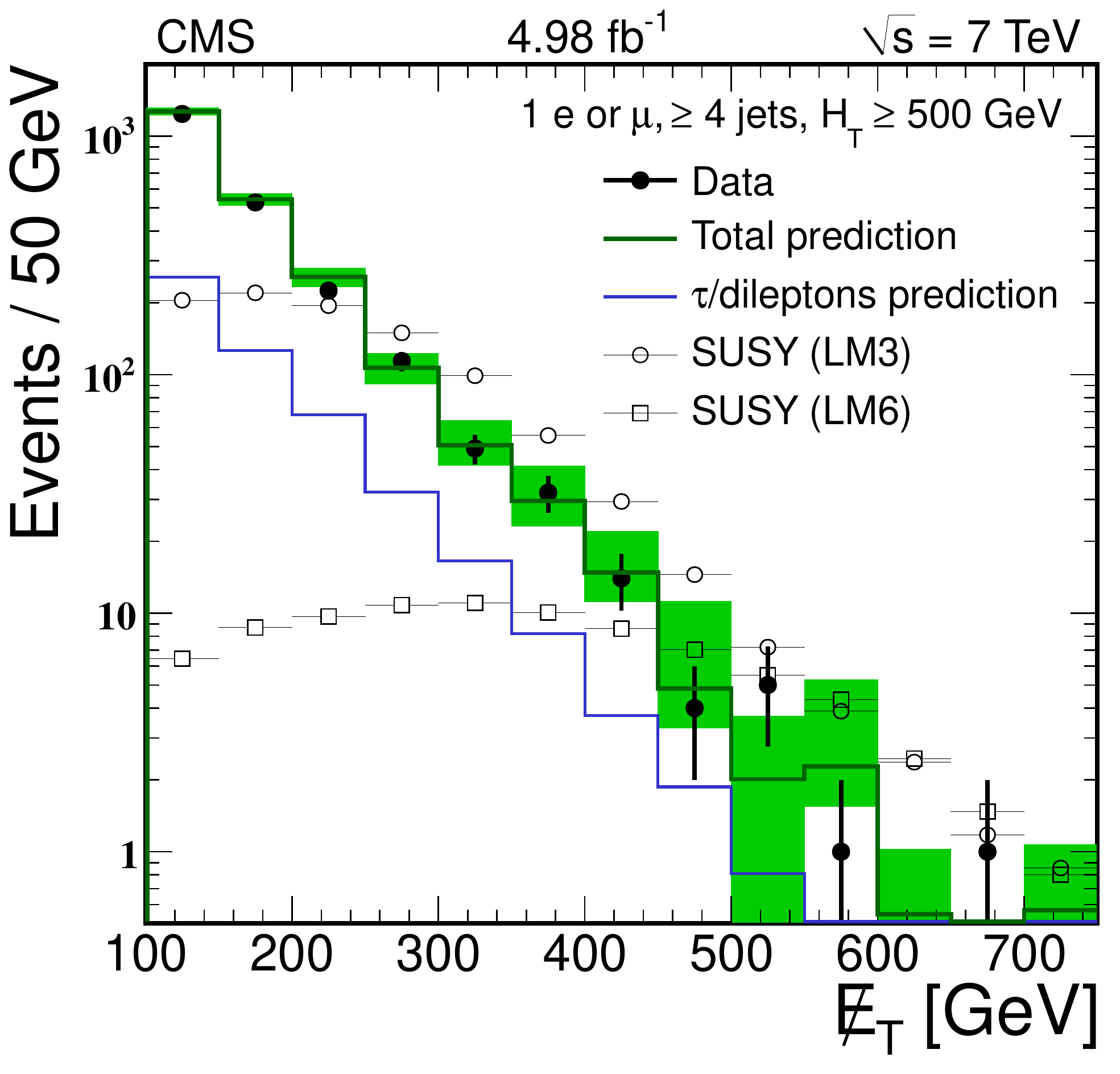}
\includegraphics[width=0.40\textwidth]{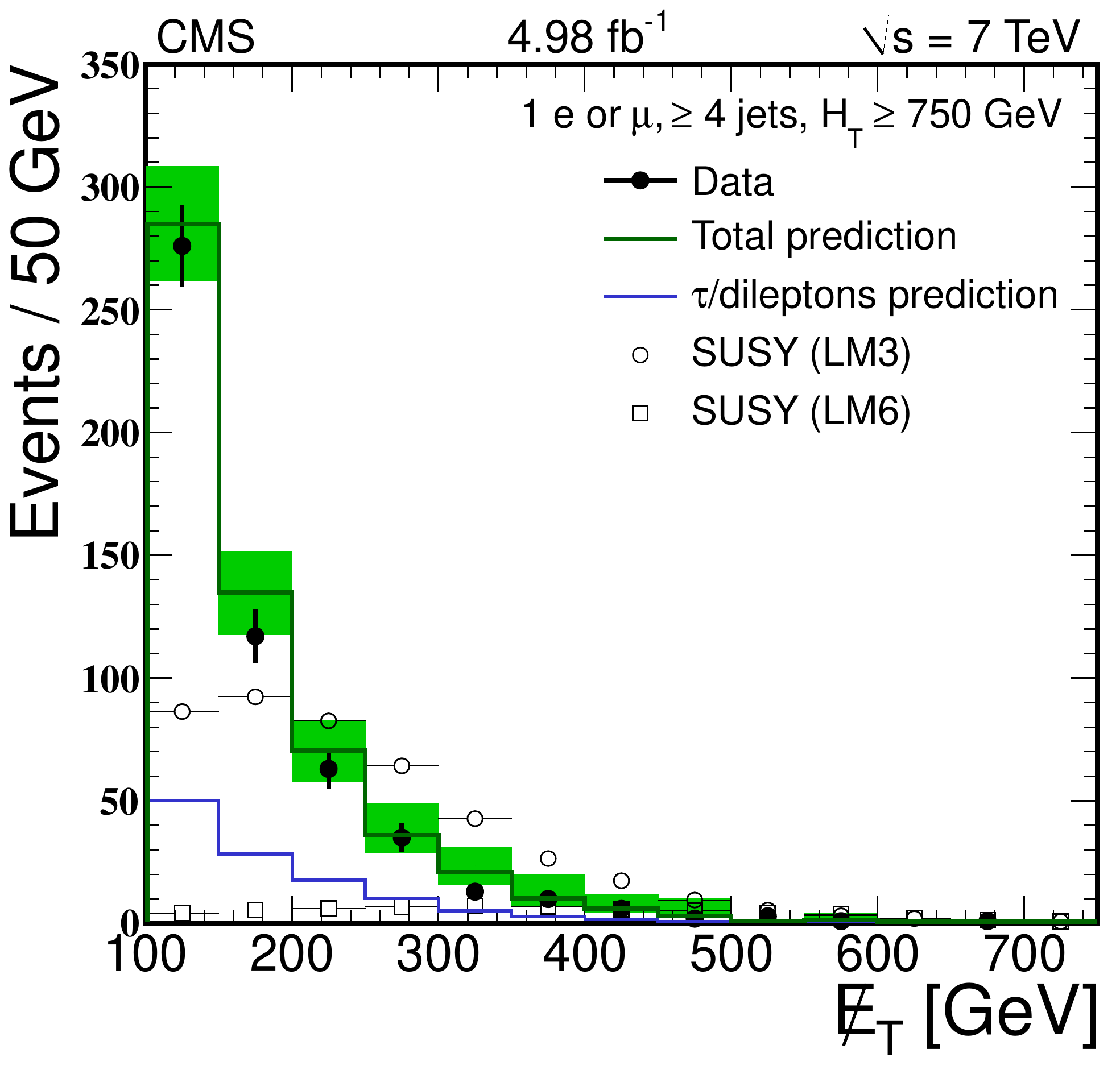}
\includegraphics[width=0.40\textwidth]{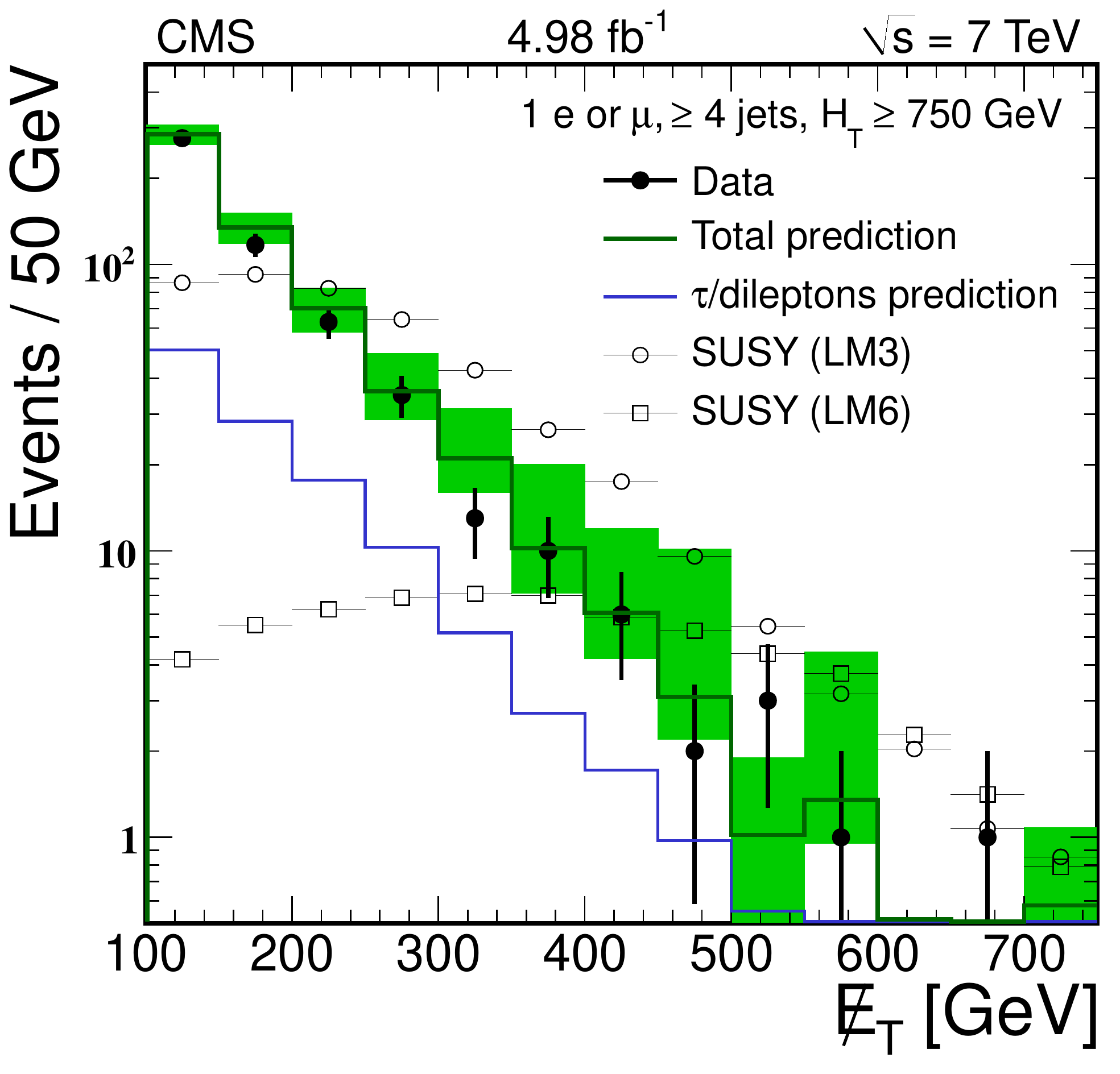}
\includegraphics[width=0.40\textwidth]{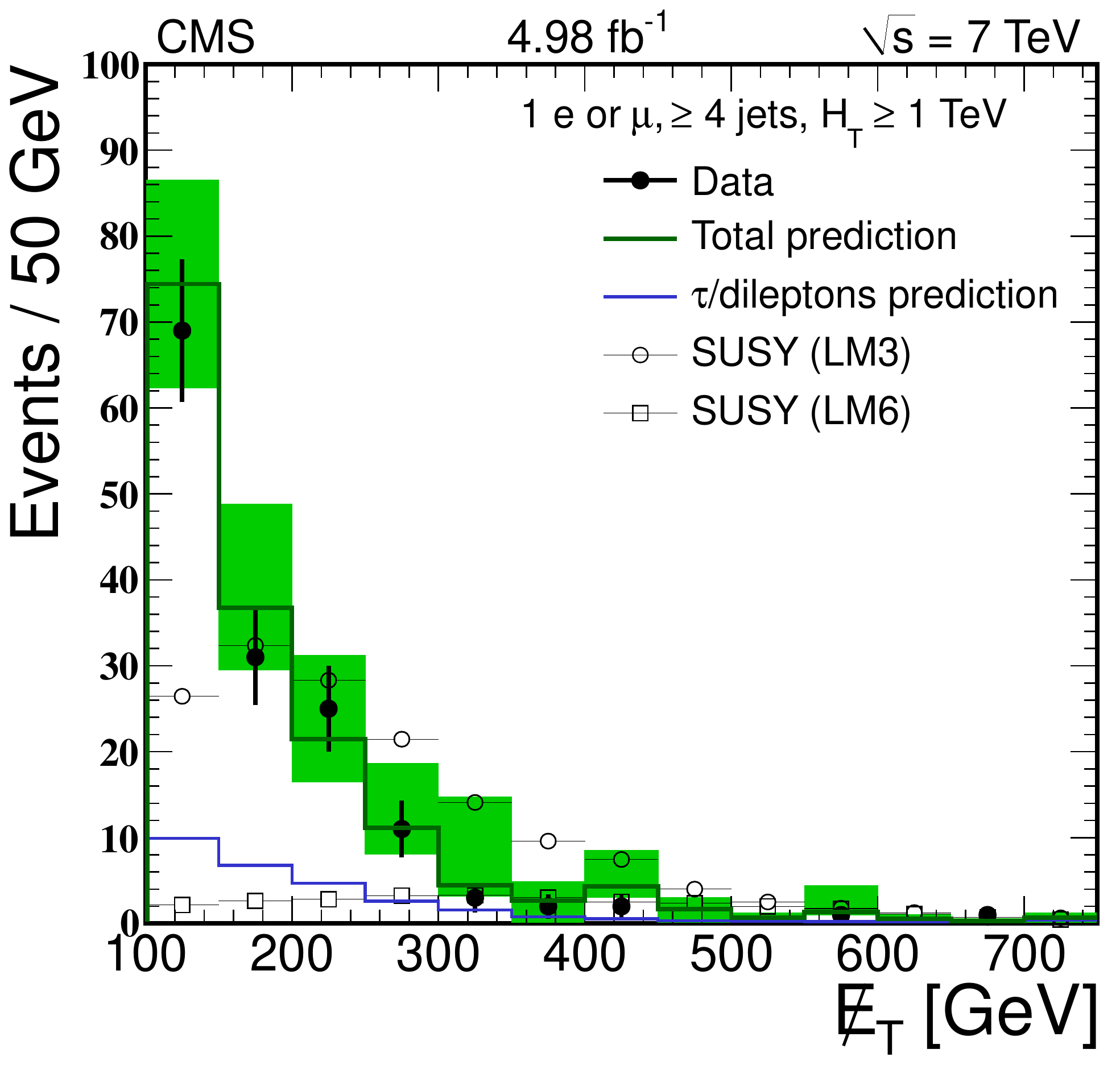}
\includegraphics[width=0.40\textwidth]{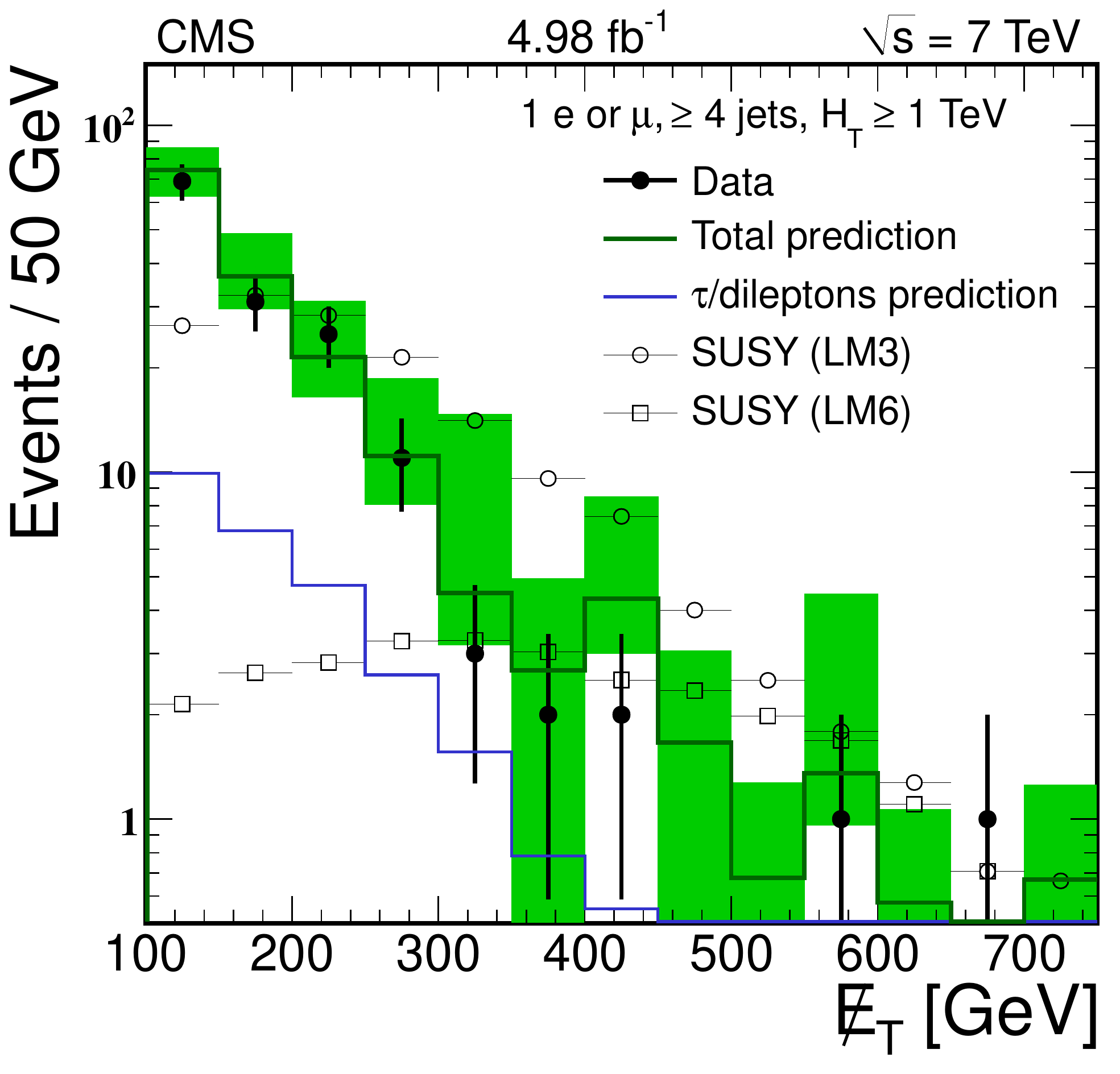}
\caption{ Lepton Spectrum method: observed \ETslash distributions in data (filled points with error bars) compared with predicted \ETslash distributions
(green bars) in the combined electron and muon channels, on linear (left) and logarithmic (right) scales.
Three different $\HT$ thresholds are applied: $\HT\ge 500$\GeV (upper row), $\HT\ge 750$\GeV (middle row), and
$\HT\ge 1000$\GeV (lower row).}
\label{fig:LeptonSpectrumResults}
\end{figure*}

Figure~\ref{fig:LeptonSpectrumResults} shows the $\ETslash$ distributions
in data for the combined muon and electron channels, with all of the
selection requirements, except that on $\ETslash$ itself. The distributions
are shown for $\HT\ge 500$\GeV, $\HT\ge 750$\GeV, and $\HT\ge 1000$\GeV, on both
linear and logarithmic scales. The predicted $\ETslash$ distribution (green-bar histogram)
is a sum over three sources:
single-lepton backgrounds (from \cPqt\cPaqt, single-top, and $\PW$+jets events), dilepton
background from \cPqt\cPaqt, and single-$\tau$ events (from both \cPqt\cPaqt\
and $\PW$+jets processes). The vertical span of the green bar corresponds to the
statistical uncertainty on the background prediction. (The systematic uncertainties
are computed in wider bins used for setting the limits and are given
in Tables~\ref{tab:table-ht500}, \ref{tab:table-ht750}, and \ref{tab:table-ht1000}.)
In each signal region, the blue histogram shows the contribution from
the dilepton and single-$\tau$ backgrounds only. It is evident that the
single-lepton background is dominant in all cases.
The $\ETslash$ distributions for the SUSY benchmark models LM3 and LM6 are
overlaid (not summed) for comparison. Systematic uncertainties
and the interpretation are presented in Section~\ref{sec:Results}.

\section{Lepton Projection method}
\label{sec:LPMethod}
\subsection{Overview of the Lepton Projection method}

The Lepton-Projection (LP) method uses the difference between SM and SUSY processes in the correlation of the lepton transverse
momentum and the missing transverse momentum.
As previously discussed, in the SM processes the \ETslash corresponds to the neutrino in the decay of
the W boson, either in W+jets or in \ttbar events. The kinematics of W decays are dictated by the
V$-$A nature of the W coupling to leptons and the helicity of the W boson, as discussed in
Section~\ref{sec:EventSamples}.  Since W bosons that are produced with high transverse momentum
in W+jets events exhibit a sizable left-handed polarization, there is a significant asymmetry in the
\pt spectra of the neutrino and charged lepton. A smaller asymmetry is expected in W bosons from
t quark ($\bar{\rm t}$ antiquark) decays, which yield W bosons which are predominantly longitudinally
polarized with smaller left-handed (right-handed) components for W$^+$ (W$^-$).

We have measured the fraction of the helicity states of the W boson using an angular
analysis of leptonic W decays~\cite{Wpol-PRL}. Since the total momentum of the W boson in these decays,
and therefore its center-of-mass frame, cannot be accurately determined because the momentum of the
neutrino along the beam axis cannot be measured, an observable that depends only on transverse quantities
is used. A variable that is highly correlated with the cosine of the polar angle in the center-of-mass frame
of the W boson is the ``lepton projection variable'':
\begin{equation}
\Lp = \frac{\ptvec(\ell) \cdot \ptvec({\PW})}{|\ptvec({\PW})|^2},
\end{equation}
where $\ptvec(\ell)$ is the transverse momentum of the charged lepton and
$\ptvec({\rm W})$ is the transverse momentum of the W boson.
The latter quantity is obtained from the vector sum of the electron transverse momentum
and the missing transverse momentum in the event.

\begin{figure}[htb]
\begin{center}
\includegraphics[width=.40\textwidth]{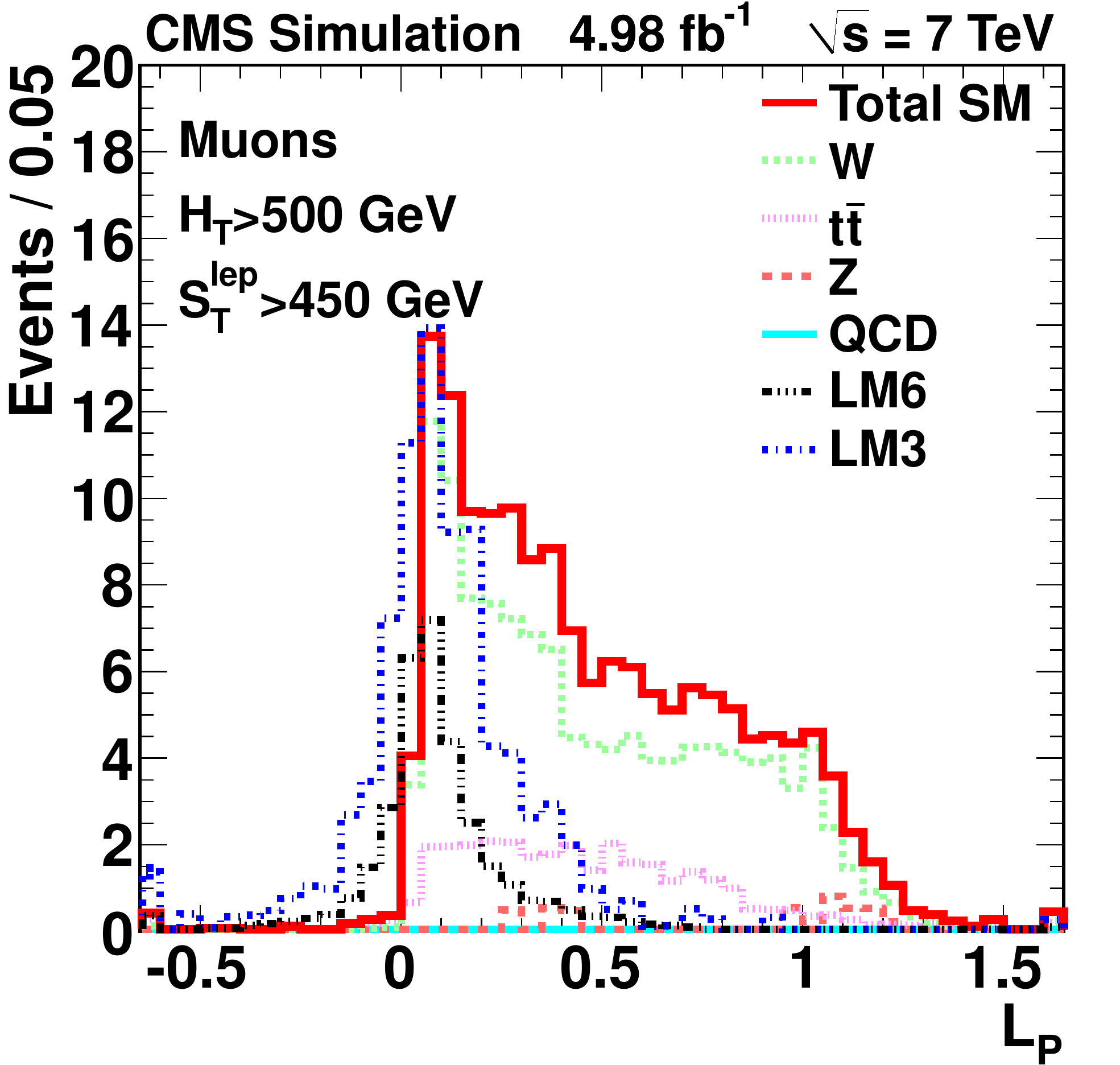}
\includegraphics[width=.40\textwidth]{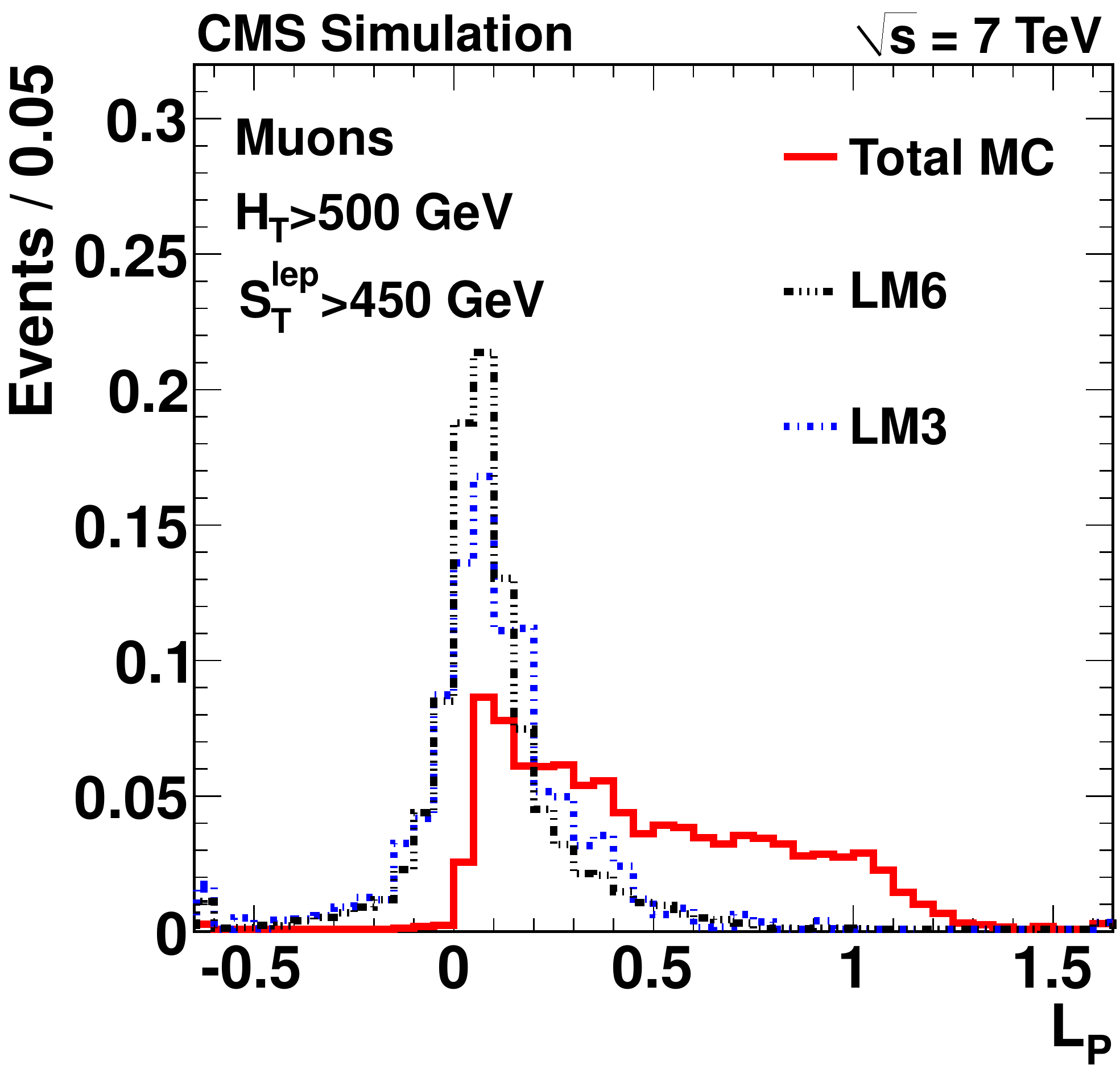}
\caption{Distribution of \Lp in SUSY and standard model processes from simulation. \cmsLLeft: all distributions are normalized to the integrated luminosity. The different contributions from SM processes are shown, whereas for SUSY two benchmark points, LM3 and LM6, are displayed. \cmsRRight: the same distributions normalized to unity.  The SM distribution is the sum of all the individual SM processes shown in the left pane. The quantity $\stlep = \pt(\ell)+\ETslash$ is discussed in the text.}
\label{fig:LP-SUSY-vs-SM}
\end{center}
\end{figure}

Since SUSY decay chains result in large values of $\ETslash$, and often result in relatively
low values of the lepton momentum as well, the \Lp distribution for SUSY events
tends to peak near zero, whereas W+jets and \ttbar yield a broad range of \Lp values.
This behavior is illustrated in Fig.~\ref{fig:LP-SUSY-vs-SM}, which compares the \Lp distribution from
both SM processes and from two representative SUSY benchmark points (LM3 and LM6).

In the \Lp method, two regions in \Lp are defined: the region with $\Lp<0.15$ is used as the signal region;
  the region with $\Lp>0.3$ is used as
the control region, i.e., a sample that is depleted in the signal expected and is instead
dominated by SM processes.  These regions are selected using simulated event samples of W+jets, Z+jets, and \ttbar,
that are collectively referred to as electroweak (EWK) processes in what follows, as well as with simulated SUSY events with SUSY particle masses near the region currently under exploration.

\subsection{Background estimation in the \texorpdfstring{\Lp}{LP} method}

The key ingredient of the analysis is the estimate of the number of events in the signal region from
the SM processes.  We define a translation factor,
\begin{equation}
\label{eqn:RCS}
R_\mathrm{CS} = \frac{N_\mathrm{MC}(\Lp<0.15)}{N_\mathrm{MC}(\Lp>0.3)},
\end{equation}
which is the ratio of the number of events in the signal and control regions for the
EWK processes.
The translation factor is obtained from MC simulation of the EWK processes, and
the uncertainties on this factor are included in the systematic uncertainty of
the background estimate.  In the case of muons, where the background from QCD multijets
is negligible, the total number of events predicted from SM processes in the signal region,
$N_\mathrm{SM}^\text{pred}(\Lp<0.15)$, can be determined directly from the number of events
observed in the data in the control region, $N_\text{data}(\Lp>0.3)$:
\begin{equation}
\label{eqn:prediction}
N_{\rm SM}^\text{pred}(\Lp<0.15) = R_\mathrm{CS} \cdot N_\text{data}(\Lp>0.3).
\end{equation}
In the case of the electrons, the presence of events from QCD multijet processes necessitates
an independent evaluation of this background prior to the application of the translation
factor for EWK processes.

The number of events estimated with this method is then compared to the number of events
observed in the data in the signal region, $N_\text{data}(\Lp<0.15)$, for indications of
an excess of events over the SM expectation. The analysis is performed in different regions
of the event mass scale.  To characterize the latter without affecting the correlation of
the charged lepton and the neutrino in SM events, the scalar sum of the lepton transverse
momentum and the missing transverse momentum, \stlep, is used: $\stlep = \pt(\ell)+\ETslash$.
For W decays, $\stlep \approx \pt ({\PW})$ at large values of $\pt({\PW})$.

In order to make the search optimization less dependent on the unknown energy
scale of a new physics signal, the  analysis is performed in disjoint ranges of \stlep
and the results in these ranges are combined. In addition, the selection is also binned in a
second dimension, the \Ht variable, defined in Eq.~(\ref{eq:HTdef}).

As indicated in Table~\ref{tab:Preselection},
the event selection used in this analysis is slightly different from the corresponding
one in the LS analysis. To increase the sensitivity to SUSY decays, this
analysis requires three or more jets. While this results in a significant increase in
W+jets events, the additional SM background is mostly concentrated in the control
region in \Lp.

\begin{table*}[tb!]
\topcaption{Expected event yields in the signal region ($\Lp < 0.15$) from simulation. These yields are for $\Ht>500$\GeV. These MC values are only listed for illustration purposes.}
\label{tab:signalExpMu}
\begin{center}
\begin{tabular}{ l D{,}{\,\pm\,}{5.3} D{,}{\,\pm\,}{4.3} D{,}{\,\pm\,}{4.3} | D{,}{\,\pm\,}{5.3} D{,}{\,\pm\,}{4.3} D{,}{\,\pm\,}{4.3}}
\hline
$\Lp<0.15$ & \multicolumn{3}{c}{  Muons: \stlep range [\GeVns{}] } &  \multicolumn{3}{c}{  Electrons: \stlep range [\GeVns{}] } \\
 & \multicolumn{1}{c}{[250--350]} & \multicolumn{1}{c}{[350--450]} & \multicolumn{1}{c}{[450--$\infty$]} & \multicolumn{1}{c}{[250--350]} & \multicolumn{1}{c}{[350--450]} & \multicolumn{1}{c}{[450--$\infty$]} \\
\hline
\ttbar ($\ell$)  & 50.0 , 1.0 & 15.3 , 0.5 & 4.8 , 0.3
                & 37.9 , 0.8 & 11.0 , 0.4 & 3.6 , 0.2\\
\ttbar ($\ell\ell$)     &  12.4 , 0.4 & 3.9 , 0.2 & 1.2 , 0.1
                                & 10.4 , 0.4 & 2.9 , 0.2 & 0.8 , 0.1\\
W               &  66.2 , 2.0 & 35.6 , 1.4 & 26.0 , 1.2
                                &  48.9 , 1.7 & 24.2 , 1.2 & 20.9 , 1.1\\
Z               &   2.1 , 1.0 & 0.4 , 0.4 & 0.0 , 0.2
                                & 1.4 , 0.8 & 0.0 , 0.2 & 0.0 , 0.2\\
Total MC                & 130.8 , 2.4 & 55.3 , 1.6 & 32.0 , 1.3
                                & 98.6 , 2.1 & 38.1 , 1.3 & 25.3 , 1.1\\
LM3                     & 136.8 , 3.8 & 89.1 , 3.1 & 53.9 , 2.4
                           & 111.7 , 3.4 & 70.8 , 2.7 & 47.0 , 2.2\\
LM6                     & 8.4 , 0.2 & 11.0 , 0.2 & 24.9 , 0.3
                           & 6.7 , 0.2 & 8.5 , 0.2 & 20.5 , 0.3\\
\hline
\end{tabular}
\end{center}
\end{table*}
The event yields in the muon and electron channels, as predicted from simulation,
are shown in Table~\ref{tab:signalExpMu}. As discussed previously, the dominant
backgrounds to the lepton plus jets and \ETslash signature arise from the production and decay of W+jets and \ttbar. The production of single W bosons in association
with jets, and with large transverse momenta, is in general the larger of the two,
especially at lower jet multiplicities. The majority of the \ttbar background arises
from semi-leptonic \ttbar decays,  with fully leptonic \ttbar decays in which a
lepton is either ignored or not reconstructed contributing about 20\% of the total \ttbar background.

A source of background, which is not listed in Table~\ref{tab:signalExpMu}, stems from
QCD multijet events in which a jet is misreconstructed as a lepton. The simulation indicates that the
magnitude of this background is small in the control region and negligible in the signal
region. Nevertheless, since the uncertainties in simulating these backgrounds can be
significant, we use control data samples to estimate the background in the muon and
electron channels.

To estimate the background from QCD multijets in the muon final state, we use the relative combined isolation, $I_\text{comb}^\text{rel}$, of the muon. Multijet events are expected to populate the region at high values of $I_\text{comb}^\text{rel}$, whereas muons from SUSY decays are isolated and thus have low values of $I_\text{comb}^\text{rel}$.  We employ an additional control data sample, which is specially selected to be enriched in QCD multijets, to determine the ratio of multijet events at low values of the relative isolation.  Using this ratio and the number of multijet events expected in the control region of the sample passing the preselection requirements, we estimate the background from multijet events in the signal region to be always smaller than 1\% of the EWK backgrounds. This level of background is negligible and is thus ignored in what follows.

The main sources of electrons in QCD multijet events are misidentified jets and
photon conversions. This background is expected to be more substantial
than the corresponding one in the muon sample, and its estimate exhibits a large
dependence on the details of the simulation. For this reason, we estimate this background from
control samples in data.  The method relies on the inversion of one
or more of the electron identification requirements in order to obtain a sample of
anti-selected events, which is dominated by jets misidentified as electrons.  We find that the inversion
of the requirements on the spatial matching of the calorimeter cluster and the
charged-particle track in pseudorapidity and azimuth leaves the relative fraction
of the different background sources in QCD multijets unchanged. Moreover, to increase
the number of events in this control sample, the requirements on $d_0$ and $d_z$ are
removed, while the isolation requirement is loosened. These changes to the event selection
have a negligible effect on the \Lp distribution in the data.  In the
simulated event samples, it is
found that the \Lp distribution from the control sample events provides a good
description of the corresponding distribution from QCD background passing all
selection requirements.

The \Lp distribution obtained with this control sample is used as a template to fit,
along with the \Lp distribution from EWK processes, the \Lp distribution
in the data. In this fit, the EWK template is taken from simulation. This approach, which provides
a template obtained from data for the QCD contamination, was applied in the measurement
of the polarization of high-\pt W bosons~\cite{Wpol-PRL}.  The fit is performed in the
control region ($\Lp>0.3$), where the possible presence of signal is highly suppressed.
The numbers of QCD and EWK events obtained by the fit are used to estimate the
total SM contamination in the signal region ($\Lp<0.15$). The method for estimating the number of SM events expected in the signal
region is applied in each range of \stlep and \Ht.

\begin{figure}[htb]
\begin{center}
\includegraphics[width=.45\textwidth]{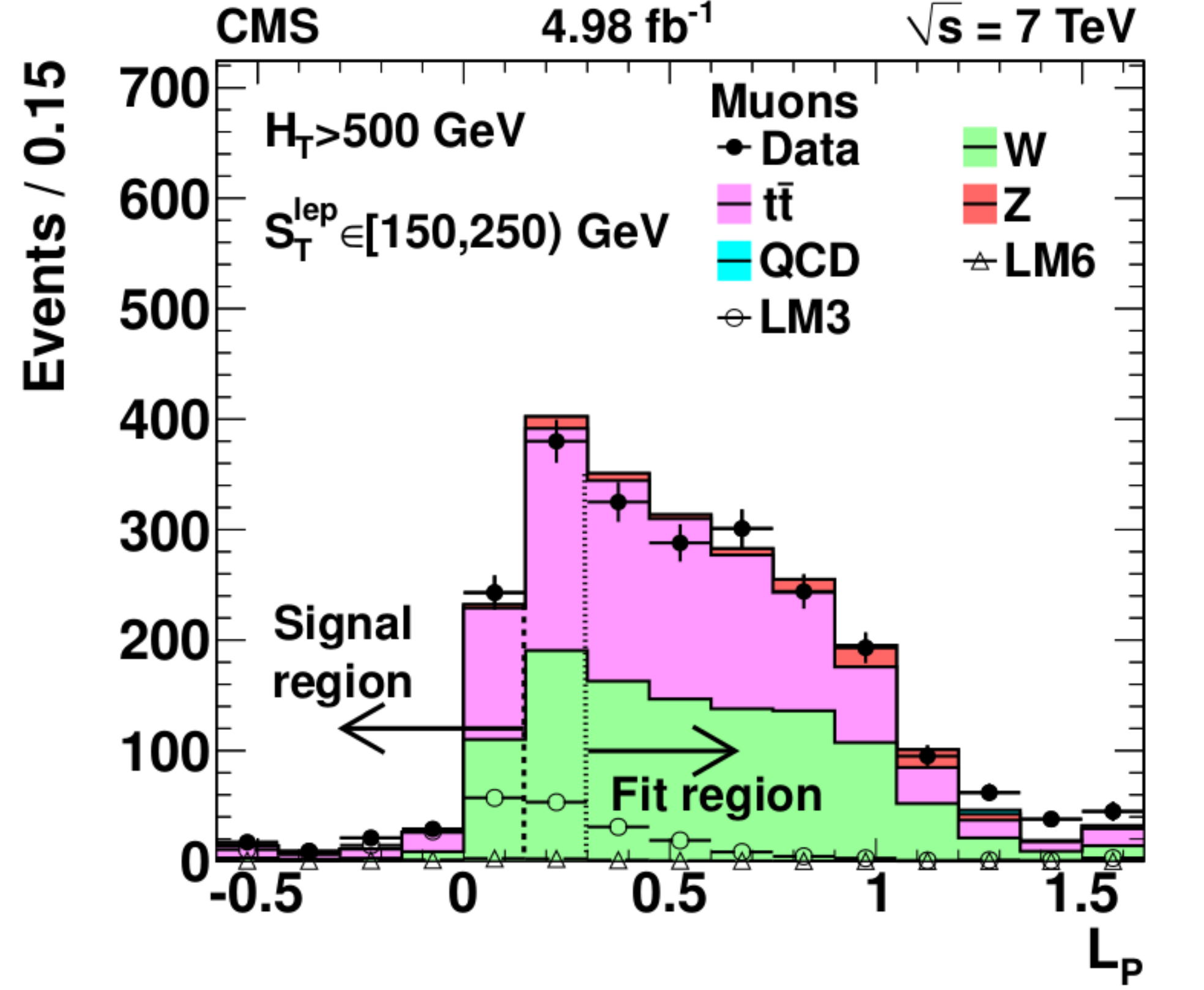}
\includegraphics[width=.463\textwidth]{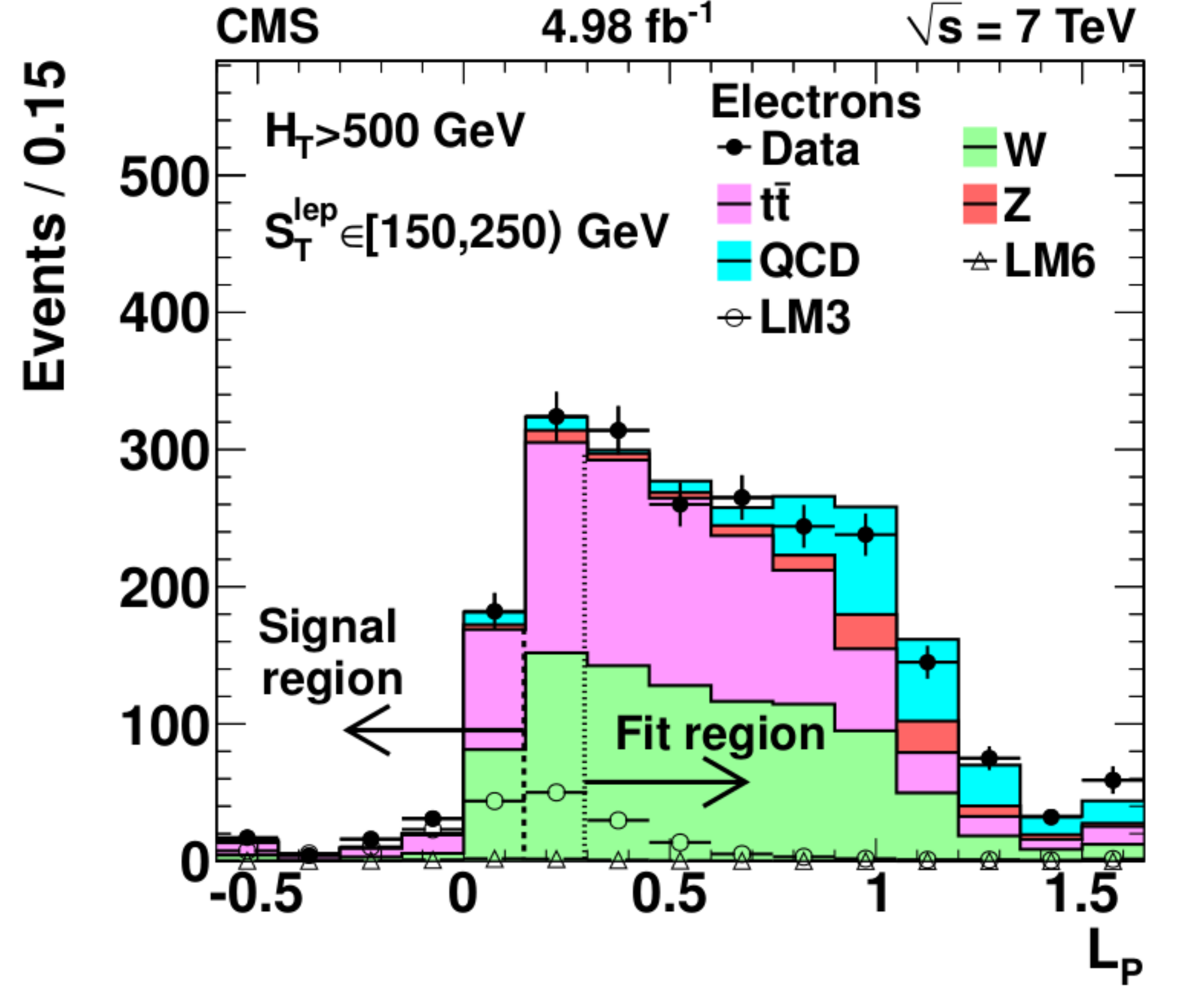}
\caption{Fit results on data for $150 < \stlep <250\gev$, in the muon (\cmsLeft) and electron (\cmsRight) search samples. The fit is performed in the control region ($\Lp>0.3$) and the result is extrapolated into the signal region ($\Lp<0.15$). }
\label{fig:QCDfit_ST}
\end{center}
\end{figure}

The method for estimating the SM expectation in the signal region is checked using two different control samples, where both the fit and signal regions have a negligible expected SUSY yield. The first sample is defined as all events satisfying the preselection requirements
but confined to low values of \stlep: $150<\stlep<250$\GeV. The method described above
is employed to predict the number of events expected in the signal region for both muons and
electrons. This prediction is found to be fully consistent with the number of events observed
in the data in signal region. The results of the fits and the yields of QCD and EWK events
in the region of low \stlep ($ <250$\GeV) are displayed in Fig.~\ref{fig:QCDfit_ST} for the electron and muon samples.
As can be seen in Fig.~\ref{fig:QCDfit_ST}, the QCD contamination in the signal region, $\Lp<0.15$, is negligible, as expected,
since low values of \Lp favor events with low-\pt leptons and high \ETslash. The second sample, used only for events with muons, is
collected with a separate trigger without any requirements on \Ht or \ETslash. The muon transverse momentum threshold
is raised to $\pt(\mu)>35$\GeV, while the \Ht threshold is lowered from 500\GeV to 200\GeV
and the jet multiplicity requirement is reversed, to be fewer than three jets.  Given these
requirements on \Ht and on the jet multiplicity, this control sample is dominated by SM
processes. It is found that the estimated background agrees well with the number of events
seen in the signal region $\Lp<0.15$.

\begin{figure*}[ht!]
\centering
\includegraphics[width=0.32\linewidth]{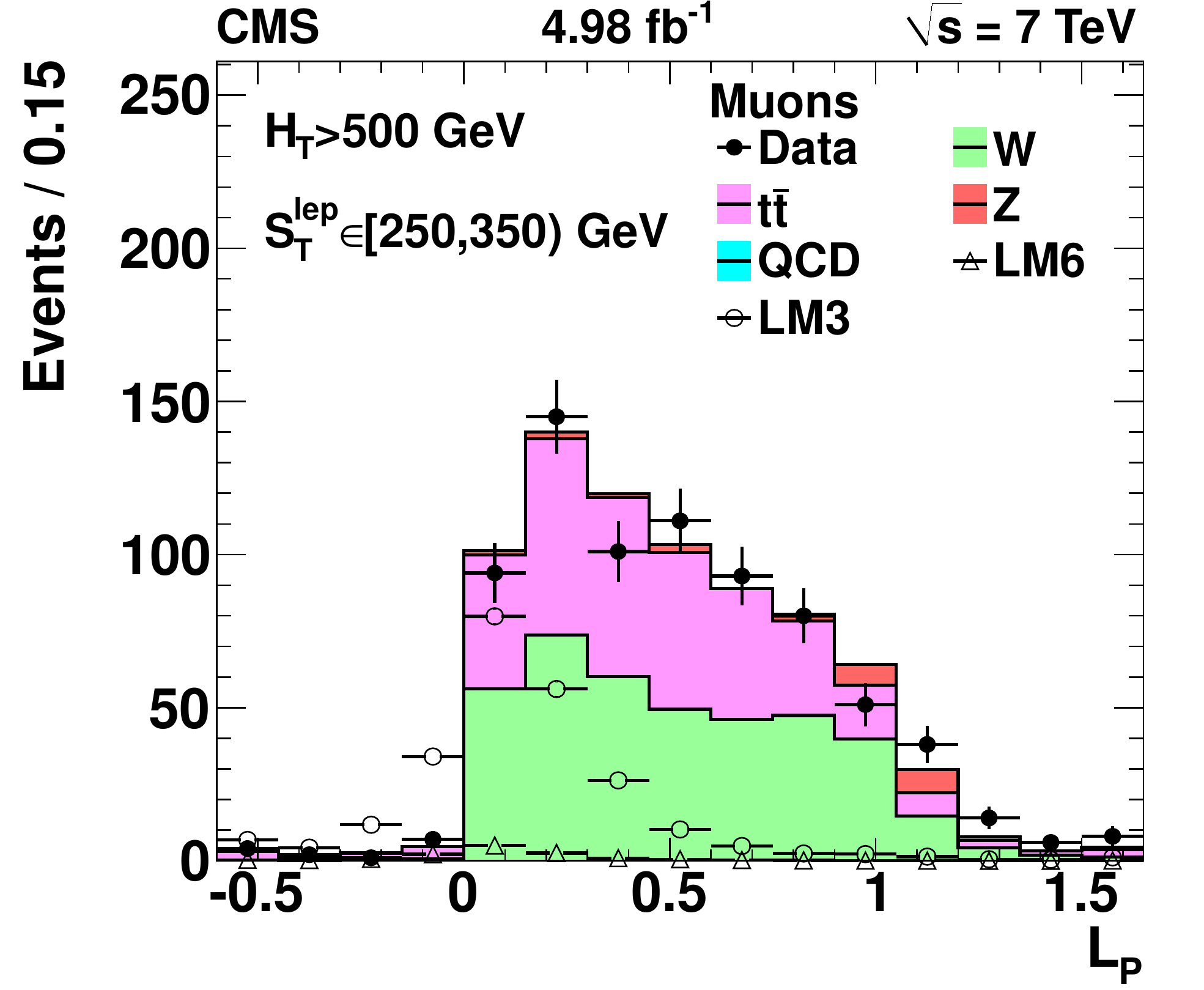}
\includegraphics[width=0.32\linewidth]{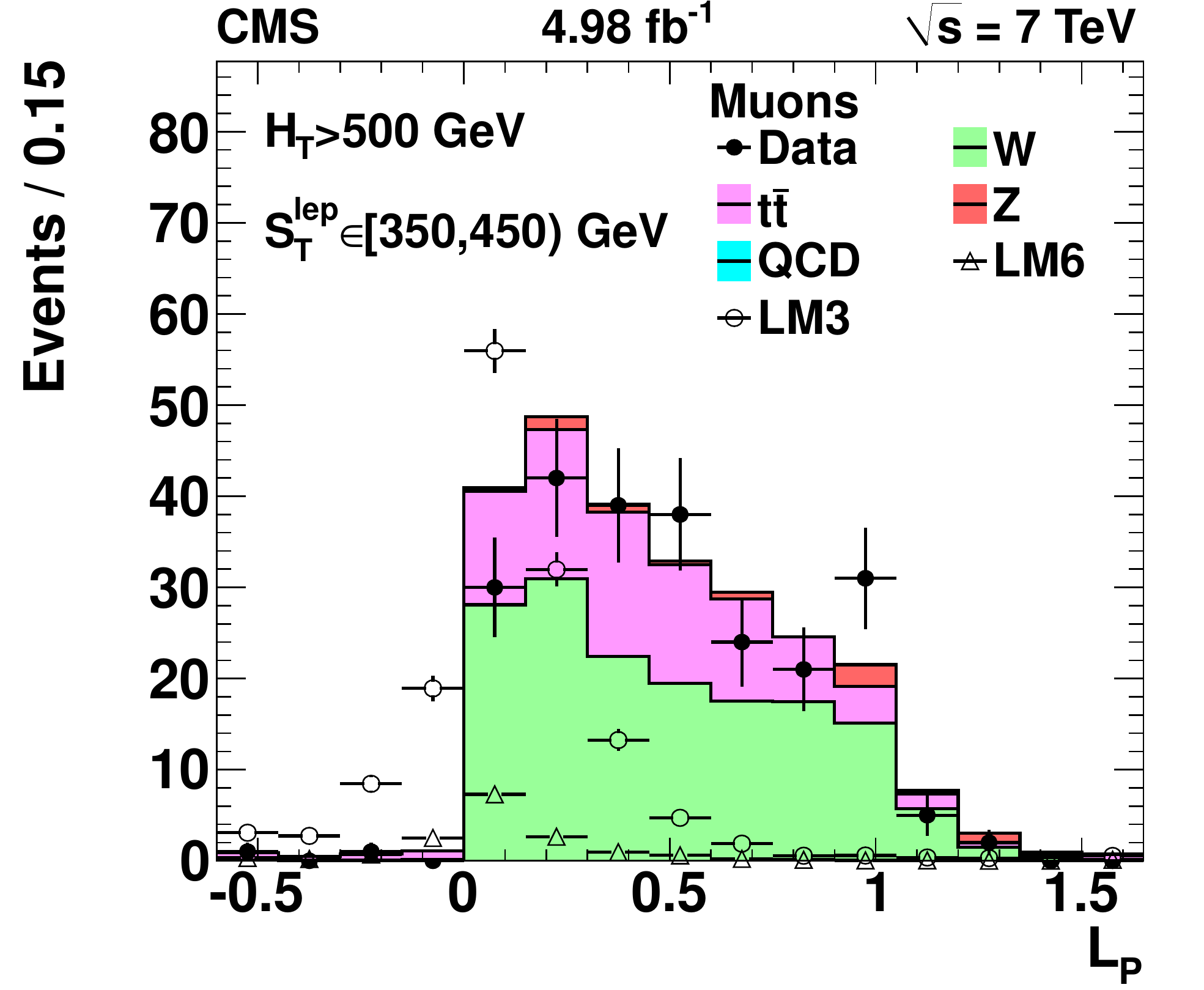}
\includegraphics[width=0.32\linewidth]{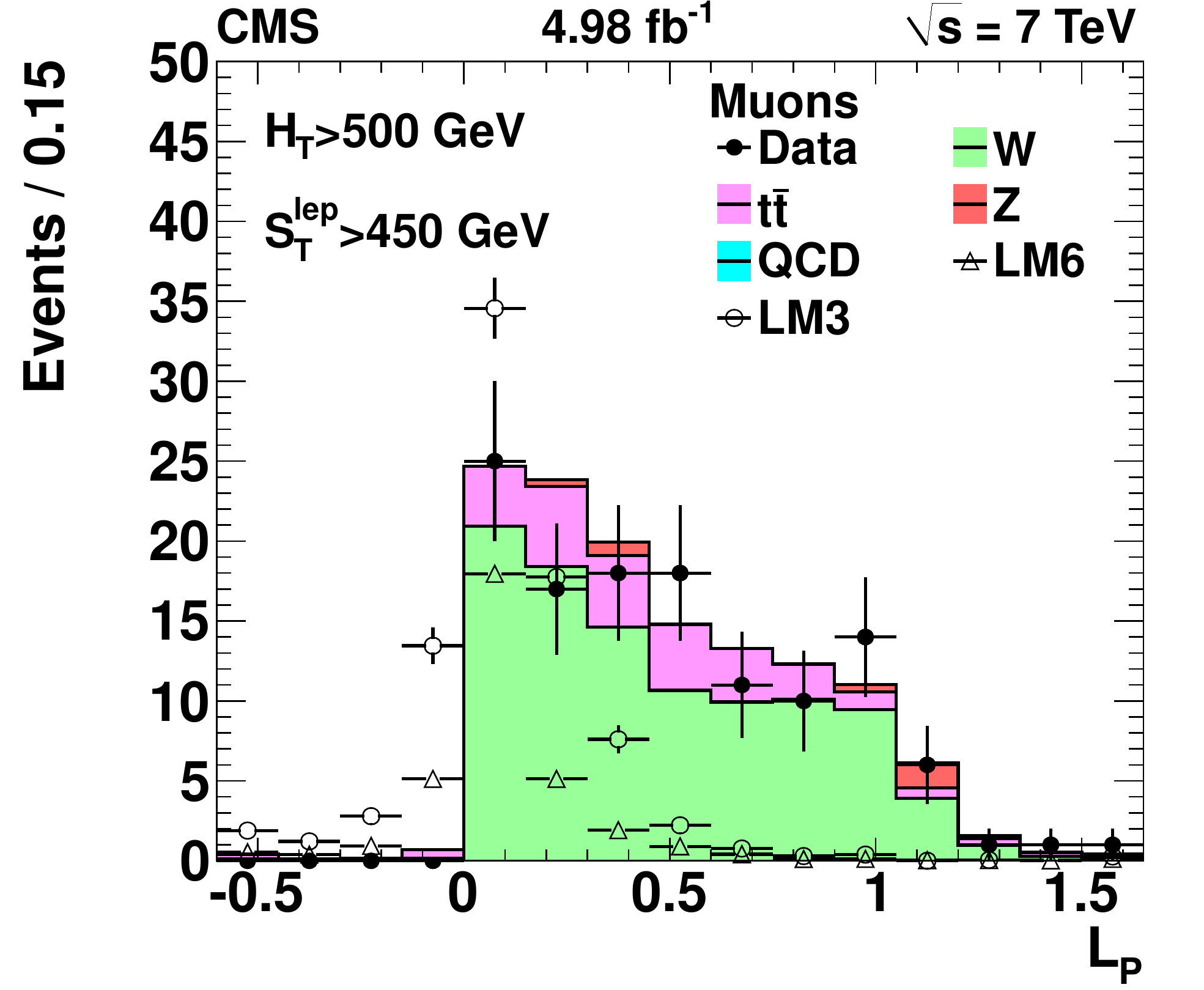}
\includegraphics[width=0.32\linewidth]{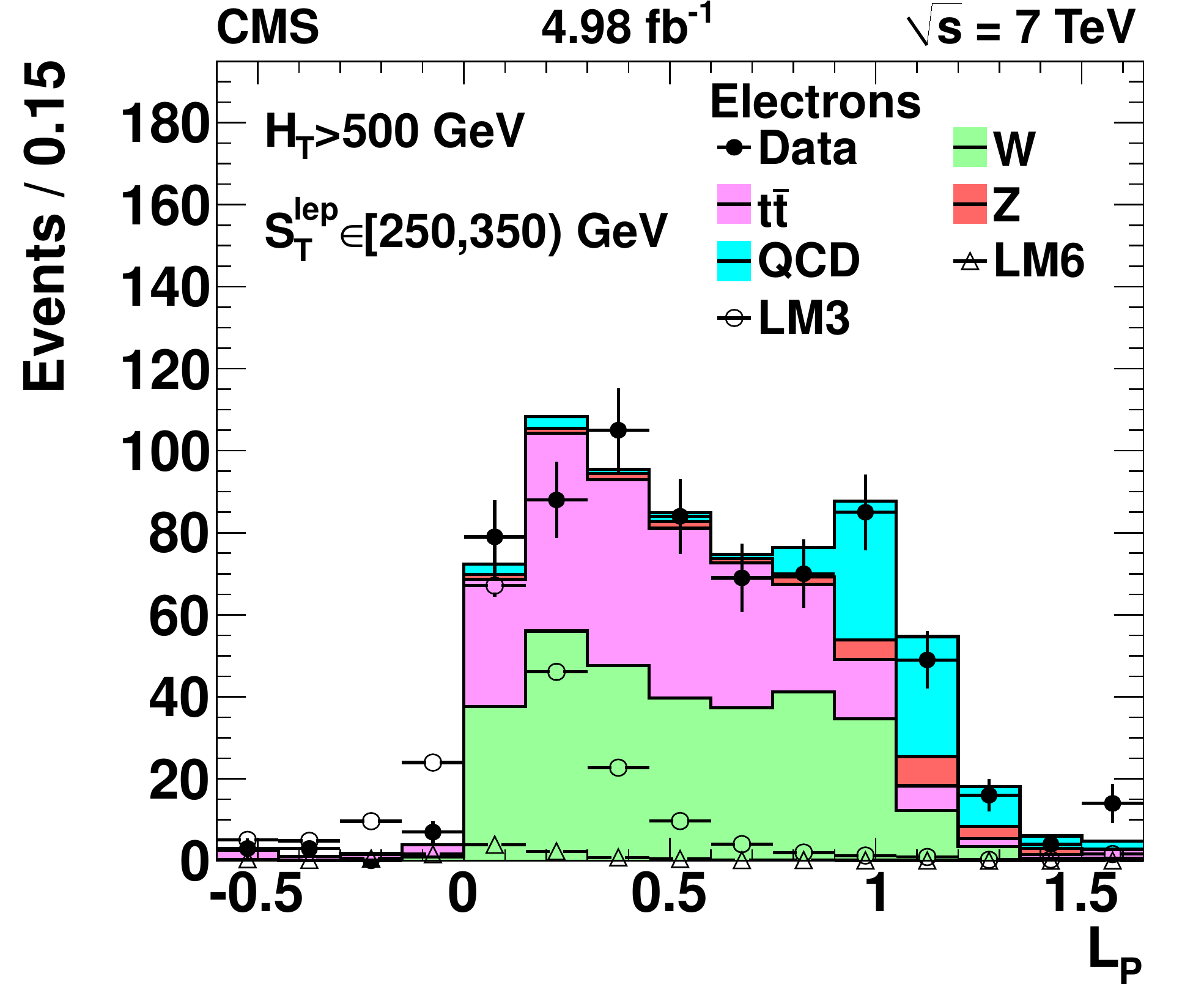}
\includegraphics[width=0.32\linewidth]{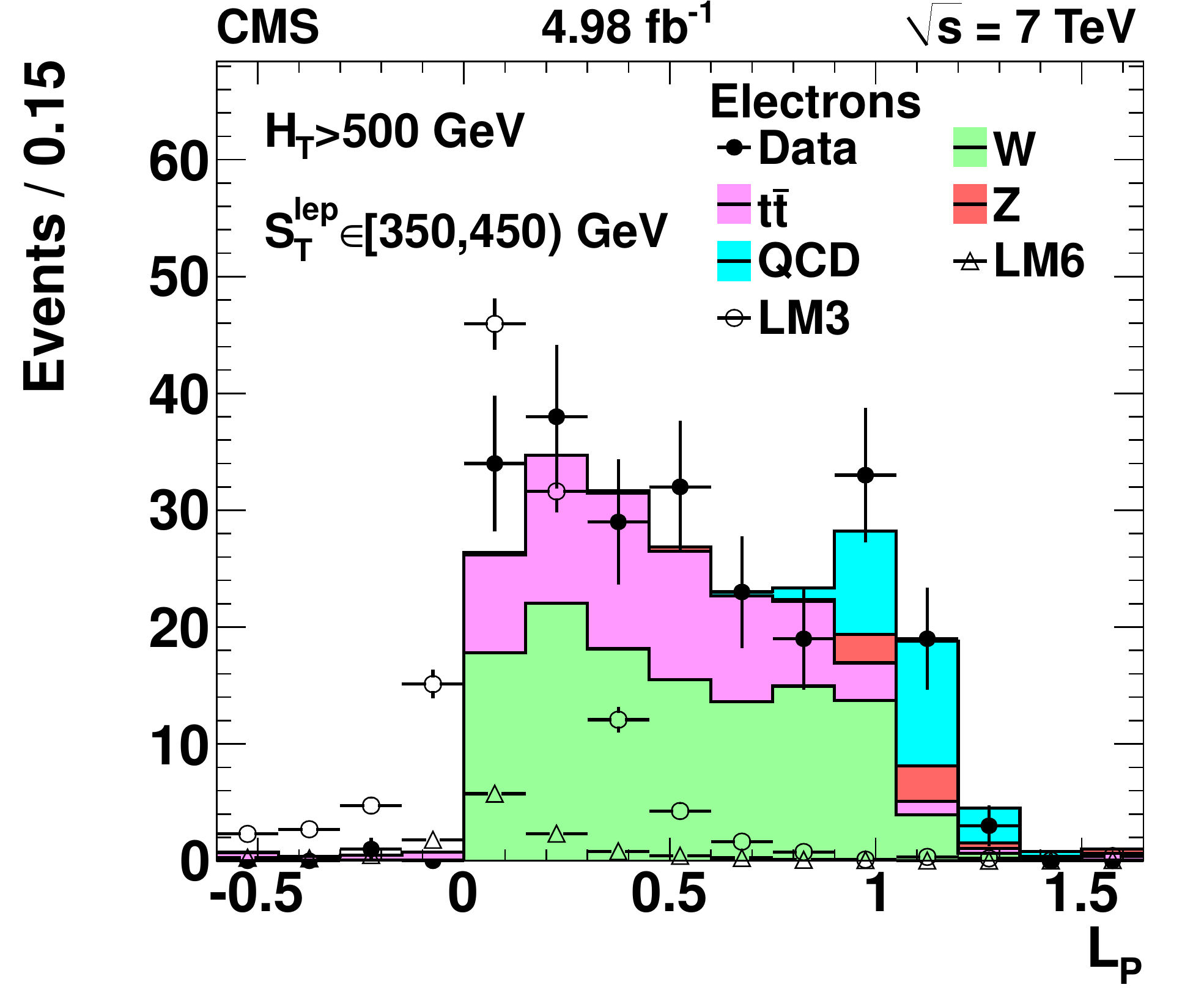}
\includegraphics[width=0.32\linewidth]{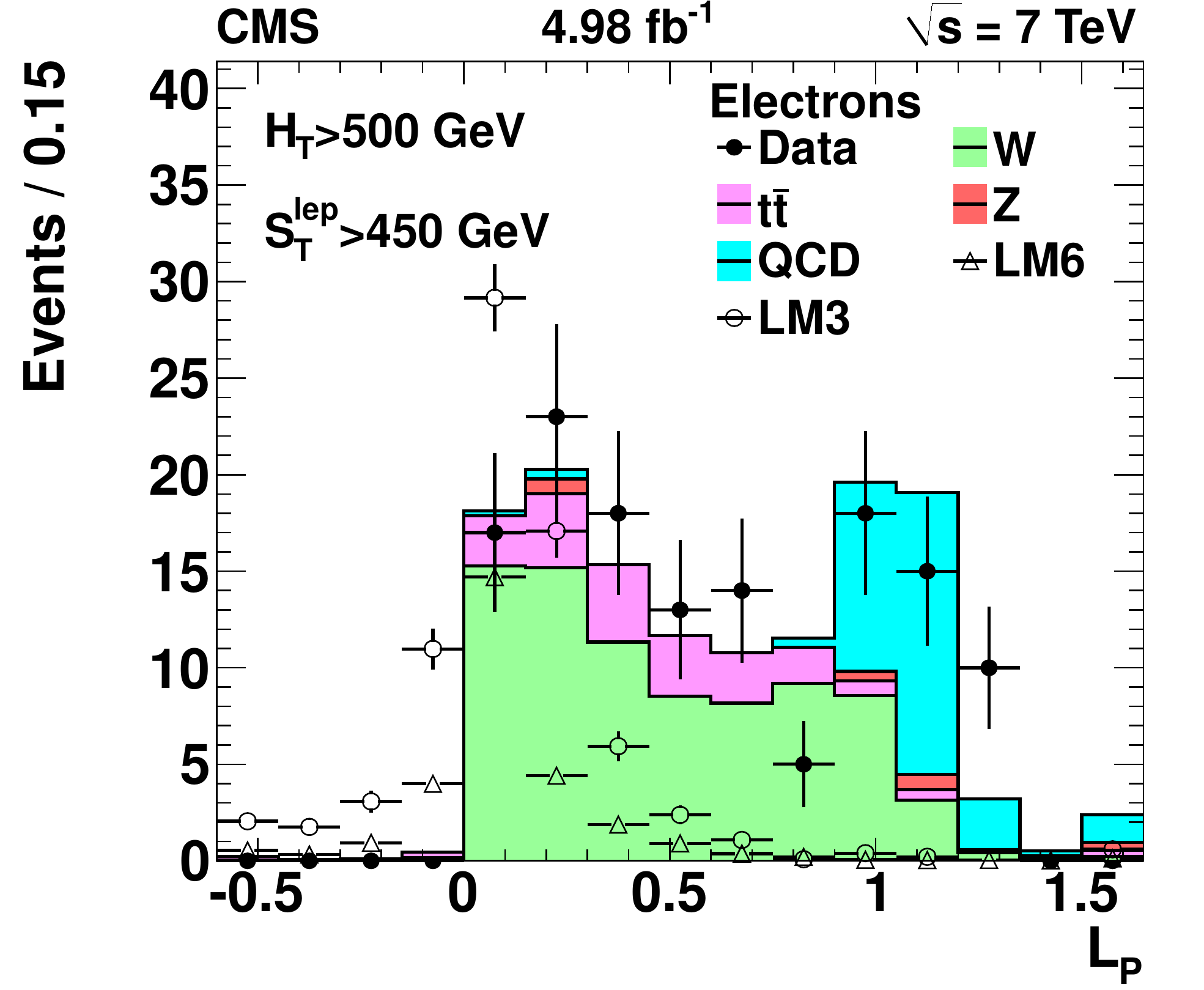}
\caption{Data and fit results for the predictions for the \Lp distribution, for events in the search sample, in different \stlep regions. Top plots for the muon channel; bottom plots for the electron channel. Left: $250<\stlep<350\gev$, center: $350<\stlep<450\gev$, and right: $\stlep>450\gev$.}
\label{fig:searchSampleLP}
\end{figure*}

\subsection{Results of the \texorpdfstring{\Lp}{LP} method}
\label{sec:Lpresults}

\begin{figure*}[h!tb]
\begin{center}
\includegraphics[width=\linewidth]{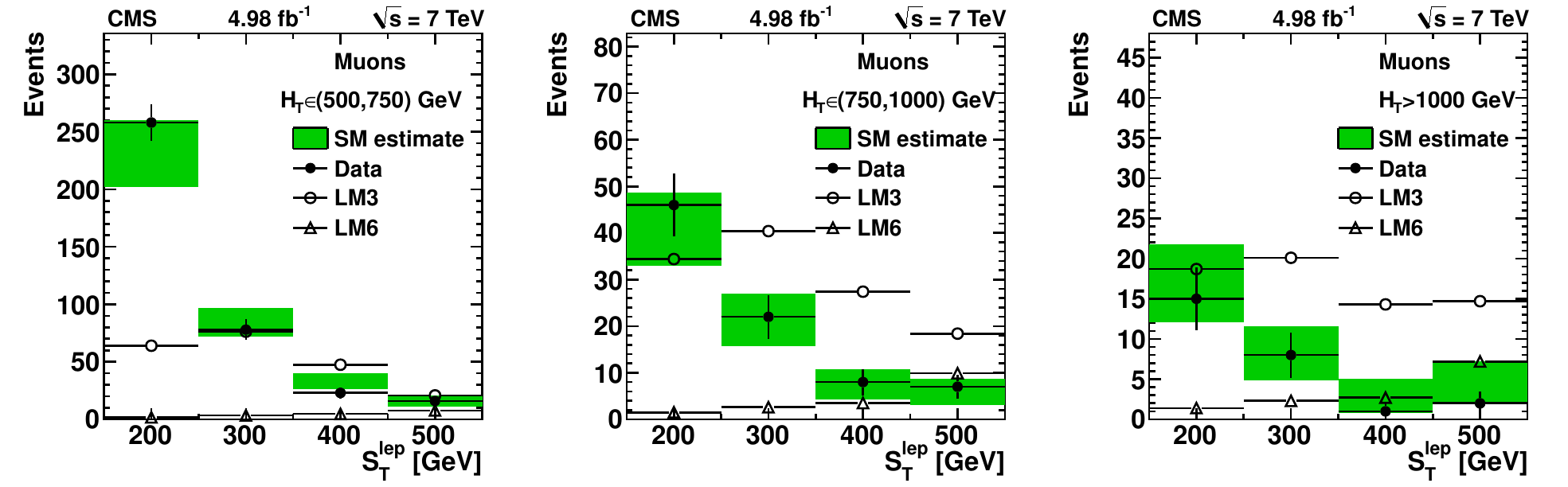}
\caption{
Comparison of the number of events observed in the data and the expectations from the background estimation methods for the muon channel, in the different \stlep bins. Left: $500<\Ht<750$\GeV; Center: $750<\Ht<1000$\GeV; Right: \Ht$>1000$\GeV. The error bars indicate the statistical uncertainty of the data only, while the green band indicates the total statistical and systematic uncertainty on the background estimate.}
\label{HTbinnedPred-Mu}
\end{center}
\end{figure*}

\begin{figure*}[tb!]
\begin{center}
\includegraphics[width=\linewidth]{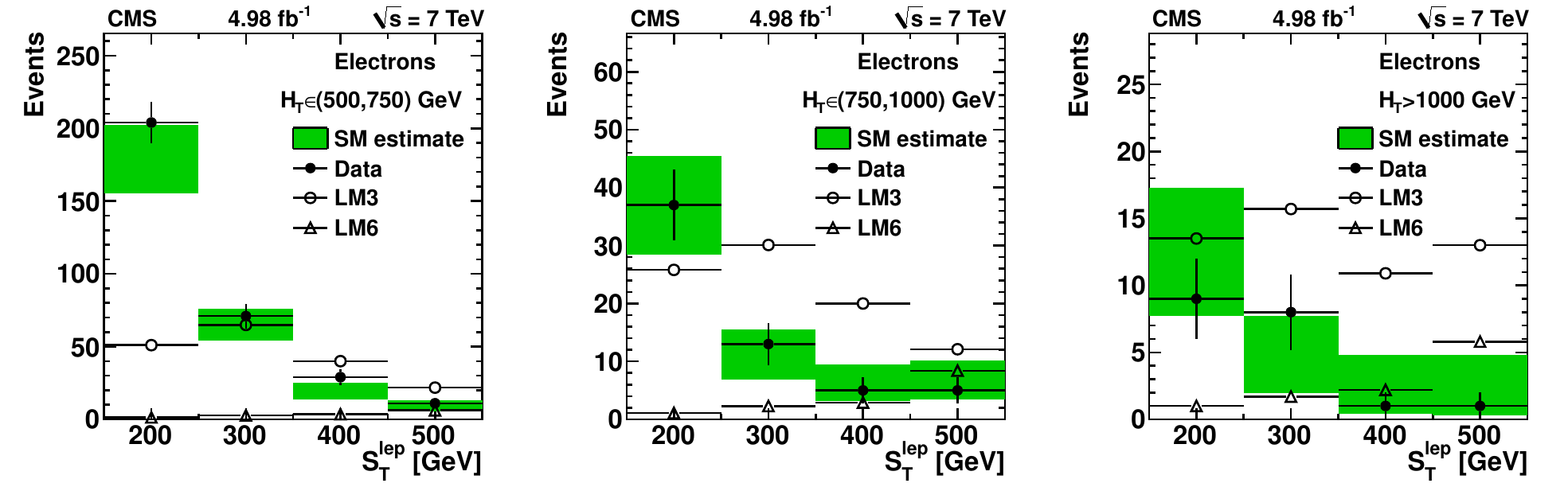}
\caption{
Comparison of the number of events observed in the data and the expectations from the background estimation methods for the electron channel, in the different \stlep bins. Left: $500<\Ht<750$\GeV; Center: $750<\Ht<1000$\GeV; Right: \Ht$>1000$\GeV. The error bars indicate the statistical uncertainty of the data only, while the green band indicates the total statistical and systematic uncertainty on the background estimate.}
\label{HTbinnedPred-El}
\end{center}
\end{figure*}

The \Lp distributions in three ranges of \stlep, are displayed in
Fig.~\ref{fig:searchSampleLP} for muons (top) and electrons (bottom).
Tables~\ref{tab:muonpredictionHT} and~\ref{tab:electronpredictionHT} list the numbers of
events observed and the number of events expected from all SM processes as presented above,
in the signal region, for the muon and electron channels, respectively. The predictions, along with
the numbers of events observed in each range of \stlep and \Ht, are also displayed graphically
in Fig.~\ref{HTbinnedPred-Mu} for muons and in Fig.~\ref{HTbinnedPred-El} for electrons.
The uncertainties quoted in Table~\ref{tab:electronpredictionHT} correspond to the statistical
uncertainty of the fit, while the predictions displayed in
Fig.~\ref{HTbinnedPred-El} include the total statistical and systematic uncertainty.

\begin{table*}[ht!]
\topcaption{
Event yields in data and MC simulation for the muon sample. The results in the columns labelled ``Total MC'' are listed for reference only.  The corresponding uncertainties statistical only.
}
\label{tab:muonpredictionHT}
\begin{center}
\begin{tabular}{l D{,}{\,\pm\,}{4.3} r | D{,}{\,\pm\,}{4.3} r@{$.$}l r}
\hline
\multicolumn{1}{l}{\stlep range}       & \multicolumn{1}{c}{Total MC} & \multicolumn{1}{c|}{Data} & \multicolumn{1}{c}{Total MC} & \multicolumn{2}{c}{SM estimate} & \multicolumn{1}{c}{Data} \\
\multicolumn{1}{l}{[\GeVns{}]} & \multicolumn{2}{c |}{Control region ($\Lp>0.3$)} & \multicolumn{4}{c }{Signal region ($\Lp<$0.15)} \\
 \hline
&\multicolumn{6}{ c }{ $500<\Ht<750$\GeV} \\[1ex]

{[150--250)}&1465 , 11 &1297& 261 , 3.2 &\multicolumn{2}{c}{261$\,\pm\,$7\hphantom{.2}$\,\pm\,$24}& 258 \\
{[250--350)}& 452 , 5.2 & 383&99.3 , 2.1&84.1$\,\pm\,$4&2$\,\pm\,$7.3  & 78\\
{[350--450)}&154 , 3.1 &128&40.2 , 1.4& 33.3$\,\pm\,$3&0$\,\pm\,$2.6  & 23 \\
$\ge$ 450  &  59.2 , 1.8 &50& 18.6 , 1.0&15.7$\,\pm\,$2&2$\,\pm\,$2.0  & 16\\
\noalign{\vspace{6pt}}
 &\multicolumn{6}{ c}{$750<\Ht<1000$\GeV}\\
{[150--250)}&280 , 4.1 &218& 52.4 , 1.6& 40.8$\,\pm\,$2&9$\,\pm\,$3.5 & 46 \\
{[250--350)}& 91.9 , 2.1&88 & 22.3 , 0.9& 21.3$\,\pm\,$2&3$\,\pm\,$2.2 & 22 \\
{[350--450)}&34.6 , 1.3&25& 10.3 , 0.6& 7.5$\,\pm\,$1&5$\,\pm\,$1.0&8 \\
$\ge$ 450  &26.7 , 1.4&18& 8.8 , 0.6& 5.9$\,\pm\,$1&4$\,\pm\,$0.7&7 \\
\noalign{\vspace{6pt}}
 & \multicolumn{6}{ c }{1000\GeV $<\Ht$} \\
{[150--250)}& 92.3 , 2.5 &76 & 20.5 , 1.0& 16.9$\,\pm\,$1&9$\,\pm\,$1.7 & 15\\
{[250--350)}& 32.9 , 1.3 &31& 8.7 , 0.8 & 8.2$\,\pm\,$1&5$\,\pm\,$1.0 & 8\\
{[350--450)}& 10.9 , 0.7 & 7 & 4.6 , 0.4 & 2.9$\,\pm\,$1&1$\,\pm\,$ 0.6 & 1\\
$\ge$ 450 & 11.9 , 0.8 &12& 4.6 , 0.5& 4.6$\,\pm\,$1&4$\,\pm\,$0.7 & 2\\
\hline
\end{tabular}
\end{center}
\end{table*}

\begin{table*}[htb!]
{\footnotesize{
\topcaption{Event yields in data and predictions of the numbers of EWK and QCD events for the electron sample in bins of $\Ht$. The sum of predicted EWK events and predicted QCD events in the control region is constrained to be equal to the total number of data events. The background estimate used in comparing to the yields in the data is the result of the procedure described earlier and is listed in the row labeled ``SM estimate''.
The uncertainties for the QCD and EWK background estimates are statistical only.
The uncertainties shown for the SM estimate are first the statistical uncertainty from the control region fit and second all other systematic uncertainties.}
\label{tab:electronpredictionHT}
\begin{center}
\begin{tabular}{l  c c c | c c c c}
\hline
\stlep range [\GeVns{}]      & QCD                   & EWK                   & Data          & QCD             & EWK                         &                       SM estimate     & Data \\
& \multicolumn{3}{c|}{Control region ($\Lp>$0.3)} & \multicolumn{4}{c}{Signal region ($\Lp<$0.15)} \\
\hline
\multicolumn{8}{ c  }{$500<\Ht<750$\GeV} \\
{[150--250)}& 184$\,\pm\,$33 & 1122$\,\pm\,$45 & 1306 & 9.1$\,\pm\,$1.6 & 170$\,\pm\,$7 & 179$\,\pm\,$7$\,\pm\,$18 & 204 \\
{[250--350)}& 66$\,\pm\,$15 & 334$\,\pm\,$22 & 400 & 2.1$\,\pm\,$0.5 & 63.3$\,\pm\,$4.1 & 65.3$\,\pm\,$4.3$\,\pm\,$5.9 & 71 \\
{[350--450)}& 26.6$\,\pm\,$7.6 & 93$\,\pm\,$11 & 120 & 0.3$\,\pm\,$0.1 & 19.2$\,\pm\,$2.3 & 19.4$\,\pm\,$2.4$\,\pm\,$2.9 & 29 \\
$\ge$ 450  & 17.1$\,\pm\,$5.1 & 33.9$\,\pm\,$6.6 & 51 & 0.2$\,\pm\,$0.0 & 9.0$\,\pm\,$1.8 & 9.2$\,\pm\,$1.9$\,\pm\,$1.7 & 11 \\
\noalign{\vspace{6pt}}
 \multicolumn{8}{ c  }{$750<\Ht<1000$\GeV}\\
{[150--250)}& 39$\,\pm\,$15 & 210$\,\pm\,$20 & 249 & 1.9$\,\pm\,$0.7 & 35.1$\,\pm\,$3.3 & 37.0$\,\pm\,$3.5$\,\pm\,$4.8 & 37 \\
{[250--350)}& 5.8$\,\pm\,$5.5 & 59.2$\,\pm\,$9.1 & 65 & 0.2$\,\pm\,$0.2 & 11.0$\,\pm\,$1.7 & 11.2$\,\pm\,$2.0$\,\pm\,$1.8 & 13 \\
{[350--450)}& 0.0$\,\pm\,$0.0 & 26.0$\,\pm\,$5.1 & 26 & 0 & 6.3$\,\pm\,$1.2 & 6.3$\,\pm\,$1.2$\,\pm\,$1.5 & 5 \\
$\ge$ 450  & 8.7$\,\pm\,$3.4 & 22.3$\,\pm\,$5.0 & 31 & 0.1$\,\pm\,$0.0 & 6.7$\,\pm\,$1.5 & 6.8$\,\pm\,$1.6$\,\pm\,$1.5 & 5 \\
\noalign{\vspace{6pt}}
\multicolumn{8}{ c }{1000\GeV $<\Ht$} \\
 {[150--250)}& 14.9$\,\pm\,$7.7 & 62$\,\pm\,$10 & 77 & 0.7$\,\pm\,$0.4 & 11.7$\,\pm\,$1.9 & 12.5$\,\pm\,$2.2$\,\pm\,$2.4 & 9 \\
{[250--350)}& 10.4$\,\pm\,$4.3 & 20.6$\,\pm\,$5.4 & 31 & 0.3$\,\pm\,$0.1 & 4.5$\,\pm\,$1.2 & 4.8$\,\pm\,$1.5$\,\pm\,$1.1 & 8 \\
{[350--450)}& 0.5$\,\pm\,$1.7 & 11.5$\,\pm\,$3.7 & 12 & 0.0$\,\pm\,$0. & 2.6$\,\pm\,$0.8 & 2.6$\,\pm\,$1.2$\,\pm\,$0.9 & 1 \\
$\ge$ 450  & 4.4$\,\pm\,$2.5 & 6.6$\,\pm\,$2.9 & 11 & 0.0$\,\pm\,$0.0 & 2.5$\,\pm\,$1.1 & 2.6$\,\pm\,$1.3$\,\pm\,$0.9 & 1 \\
\hline
\end{tabular}
\end{center}
}
}
\end{table*}

All estimates of the total contribution expected from SM processes in the various bins in (\stlep,\Ht) are
consistent with the numbers of events observed in the data, with no visible excess from a potential SUSY
signal. The result is interpreted as a limit in SUSY parameter space in the context of the CMSSM in Section~\ref{sec:Results}.

\section{The Artificial Neural Network method}
\label{sec:ANN}

\subsection{Overview of the method}
The Artificial Neural Network (ANN) method uses a multi-variate analysis to combine several event characteristics, other than \ETslash, into a single variable that distinguishes signal from background.   Signal events then preferentially populate a signal region in the two-dimensional plane of the ANN output (\zANN) and \ETslash,
and the sidebands in this plane provide an estimate of the residual background.

Four input variables drive the ANN.  The first two are $n_{\mathrm{jets}}$, the number of jets with $\pt>40$\GeV,
and \HT, the scalar sum of the \pt of each jet with $\pt > 40$\GeV.  The SUSY signal typically has heavy particles
decaying via complex cascades, and as such, is likely to produce more jets and larger \HT
than SM backgrounds.  The third variable is $\Delta\phi(\mathrm{j}_1,\mathrm{j}_2)$,
the angle between the two leading \pt jets in the transverse plane, which makes
use of the greater likelihood that the two highest \pt jets are produced back-to-back in
SM than in SUSY events. The final variable is \mT, the transverse mass of the lepton and
\ETslash system. In \ttbar and W+jets events, the lepton and \ETslash
generally arise from the decay of a W boson, and as a result, \amT peaks
near the W boson mass, with larger values arising only when there are
additional neutrinos from $\tau$ or semileptonic decays. By contrast,
in SUSY events,  \amT tends to be greater than the W mass because of
\ETslash due to undetected LSPs.

\begin{figure*}[tbp!]
  \centering
\includegraphics[width=0.45\textwidth]{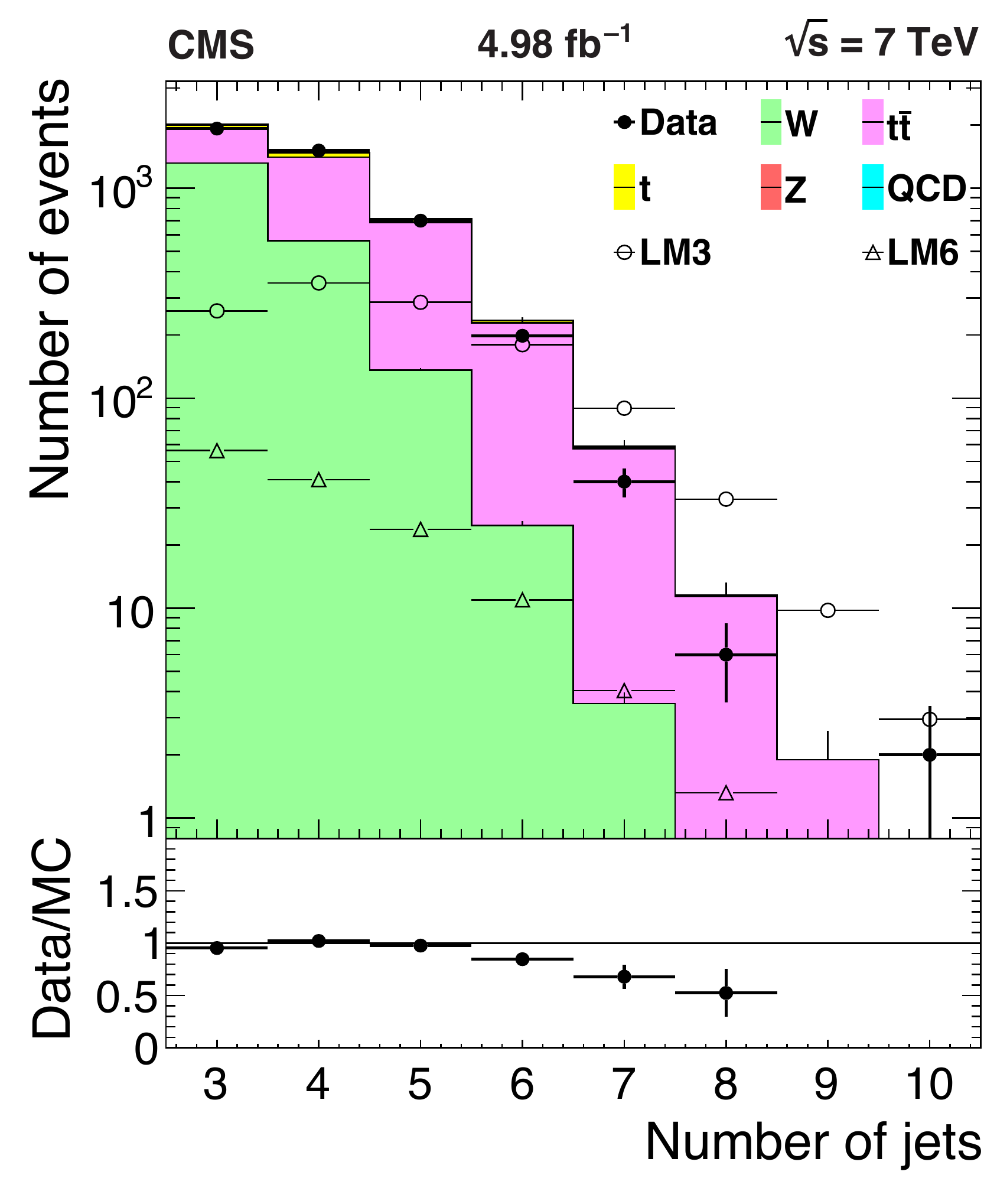}
\includegraphics[width=0.45\textwidth]{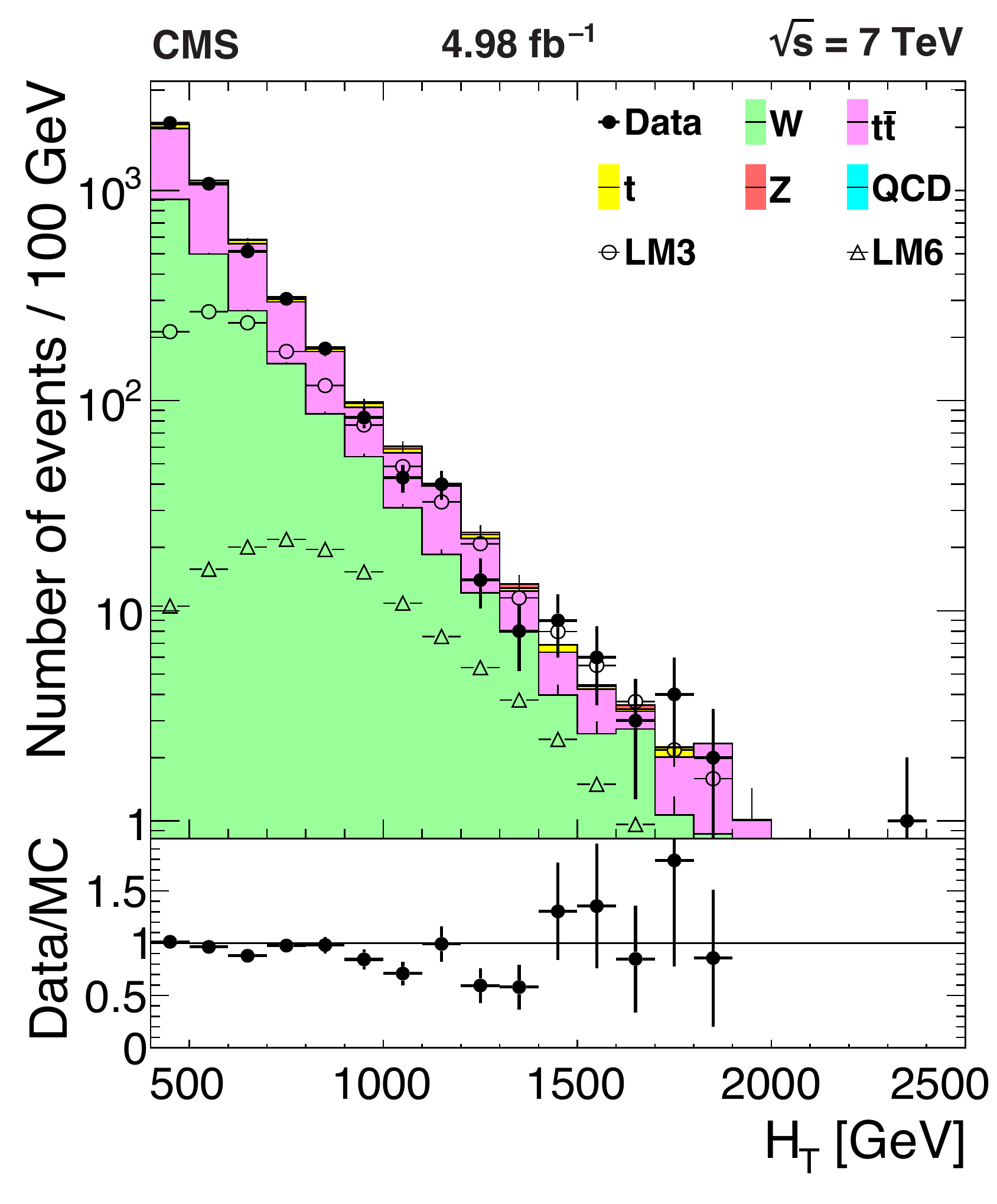}
\includegraphics[width=0.45\textwidth]{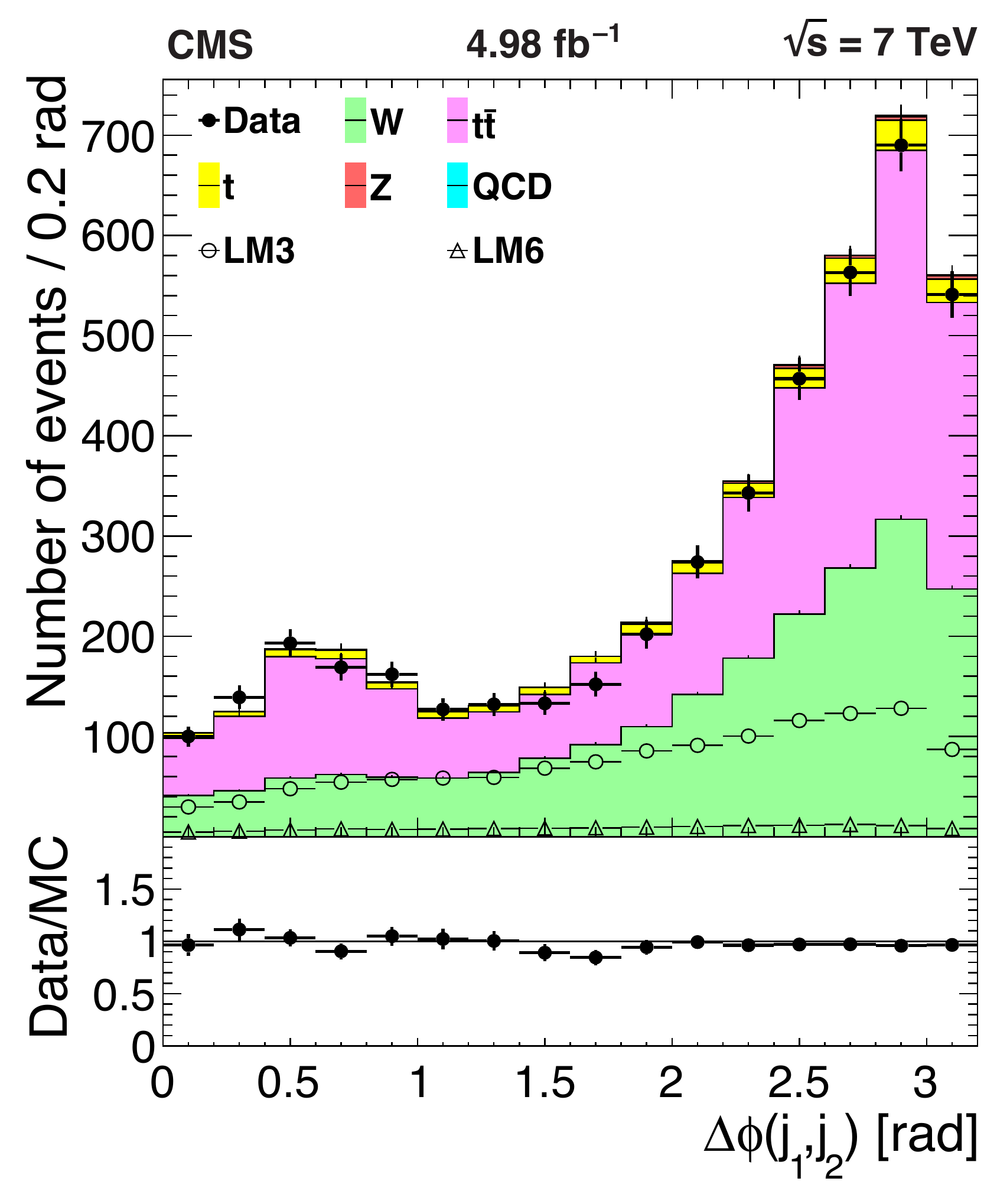}
\includegraphics[width=0.45\textwidth]{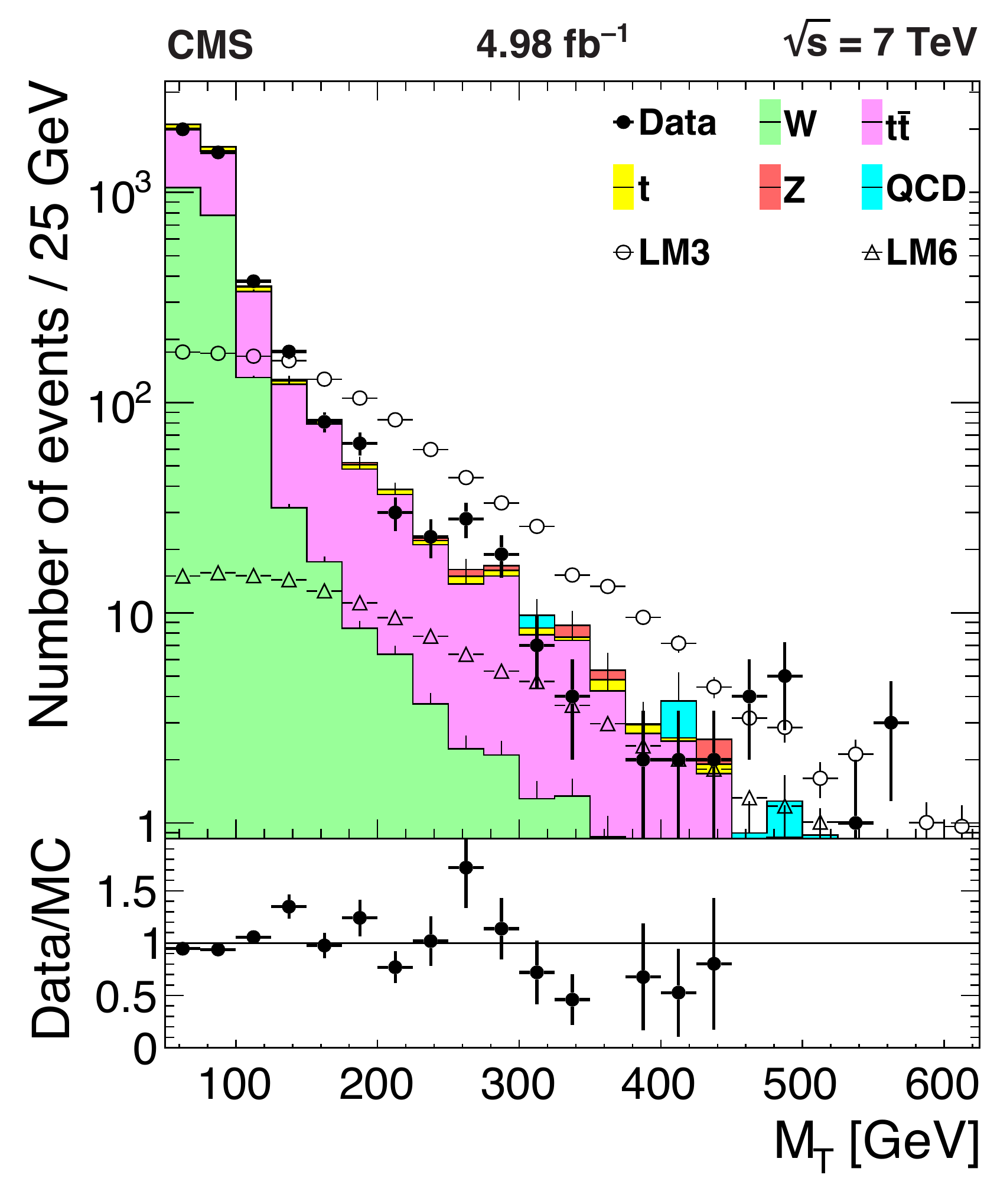}
  \caption{The distributions of $n_{\mathrm{jets}}$, \HT, $\Delta \phi$, and \mT\ for data (solid circles), simulated SM (stacked shaded histograms), LM3 (open circles), and LM6 (open triangles) events after preselection.  The small plot beneath each distribution shows the ratio of data to simulated SM yields. The muon and electron channels are combined.}
  \label{fig:datamc_combined}
\end{figure*}

Figure~\ref{fig:datamc_combined} shows the distributions of these variables for simulated
SM and SUSY events. The most powerful input variable is \mT; $n_{\mathrm{jets}}$ and \HT
also have considerable discriminating power. The $\Delta\phi(\mathrm{j}_1,\mathrm{j}_2)$
variable is weaker, but it still improves the sensitivity of the search.
Lepton \pt also discriminates between the SM and SUSY, but it is not included in
the ANN because its strong correlation with \ETslash in the SM would spoil
the background estimate. Additional variables either do little to improve sensitivity
or introduce a correlation between \zANN and \ETslash. The input variables
have similar distributions in the muon and electron channels, so we choose to
train the ANN on the two channels combined, and use the same ANN for both.
In general, the SM simulation describes the data adequately apart from a possible small
structure near 130\GeV in the \amT distribution. Reweighting the simulation to match
the \amT distribution in data does not affect the results of the analysis.

The ANN infrastructure uses standard {\sc Root} utilities~\cite{TMVA_FIXME}. During training, weights are
determined  that  minimize the root-mean-square deviation of background events from zero and
signal events from unity.  For the SUSY parameter space under study, our sensitivity depends
only mildly on the details of the signal sample that trains the ANN.  Specifically, for
 LM points 0 through 13~\cite{PTDR2}, the sensitivity is comparable (less than ~30\% variation)
whether the ANN is trained on LM0, LM6 or LM9, even though these three training samples
have rather different characteristics.  We select LM0 for training because it gives the best overall performance.
The SM simulation provides the background sample.

\begin{figure}[tbp!]
  \centering
   \includegraphics[width=0.45\textwidth]{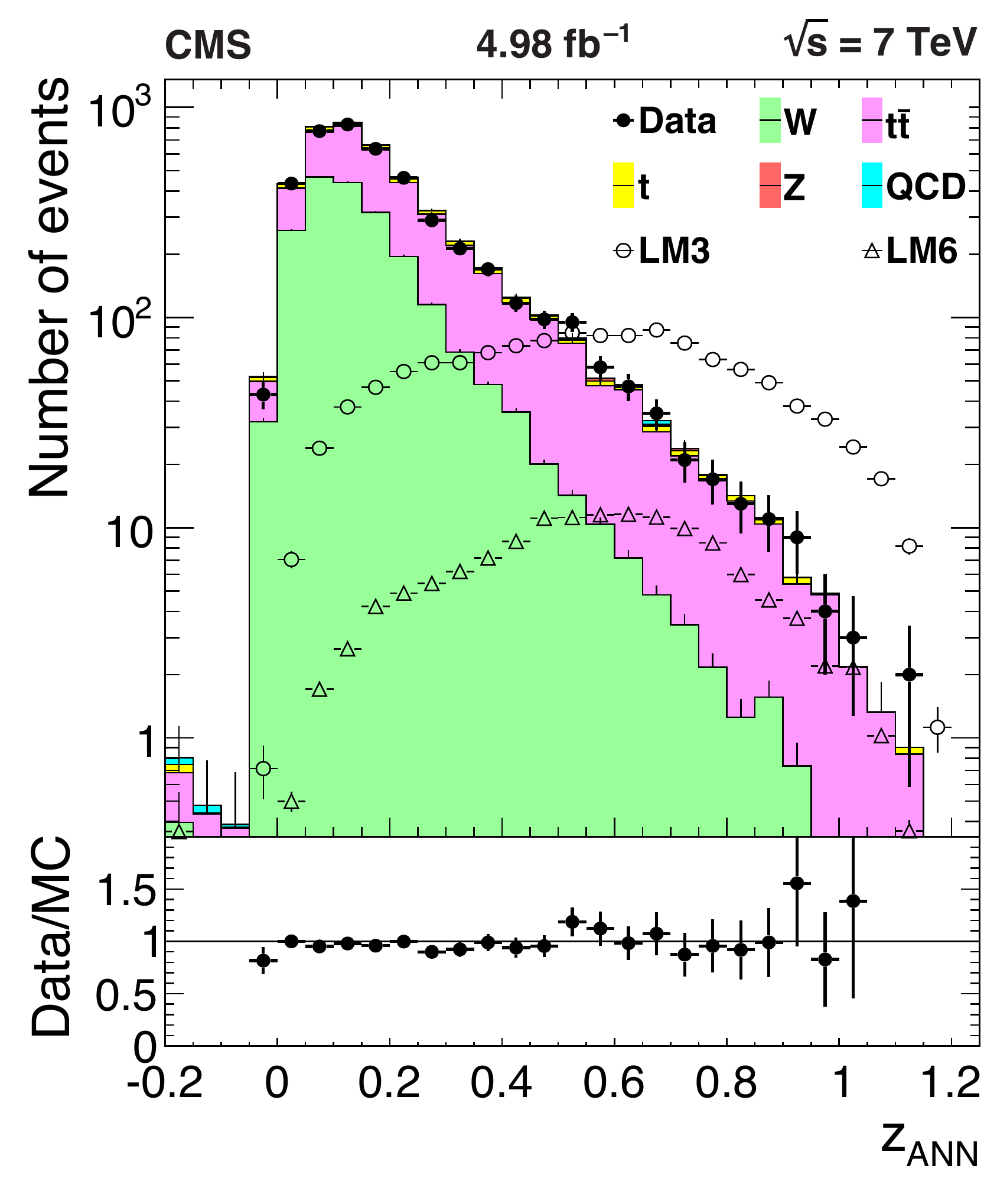}
  \caption{The \zANN distribution of the data (solid circles) and simulated SM (stacked shaded histograms), LM3 (open circles), and LM6 (open triangles) events, after preselection. The small plot beneath shows the ratio of data to simulated SM yields.}
  \label{fig:annOutput_dataMC}
\end{figure}

Figure~\ref{fig:annOutput_dataMC} compares the distributions of \zANN for data and SM
simulation for all events surviving the preselection. The two distributions are consistent within the
uncertainties.
The SM contribution is concentrated at small values of \zANN, while the LM3 and LM6 SUSY distributions,
which are also shown, extend to high values of \zANN where the SM is suppressed.

We define two signal regions in the two-dimensional \ETslash and \zANN plane. One region,
referred to as the ``low-\ETslash'' signal region, has $\zANN > 0.4$ and  $350<\ETslash<500$\GeV,
while the other, the ``high-\ETslash'' signal region, has the same \zANN range,
but $\ETslash > 500\GeV$. The high-\ETslash signal region minimizes the probability
that the expected background fluctuates up to a LM6 signal when signal contamination
is taken into account. We observe 10 events in the low-\ETslash signal region and 1 event in the high-\ETslash signal region.

\subsection{Background estimation using the ANN sidebands}

\begin{figure*}[tbp!]
  \centering
\includegraphics[width=0.90\textwidth]{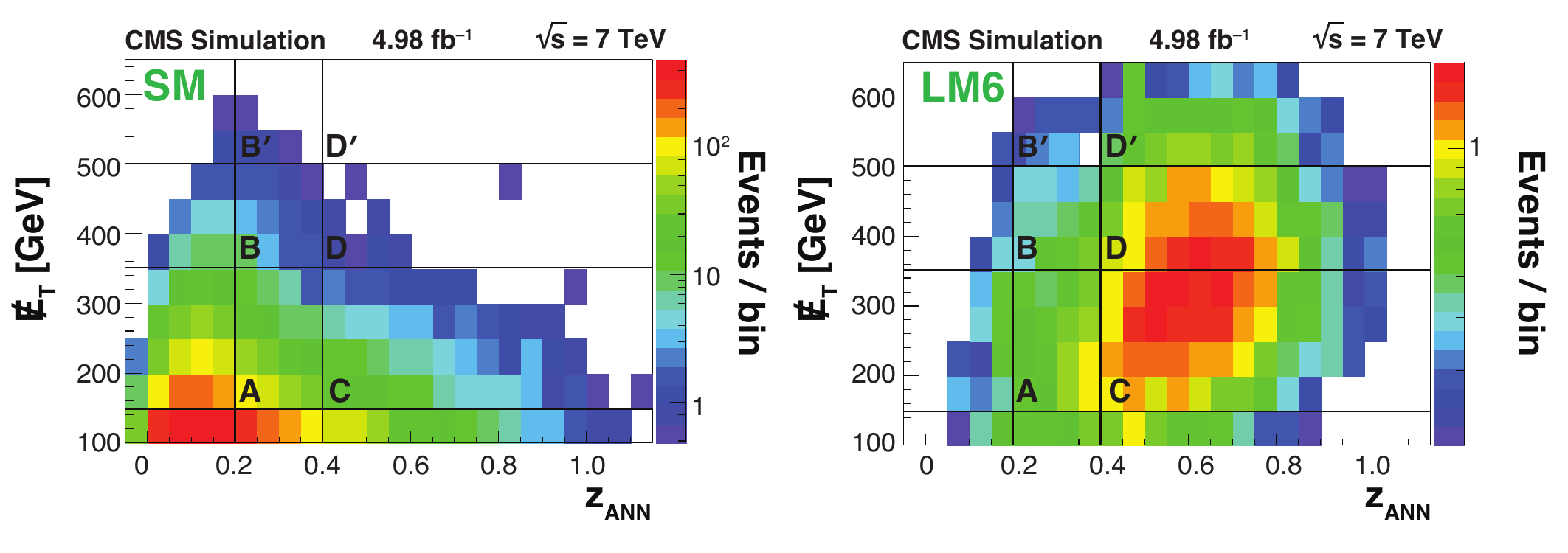}
  \caption{The yields of simulated SM (left) and LM6 (right) events in the \ETslash versus \zANN plane.  The regions D and D' are the low-\ETslash and high-\ETslash signal regions.  The sideband regions  are also indicated. }
  \label{fig:abcd}
\end{figure*}

The sidebands in the two dimensional plane of \ETslash and \zANN provide a strategy for estimating the background. The signal and sideband regions are shown in Fig.~\ref{fig:abcd} and are denoted A, B, C, and D for the low-\ETslash signal region and A, B$^\prime$, C, and D$^\prime$ for the high-\ETslash signal region. The choice of boundaries for the sideband regions balances the competing needs of statistics and insensitivity to signal contamination against preserving similar event compositions in the signal and sideband regions.

\begin{figure}[tbp!]
  \centering
     \includegraphics[width=0.45\textwidth]{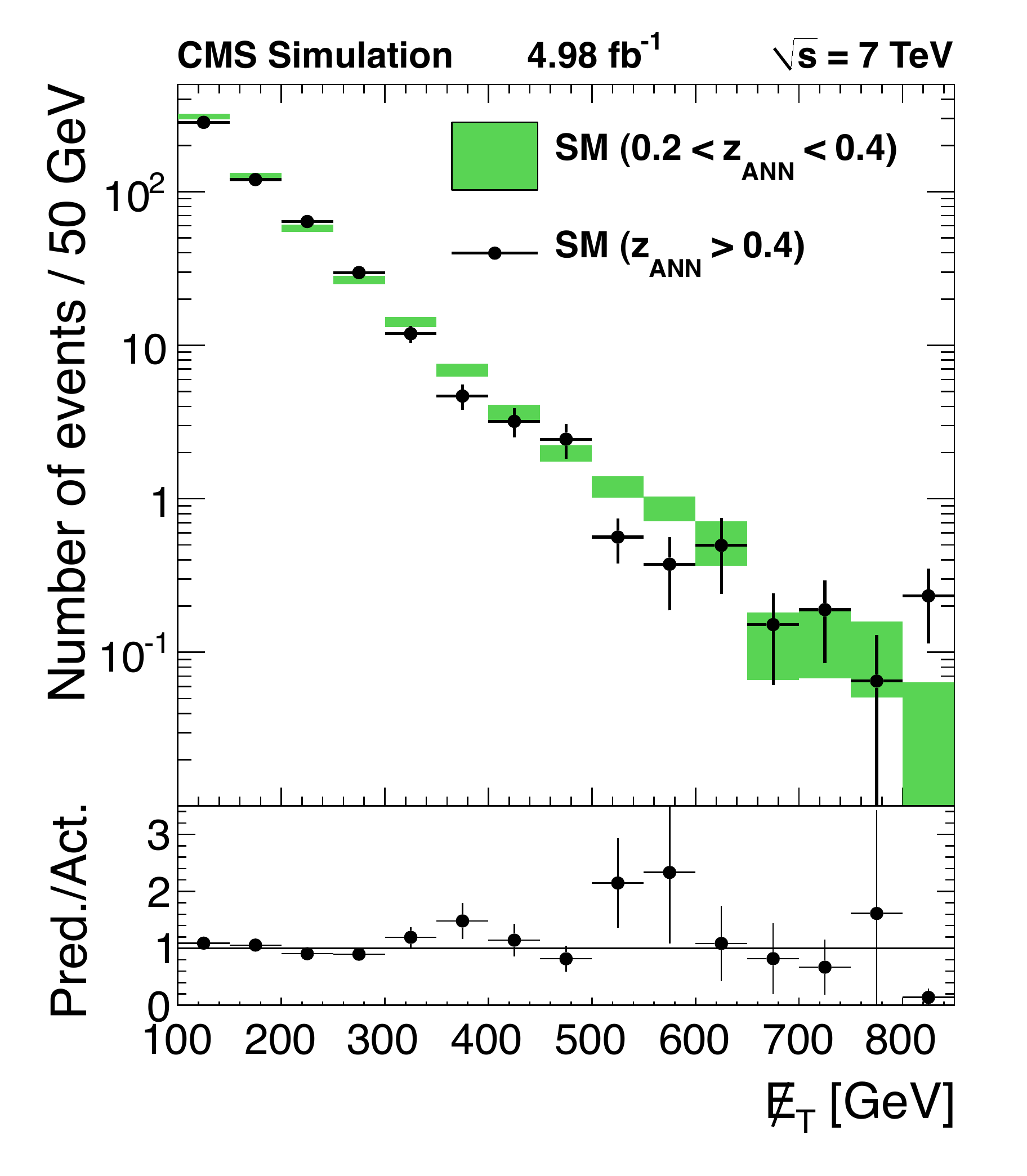}
     \includegraphics[width=0.45\textwidth]{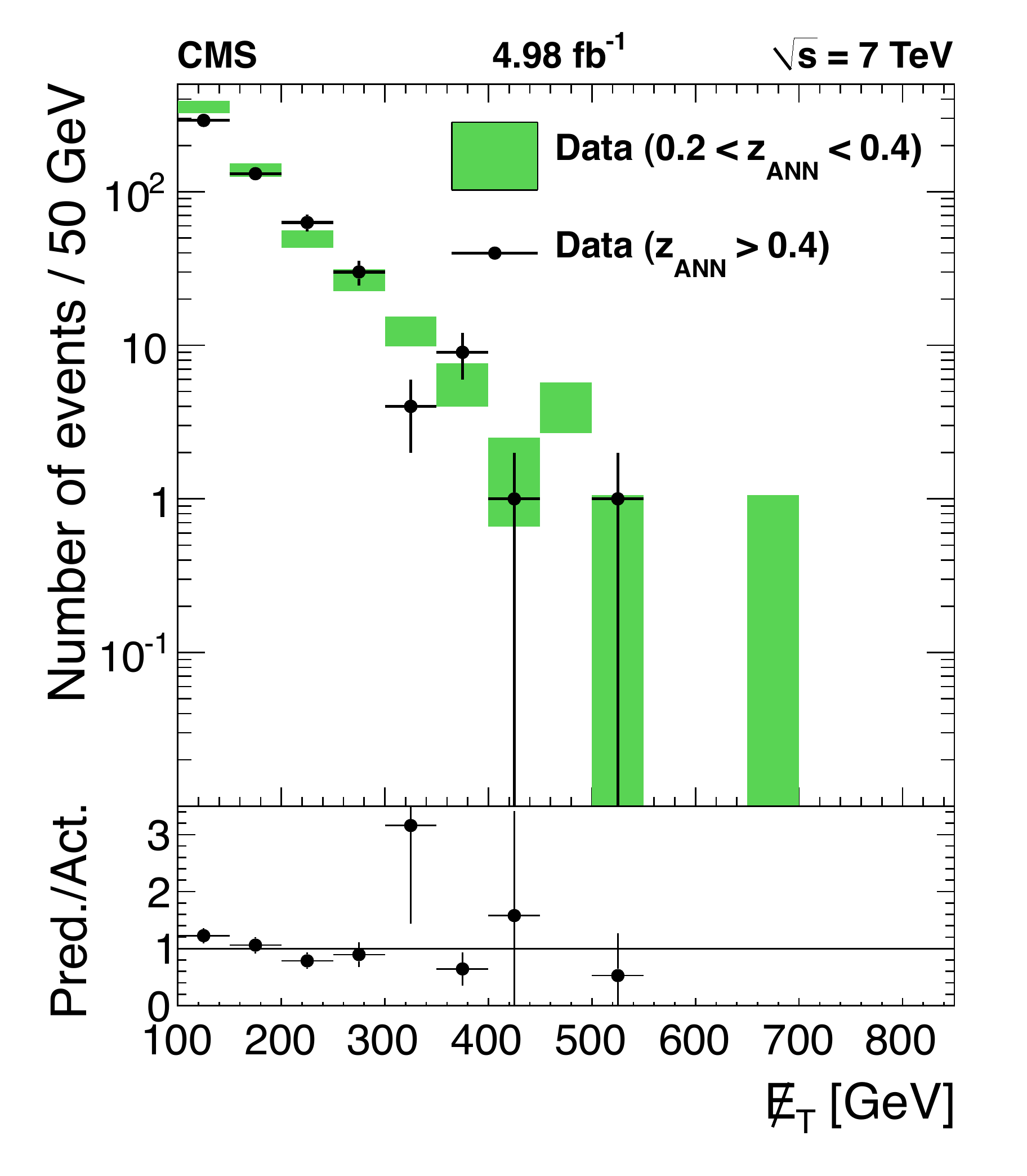} \caption{The \ETslash distributions of events in the \zANN signal region (solid circles) and sideband (green bars) for simulated SM (\cmsLeft) and data (\cmsRight) events. The distributions are normalized in the \ETslash sideband, $150<\ETslash< 350$\GeV (regions A and C for the two distributions respectively). The rightmost histogram bin includes overflow.   The small lower plots show the ratio of normalized sideband to signal yields.}
       \label{fig:bkgdPred}
\end{figure}

The predicted yield in region D is given by
\begin{equation}
N_{\mathrm{D,pred}} = \frac{ N_{\mathrm{B}} \times N_{\mathrm{C}} }{ N_{\mathrm{A}}  },
\end{equation}
where $N_i$ is the yield in region $i$, and the predicted yield in region D$^\prime$ is defined similarly. This procedure is
equivalent to using the \ETslash distribution of the $z_{\mathrm{ANN}}$ sideband regions
(A, B, and B$^\prime$) as a template for the \ETslash distribution of events with
high \zANN (C, D and D$^\prime$), normalized using the yields in regions A and C.
We test this estimation procedure using SM simulation:
Fig.~\ref{fig:bkgdPred} (\cmsLeft) shows that the \ETslash distributions for low and high \zANN are similar.

If a signal is present, it enters primarily in the signal regions D and D$^\prime$, but there are also significant contributions relative to the SM in regions B and B$^\prime$, somewhat increasing the predicted backgrounds in D and D$^\prime$.   This effect is accounted for in the final results.

Table~\ref{tab:abcdCounts} summarizes the event yields in the sideband subtraction regions for the various components of the SM background.  The W+jets and \ttbar dominate in all the regions, though their relative proportion varies.  The W+jets events are most important at low \zANN since \mT, which largely drives \zANN, tends to peak near the W-boson mass.  Because the W bosons (and hence their daughters) can be highly boosted, these events extend to very high values of \ETslash. As seen in Fig.~\ref{fig:annOutput_dataMC}, \ttbar events are more likely to have high values of \zANN than are W+jets events; this is because of the presence of dilepton \ttbar events, in which both W bosons (from the top quark pair) decay leptonically, but only one lepton is identified (dilepton ($\ell$)), giving large \mT. There is also a small contribution from events in which the lepton comes from the decay of a $\tau$ produced from a top quark decay, with the other top quark decaying either leptonically (dilepton ($\tau \rightarrow \ell$)) or hadronically (single $\tau$). The remaining small backgrounds come from single-top-quark, QCD multijet and Z+jets events.

There are too few events in the simulated QCD multijet and Z+jets samples to populate the high \ETslash regions (B, B$^\prime$, D and D$^\prime$). For the results quoted in Table~\ref{tab:abcdCounts} for QCD multijet and Z+jets events, we employ an extrapolation technique based on loosening the $z_{\mathrm{ANN}}$ and \ETslash requirements. The extrapolated numbers for all the regions are consistent with those obtained from the simulated samples. The simulated yields in the sideband and signal regions indicate that QCD multijet and Z+jets events are negligible.

\begin{figure*}[tbp!]
  \centering
    \includegraphics[width=0.45\textwidth]{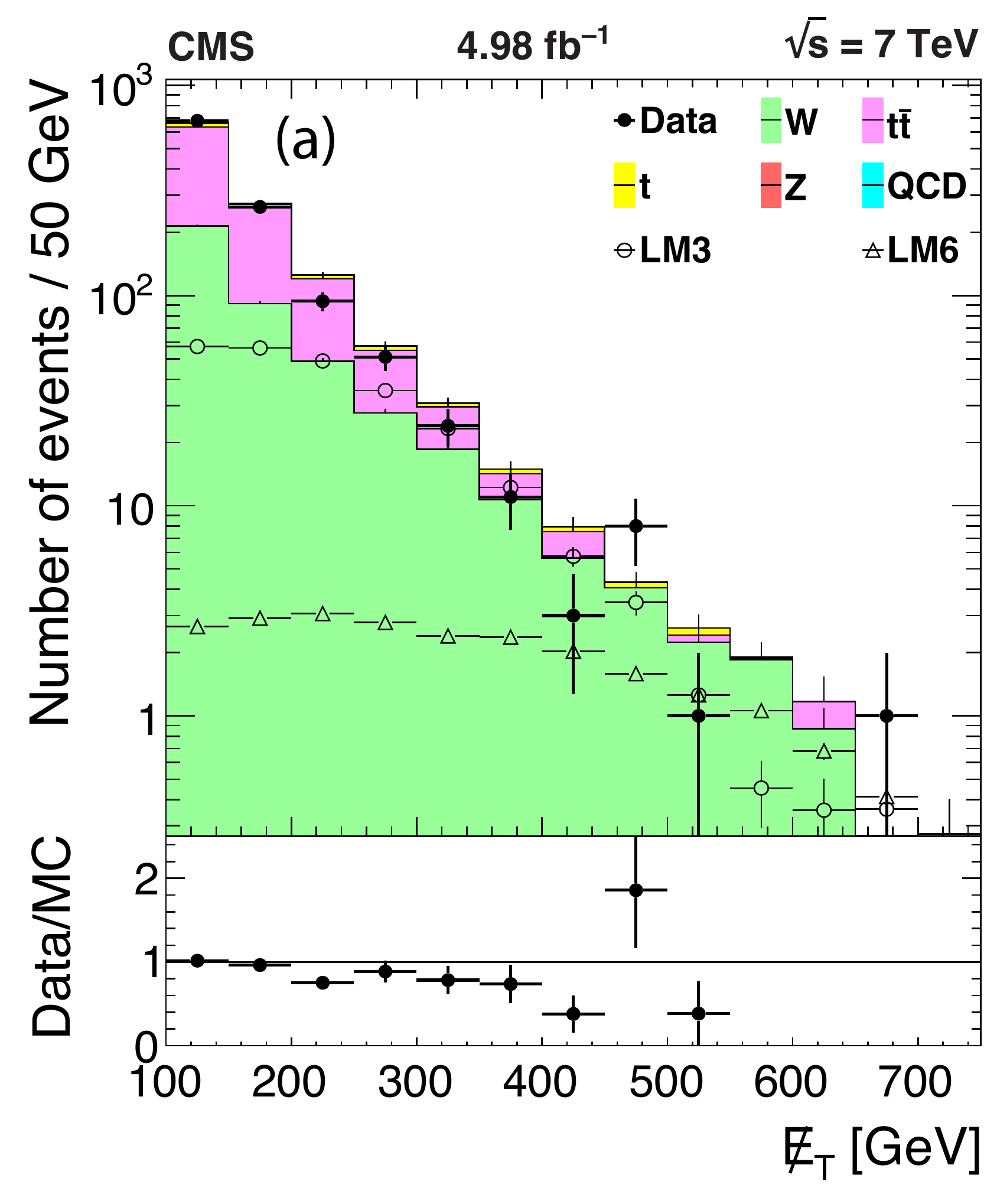}
    \includegraphics[width=0.45\textwidth]{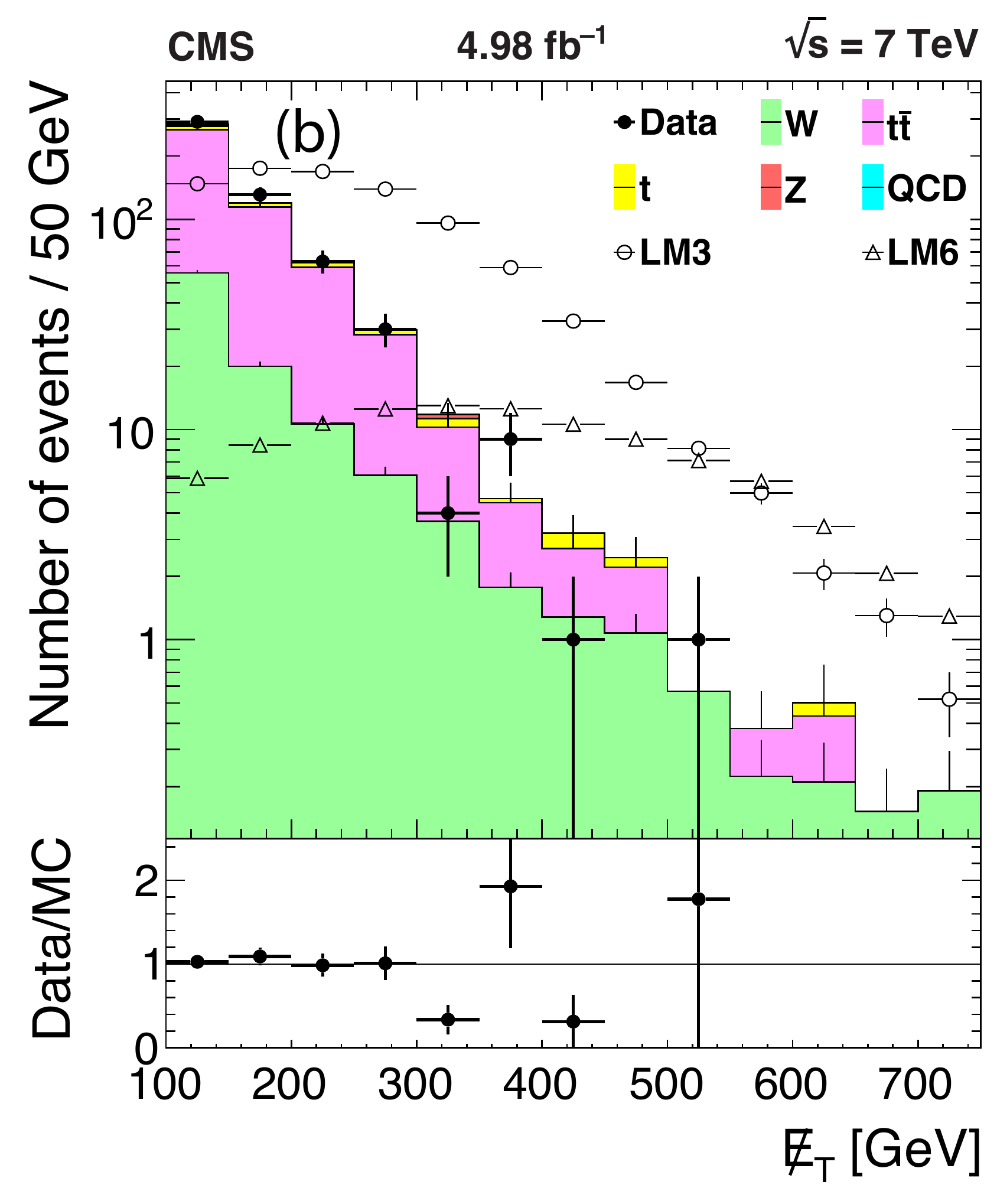}
    \includegraphics[width=0.45\textwidth]{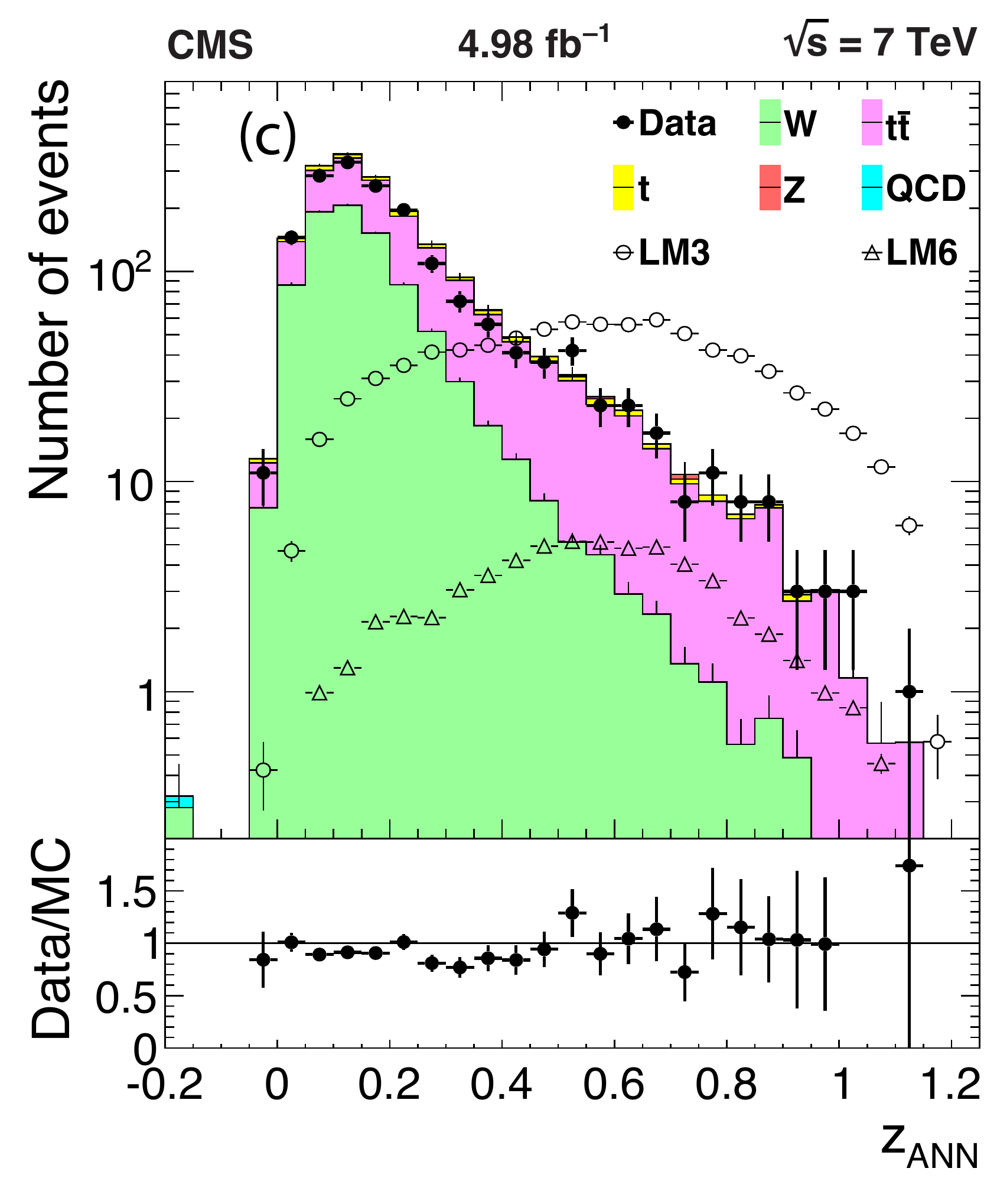}
    \includegraphics[width=0.45\textwidth]{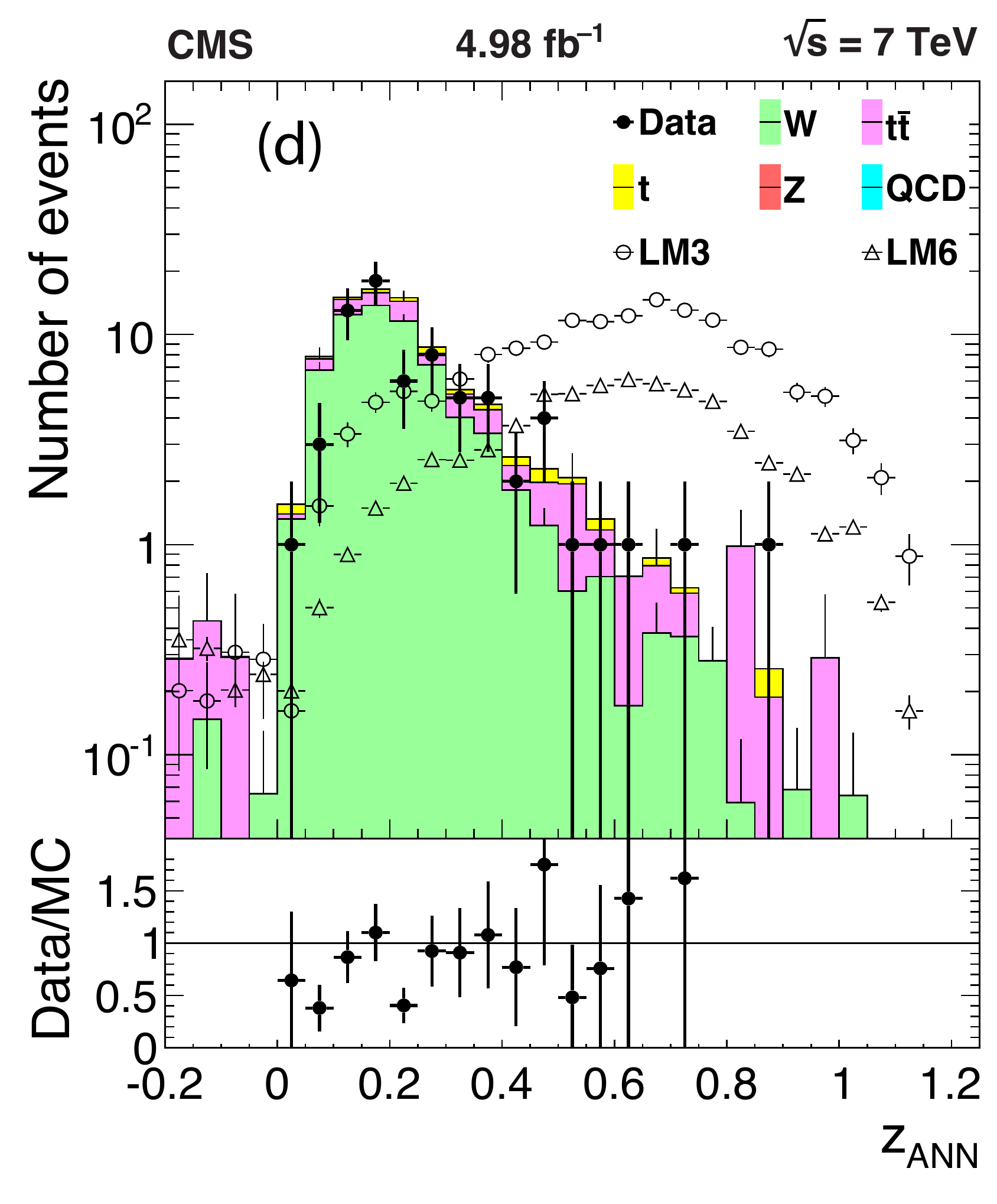}
  \caption{Distributions of \ETslash for (a) $0.2 < \zANN < 0.4$ and (b) $z_{ANN} >$ 0.4, and distributions of $z_{\mathrm{ANN}}$ for (c) $150<\ETslash< 350$\GeV and (d) $\ETslash>350$\GeV. The samples shown are data (solid circles), simulated SM (stacked shaded histograms), LM3 (open circles), and LM6 (open triangles) events. The small plot beneath each distribution shows the ratio of data to simulated SM yields.}
  \label{fig:datamc_met_lowhighANN}
\end{figure*}

The total SM simulation yields agree well with data in all regions, suggesting that the data share the main features described above.   The \zANN and  \ETslash  distributions are shown in Fig.~\ref{fig:datamc_met_lowhighANN}.

\begin{table*}
  \centering
  \topcaption{Event yields for the sideband (SB) and signal regions used in the ANN method.  The uncertainties listed are statistical only. }
\small{
\begin{tabular}{ l  D{,}{\,\pm\,}{4.3}   D{,}{\,\pm\,}{4.3}   D{,}{\,\pm\,}{4.4}   D{,}{\,\pm\,}{4.3}   D{,}{\,\pm\,}{4.3}   D{,}{\,\pm\,}{4.3} }
\hline
Sample type & \multicolumn{1}{c}{A} &  \multicolumn{1}{c}{B} &  \multicolumn{1}{c}{B$^\prime$} &  \multicolumn{1}{c}{C} &  \multicolumn{1}{c}{D} &  \multicolumn{1}{c}{D$^\prime$} \\
\hline
  &  \multicolumn{1}{c}{\zANN\ SB}& \multicolumn{1}{c}{\zANN\ SB}& \multicolumn{1}{c}{\zANN\ SB}& \multicolumn{1}{c}{\zANN\ signal }& \multicolumn{1}{c}{\zANN\ signal}& \multicolumn{1}{c}{\zANN\ signal}\\
  &  \multicolumn{1}{c}{\ETslash SB}&  \multicolumn{1}{c}{Low \ETslash} &  \multicolumn{1}{c}{High \ETslash} &  \multicolumn{1}{c}{\ETslash SB}&  \multicolumn{1}{c}{Low \ETslash} &  \multicolumn{1}{c}{High \ETslash} \\[1ex]

  \ttbar single lepton & 210  ,  8 & 4.8  ,  1.1 & 0.2  ,  0.2 & 55  ,  4 & 1.7  ,  0.7 &  \multicolumn{1}{c}{$0.0^{+0.2}_{-0.0}$}  \\
  \ttbar dilepton ($\ell$) & 56  ,  4 & 0.3  ,  0.3 & 0.01  ,  0.01 & 109  ,  5 & 3.6  ,  1.0 &  0.2  ,  0.2 \\
  \ttbar dilepton ($\tau \rightarrow \ell$) & 3.9  ,  1.1 & 0.01  ,  0.01 & 0.3  ,  0.3 & 4.3  ,  1.0 & \multicolumn{1}{c}{$0.0^{+0.2}_{-0.0}$} & 0.2  ,  0.2 \\
  \ttbar single $\tau$ & 9.4  ,  1.7 & 0.3  ,  0.3 & \multicolumn{1}{c}{$0.0^{+0.2}_{-0.0}$} & 2.6  ,  0.8 & \multicolumn{1}{c}{$0.0^{+0.2}_{-0.0}$} & \multicolumn{1}{c}{$0.0^{+0.2}_{-0.0}$} \\

 Total \ttbar & 279  ,  9 & 5.4  ,  1.2 & 0.5  ,  0.3 & 171  ,  7 & 5.3  ,  1.2 & 0.4  ,  0.3 \\
  W+jets & 186  ,  3 & 20.4  ,  1.1 & 5.8  ,  0.6 & 40  ,  2 & 4.1  ,  0.5 & 1.6  ,  0.3  \\
  Single top quark  & 20  ,  1 & 1.5  ,  0.3 & 0.2  ,  0.1 & 11  ,  1 & 0.9  ,  0.2 & 0.1  ,  0.1 \\
  Z+jets & 2.1  ,  0.3 & \multicolumn{1}{c}{$0.07^{+0.12}_{-0.07}$} & \multicolumn{1}{c}{$0.07^{+0.12}_{-0.07}$} & 0.8  ,  0.1 & \multicolumn{1}{c}{$0.03^{+0.05}_{-0.03}$} &  \multicolumn{1}{c}{$0.03^{+0.05}_{-0.03}$}  \\
  QCD multijet & \multicolumn{1}{c}{$0.3^{+0.4}_{-0.3}$} & \multicolumn{1}{c}{$0.00^{+0.04}_{-0.00}$} & \multicolumn{1}{c}{$0.00^{+0.04}_{-0.00}$} & 0.1  ,  0.1& \multicolumn{1}{c}{$0.00^{+0.02}_{-0.00}$} & \multicolumn{1}{c}{$0.00^{+0.02}_{-0.00}$} \\

  Total SM & 487  ,  9 & 27.3  ,  1.8 & 6.6  ,  0.7 & 224  ,  7 & 10.3  ,  1.3 & 2.1  ,  0.4 \\

  Data & 433 & 22 & 2 & 228 & 10 & 1 \\
  LM3 & 164  ,  3 & 21  ,  1 & 2.9 , 0.4 & 579  ,  6 & 108  ,  3 & 17.8  ,  1.1\\
  LM6 & 11.2  ,  0.3 & 6.0  ,  0.2 & 3.9  ,  0.1 & 44.6  ,  0.5 & 32.1  ,  0.4 & 21.0  ,  0.3\\
\hline
\end{tabular}
}
  \label{tab:abcdCounts}
\end{table*}

\subsection{Results of the ANN method}

Figure~\ref{fig:bkgdPred} (\cmsLeft) shows the results of applying the background estimation method  to the SM simulation.
We find that the method correctly predicts the background within a factor of
$\kappa = \mathrm{D}^{\prime}/ \mathrm{D}^{\prime}_{\text{pred}}$ of $0.82 \pm 0.12\stat$
in the low-\ETslash signal region and $0.69 \pm 0.16\stat$ in the high-\ETslash signal region.
The modest deviation from unity results from a
correlation between \zANN and \ETslash that arises because the W+jets background, which extends to large \ETslash values,
dominates in the \zANN sideband (because it tends to have \mT\ near the W mass), whereas dileptonic \ttbar events,
with their somewhat softer \ETslash spectrum, dominate in the \zANN signal region.

Figure~\ref{fig:bkgdPred} (\cmsRight) shows the \ETslash distributions of the data in the high and low \zANN regions,
after normalizing in the region $150 <\ETslash<350$\GeV (A and C).  Because the SM simulation
appears to describe the data well, with, for example, consistent exponential decay constants describing
the \ETslash  distributions in the ANN sidebands, we choose to scale the background prediction of the data by $\kappa$.
The uncertainty in the background from the relative cross sections of SM processes and other effects
is quantified in Section~\ref{sec:SystematicUncertainties}.  In the low-\ETslash signal region,
we expect $9.5\pm 2.2\stat$ events, and in the high-\ETslash signal region
$0.7 \pm 0.5\stat$ events.  The observed yields are 10 and 1 events, respectively,
consistent with the background prediction.
\section{Systematic uncertainties}
\label{sec:SystematicUncertainties}
\label{ssec:SystematicsLSMethod}
\label{ssec:SystematicsLPMethod}

Systematic uncertainties affect both the background estimates and the signal efficiencies.
The sources of systematic uncertainty in the background predictions vary among the three
methods, both because the final event selections differ and because the background estimation
methods themselves differ. The systematic uncertainties stem from lack of perfect knowledge
of the detector response and from uncertainties in the properties of the SM backgrounds.
Common uncertainties for all methods are described in Section~\ref{ssec:Uncertainties},
while details that are specific to each method are given in Sections~\ref{ssec:LSUncertainties}, \ref{ssec:LPUncertainties},
and~\ref{ssec:ANNUncertainties} for the LS , \Lp, and ANN methods, respectively.
Tables~\ref{tab:LeptonSpectrumSys}, \ref{tab:LP_Systematics}, and \ref{tab:systUncSum} list the main uncertainties associated with each method.
The systematic uncertainties affecting the signal efficiency and luminosity,
which are largely common to all methods, are described
in Section~\ref{ssec:SignalEfficiencyUncertainties}.

\subsection{Common uncertainties in the background predictions}
\label{ssec:Uncertainties}

\begin{table*}[thb]
\topcaption{Sources of systematic uncertainties for the LS method and their effects on the background prediction in bins of \ETslash.
The full list of systematic uncertainties is given for $\Ht>750$\GeV, and the total uncertainties are shown for
$\HT > 500$\GeV and $\HT > 1000$\GeV.
Each uncertainty is expressed as a change in the ratio of the predicted to
the true number of events (evaluated with simulation). Uncertainties
associated with the dilepton and QCD backgrounds are discussed in the text.  The total uncertainty is the individual uncertainties summed in quadrature.}
\label{tab:LeptonSpectrumSys}
\begin{center}
\begin{tabular}{l c c c c}
\hline
\ETslash [\GeVns{}]:  &   [250--350)  &  [350--450) & [450--550) & $\ge$550 \\
 & (\%) & (\%) & (\%) & (\%)\\
\hline
\multicolumn{5}{ c }{$\HT>750$\GeV } \\
Jet and \ETslash energy scale          &   11  & 13  &   14     & 16  \\
Lepton efficiency &  1   & 1  &   1     & 1  \\
Lepton $\pt$ scale            &  1   & 2  &   6     & 2  \\
$\sigma(\cPqt\cPaqt)$ and $\sigma(\PW)$   &  1   & 1  &   4     & 4  \\
\PW\ polarization in \cPqt\cPaqt      &   1   & 1  &   1     & 1  \\
\PW\ polarization in \PW+jets       &   3   & 4  &   12    & 11  \\
\cPZ+jets background           &  4   & 4  &   4     & 4  \\
SM simulation statistics (K-factors)           &  4   & 7  &   12    & 17 \\
Total systematic uncertainty                      &  13  & 16  &   24     & 27  \\
\noalign{\vspace{6pt}}
\multicolumn{5}{c}{$\HT>500$\GeV } \\
Total systematic uncertainty                      &  16  & 18  &   29     & 30  \\
\noalign{\vspace{6pt}}
\multicolumn{5}{ c}{$\HT>1000$\GeV } \\
Total systematic uncertainty                      &  15  & 18  &   28     & 32  \\
\hline
\end{tabular}
\end{center}
\end{table*}

\begin{table*}[thb]
\topcaption{
Sources of systematic uncertainties for the \Lp method and their effects on
the background prediction in bins of \stlep for the muon and electron channels.  The full list of systematic
uncertainties are given for the range $500<\Ht<750$\GeV, and the total uncertainties are shown for
the two ranges $750<\HT<1000$\GeV and $\Ht>1000$\GeV.
The total uncertainty is the individual uncertainties summed in quadrature.}
\label{tab:LP_Systematics}
\centering
\begin{tabular}{l cc | cc | cc | cc}
\hline
\stlep range [\GeVns{}]: & \multicolumn{2}{c|}{[150--250)} & \multicolumn{2}{c|}{[250--350)} & \multicolumn{2}{c}{[350--450)} &  \multicolumn{2}{c}{$\ge$450} \\
 & \multicolumn{2}{c|}{(\%)} & \multicolumn{2}{c|}{(\%)} & \multicolumn{2}{c|}{(\%)} &  \multicolumn{2}{c}{(\%)} \\
Channel &\hspace{.25cm}$\mu\hspace{.25cm}$&\hspace{.25cm}$\Pe\hspace{.25cm}$&\hspace{.25cm}$\mu\hspace{.25cm}$&\hspace{.25cm}$\Pe\hspace{.25cm}$&\hspace{.25cm}$\mu\hspace{.25cm}$&\hspace{.25cm}$\Pe\hspace{.25cm}$&\hspace{.25cm}$\mu\hspace{.25cm}$&\hspace{.25cm}$\Pe\hspace{.25cm}$\\
\hline
\multicolumn{9}{c}{$500<\Ht<750$\GeV} \\
Jet and \ETslash energy scale        &6&6&4&5&5&9&9&9\\
Lepton efficiency  &5&5&5&2&3&1&1&2\\
Lepton $\pt$ scale        & 0 &-&1 &-& 1&- & 2&-\\
$\sigma(\cPqt\cPaqt)$ and $\sigma(\PW)$ & 3&1&1&1&1&2&1&1\\
\PW\ polarization in \cPqt\cPaqt     &0&1&1&1&1&1&1&2\\
\PW\ polarization in \PW+jets      &2&1&2&1&2&3&3&4\\
\ETslash resolution  &2&2&1&1&1&2&4&4\\
\cPqt\cPaqt($\ell\ell$)  &5&5 & 5&5 & 3&3 & 1&1\\
SM simulation statistics & 1&1&2&2&4&5&6&7\\
Total systematic uncertainty  & 11 &10&9&8&8&12&13&13 \\
\noalign{\vspace{6pt}}
\multicolumn{9}{c}{$750<\HT<1000$\GeV}\\
Total systematic uncertainty  &  9  &  12   & 10 &  11  & 13  & 13 &   12 &13 \\
\noalign{\vspace{6pt}}
\multicolumn{9}{c}{$\HT>1000$\GeV } \\
Total systematic uncertainty  & 10& 15 &  13 & 15 & 20 & 18 &  16 & 20 \\
\hline
\end{tabular}
\end{table*}

\begin{table}[htb]
  \centering
  \topcaption{Sources of systematic uncertainties for the ANN method and their effects on the background prediction in bins
 of \ETslash.
The total uncertainty is the individual uncertainties summed in quadrature.}
\begin{tabular}{ l  c  c }
\hline
\ETslash range [\GeVns{}]: & [350--500) & $\ge$500\\
  & (\%) & (\%) \\
\hline
  Jet and \ETslash energy scale & 3 & 4\\
 Lepton $\pt$ scale & 3 & 5 \\
  Lepton efficiency & 0.3  & 0.4   \\
$\sigma(\cPqt\cPaqt)$ and $\sigma(\PW)$ & 3 & 2 \\
 \PW\ polarization in \PW+jets & 1  & 3  \\
 W boson \pt spectrum in \PW+jets & 10  & 2  \\
 \cPqt\cPaqt($\ell\ell$) & 1  & 7  \\
 Other backgrounds & 1 & 1   \\
  SM simulation statistics & 15 & 23  \\

  Total systematic uncertainty &19 &26  \\
\hline
\end{tabular}
  \label{tab:systUncSum}
\end{table}

The jet energy scale (JES) and its effect on \ETslash in the event can affect the \Ht\ and \ETslash distributions
and can also lead to differences between the lepton $\pt$ spectrum and \ETslash spectrum. To understand
the effects of energy-scale variations, we vary the jet energy scale as a function of $\pt$ and $\eta$
by amounts determined in independent studies of jet energy scale uncertainties~\cite{JES}, and corresponding to 2\GeV or less for jets with $\pt>40$\GeV, and then
recompute \HT\ and \ETslash. We also vary the energy scale of ``unclustered'' calorimeter
deposits by 10\% to take into account energy not clustered into jets (this effect is very small).

The uncertainty in the lepton efficiency accounts for differences between data and simulation
and uncertainties in the trigger efficiencies.   The lepton efficiencies are studied using a sample of lepton pairs with invariant mass close to the Z peak, in which one lepton satisfies tight selection criteria, and the second, reconstructed with relaxed criteria, serves as a probe of the tighter reconstruction and isolation requirements (``tag-and-probe'' method~\cite{tagandprobe}).  Discrepancies between the data and simulation for electrons are maximal at low $\pt$
(10\% effect at around 20\GeV), and we reweight events as a function of lepton $\pt$ to quantify the effect.  The total lepton efficiency in data is described by simulation with an accuracy of 3\%.  Studies of the trigger that separately determine the efficiencies of the $\HT^\text{trigger}$,  $\ETslash^\text{trigger}$, and lepton requirements show that the lepton inefficiencies dominate, and amount to 2\% to 3\% for leptons that are reconstructed successfully offline.
Muon $\pt$ scale uncertainties are obtained from the study of the $q/\pt$ (transverse curvature with sign given by the electric charge $q$) distribution of muons in $Z$ events in data. By
comparing the $q/\pt$ distribution of positive and negative muons it is possible to quantify the amount of bias in the measurement of $q/\pt$.

The relative amount of \cPqt\cPaqt\ and  \PW+jets background affects each analysis method through corrections obtained from simulation.
The contributions from \cPqt\cPaqt\ and  \PW+jet have not been specifically measured in the narrow region of phase space studied in this analysis and
their relative contribution must be evaluated.
The \cPqt\cPaqt\ cross section is validated using an algorithm based on
the reconstructed top-quark masses for both the hadronic and the
leptonic top-quark decays.  The uncertainty in the \cPqt\cPaqt\ cross section is determined by comparing yields in data and simulation
after a selection based on top mass variables.
The \PW+jets cross section is validated by comparing event yields between data and simulation
in $Z$+jets events in a dedicated dilepton event selection with similar kinematics. We assign an uncertainty to the \PW+jets cross section based
on the agreement of the data and simulation in the $Z$+jets sample.
Using the uncertainties obtained for the \cPqt\cPaqt\ and $W$+jets cross sections, we probe different relative contributions of
\cPqt\cPaqt\ and $W$+jets events in our sample and the effect on our background predictions.

Uncertainties in the polarization fraction for the \PW~boson, either in  $\cPqt\cPaqt$ or \PW+jets events, must be taken into account.
For the \PW~polarization in $\cPqt\cPaqt$ events, the theoretical uncertainties are very small (see Section~\ref{sec:EventSamples}) and have negligible effect on the background predictions.
The \PW~polarization in \PW+jets events, which is described in more detail in Section~\ref{sec:EventSamples}, is more complicated than in $\cPqt\cPaqt$ production.
In this case, we consider the effect of conservative variations of
the helicity fractions in bins of \PW-boson $\pt$ and $\eta$ with
respect to the theoretical NLO calculations~\cite{ref:Blackhat}.

For the dilepton $\cPqt\cPaqt$ background,  \cPqt\cPaqt($\ell\ell$), the uncertainties are evaluated somewhat differently for the different methods. In the \Lp and ANN methods this
background is evaluated together with the same control sample as for the main single-lepton background prediction.
Uncertainties in the prediction can arise from finite detector acceptance, inefficient lepton identification, and cross section uncertainties.
In the LS method the dilepton $\cPqt\cPaqt$ background is not predicted using the single-lepton background prediction and separate
control samples must be used. Thus the uncertainties for the dilepton $\cPqt\cPaqt$ background are estimated separately and described in the next section.

The small residual QCD multijet background is probed by inverting the requirement on $I_\text{rel}^\text{comb}$ or the electron selection criteria to obtain QCD dominated control samples. Contamination from \cPqt\cPaqt\ and \PW+jet\ events in these control samples must be considered and the uncertainties on their cross sections are the dominant uncertainty for these methods.

The $Z$+jets contribution to the signal regions is very small and uncertainties on this background prediction come from lepton efficiency and cross section uncertainties. In addition, for the LS method there is a small $Z$+jets contamination to the single-lepton control sample, which must be subtracted, and lepton efficiency and cross-section uncertainties are considered for this as well.

\subsection{Lepton Spectrum method background prediction uncertainty}
\label{ssec:LSUncertainties}

For the LS method, the systematic uncertainties for each of the different background predictions from control samples in data (1~$\ell$, dilepton, 1 $\tau$, QCD, and Z+jets) are included in
Tables~\ref{tab:table-ht500}, \ref{tab:table-ht750}, and \ref{tab:table-ht1000}.
To determine the systematic uncertainties for the largest source of background, 1-$\ell$ events
(arising from \cPqt\cPaqt, \PW+jets, and single-top processes), we evaluate deviations for the \ETslash-dependent
correction factor, which is determined from simulation and applied to the 1-$\ell$
background prediction (see Section~\ref{ssec:SingleLeptonBackgroundsLSMethod}).
Table~\ref{tab:LeptonSpectrumSys} gives a breakdown of the contributions of the
systematic uncertainties for the 1-$\ell$ prediction in bins of \ETslash and for $\HT\ >750$\GeV.
The uncertainties in the 1-$\ell$ prediction for the $\HT\ > 500$\GeV and $\HT\ >1$\TeV signal
regions are similar to those listed in Table~\ref{tab:LeptonSpectrumSys}.
The largest source of uncertainty arises from the potential difference in
the muon $\pt$ and the \ETslash scales, because the muon $\pt$ spectrum is used to
predict the \ETslash spectrum. The statistical uncertainties in the correction factors
(denoted as K-factors in Table~\ref{tab:LeptonSpectrumSys}) for the 1-$\ell$ method
are slightly smaller than the combined systematic uncertainty of the correction factor.
Table~\ref{tab:LeptonSpectrumSys} does not include an uncertainty from jet resolution effects
because this is taken into account by the smearing of the lepton $\pt$ spectra by QCD
multijet \ETslash templates (described in Section~\ref{ssec:SingleLeptonBackgroundsLSMethod}). For the purposes of setting limits, the total
systematic uncertainty in the 1-$\ell$ background prediction is treated as correlated across all bins in \ETslash.

Tables~\ref{tab:table-ht500}, \ref{tab:table-ht750}, and \ref{tab:table-ht1000} also list the non-single-lepton backgrounds, which account for about 25\% of
the total, with a relative uncertainty of 5--10\% in the lowest-\ETslash bin and about 30\% in the highest-\ETslash bin.
For the dilepton prediction of lost and ignored leptons (described in Section~\ref{ssec:NonSingleLeptonBackgroundsLSMethod})
the main sources of systematic uncertainty arise from the lepton reconstruction and identification efficiencies and the top-quark $\pt$ spectrum.
The uncertainties on the lepton efficiencies are described in Section~\ref{ssec:Uncertainties}, and the
uncertainty associated with the top-quark $\pt$ spectrum is determined from varying
the fraction of events in the tail of this distribution in simulation
in a manner consistent with the uncertainty in this tail as observed in data. This uncertainty
is then propagated through the background determination procedure.

\subsection{Lepton Projection method background prediction uncertainty}
\label{ssec:LPUncertainties}

For the \Lp\ method, the estimate of the total number of events expected from SM processes in the signal
region, $\mathrm{N}_\mathrm{SM}^\text{pred}(\Lp<0.15)$, relies on the knowledge of the translation factor,
$R_\mathrm{CS}$, as well as the number of events observed in the control region, subtracted for the QCD
background, $\mathrm{N}_\text{data}(\Lp>0.3)$.  There are, therefore, two sources of uncertainty in
this estimate: uncertainties in the number of events from EWK processes in the control region
and uncertainty in $R_\mathrm{CS}$.  The relative change on the predicted background from each source of systematic uncertainty is listed in
Table~\ref{tab:LP_Systematics} for both muons and electrons.
The largest uncertainty for high \stlep bins
is the statistical uncertainty in the data in the control region. The second largest uncertainty comes from
the JES uncertainty. The effect from the JES uncertainty is larger in the electron channel,
since the JES affects also the shape of the \Lp\ distribution used in the fit of the control region.
The uncertainty in the resolution of the measurement of the hadronic energy recoiling against the lepton and $\ETslash$
is evaluated conservatively by smearing the total recoil energy in simulation by an additional 7.5\% along the direction 
of the recoil and by 3.75\% in the direction orthogonal to the recoil.
This decreases the resolution more than 10\% for the high recoils
(above 250\GeV) of the signal region and thus covers the difference between data and simulation.

\subsection{ANN method background prediction uncertainty}
\label{ssec:ANNUncertainties}

For the ANN method, the systematic uncertainty in the background prediction is dominated by the statistics of the simulation, which probes for bias in the background estimation.   Another important uncertainty comes from the \pt spectrum of the \PW\ boson in \PW+jets events, since it affects the \ETslash distribution of these events, which preferentially populate the \zANN sideband. To assess the impact, we reweight the \pt\ spectrum of \PW\ boson events, using the differences in the \pt\ spectra of Z bosons in data and simulation as a guide. This uncertainty is driven by the statistics of the Z+jets sample.   The relative proportions of \PW+jets and $\cPqt\cPaqt$ events differ in the \zANN signal and sideband regions so the background prediction depends on their relative cross sections.  Those $\cPqt\cPaqt$ events with two leptons in the final state, only one of which is observed, have large \ETslash and are the source of most SM events in the signal region.  In addition to the $\cPqt\cPaqt$ cross section, this background depends on lepton acceptance and identification inefficiencies.  Additional  sources of systematic uncertainty are the hadronic
and leptonic energy scales.
Table~\ref{tab:systUncSum} summarizes these uncertainties.

\subsection{Signal efficiency and other multiplicative uncertainties}
\label{ssec:SignalEfficiencyUncertainties}

The systematic uncertainty in the signal yield arises from the uncertainty in the signal efficiency.
In general, this uncertainty is correlated across \ETslash or \stlep\ bins.  The JES component of
the signal efficiency uncertainty is computed separately for each model point in CMSSM
and simplified model parameter space and is correlated with the JES uncertainty in the
single-lepton background prediction.  The systematic uncertainties in the signal efficiency
associated with lepton reconstruction and the trigger amount to 3\%.  The uncertainty in
the integrated luminosity is 2.2\%~\cite{Luminosity}.   The systematic uncertainty in the signal efficiency,
not including the JES component, is 6\% for each of the analyses.

\section{Results and interpretation}
\label{sec:Results}

The LS, \Lp, and ANN methods
each yield SM background predictions that are consistent with the
number of events observed in data. We therefore proceed to
set exclusion limits on SUSY model parameters.
All limits are computed using the modified-frequentist CLs method~\cite{Junk:1999kv} with
a one-sided profile likelihood test statistic.
To interpret the absence of an observed signal, three complementary
approaches are used.

\subsection{Constraints on CMSSM parameter space}
First, we scan over models in the CMSSM and determine whether
the number of events predicted at each model point in parameter
space can be excluded by the measurements. This procedure
relies on the fact that the CMSSM parameter space can be described with just five
parameters, and we fix three of them to commonly used values ($A_0=0$\GeV, $\mu>0$, $\tan\beta=10$).
Each model point has a complete SUSY particle spectrum and a well defined cross section,
which typically involves several production subprocesses.
The CMSSM simulated samples are initially generated using leading-order
cross sections. At each point in CMSSM parameter space,
the predicted yields for each production subprocess (e.g., $\Pg\Pg\to \PSg\PSg$)
are corrected using the NLO cross sections discussed in Ref.~\cite{Kramer:2012bx}.
Using the observed yield in data and the predicted background, we
determine whether the CMSSM yield for the particular model point
can be excluded at the 95\% confidence level (CL). This procedure is complicated by the
fact that the control regions in data could potentially be contaminated by
signal events. This effect is taken into account for each model
by removing the expected contribution to the predicted background arising
from signal contamination of the control regions.

\begin{figure}[tbp!]
\begin{center}
\includegraphics[width=\cmsFigWidth]{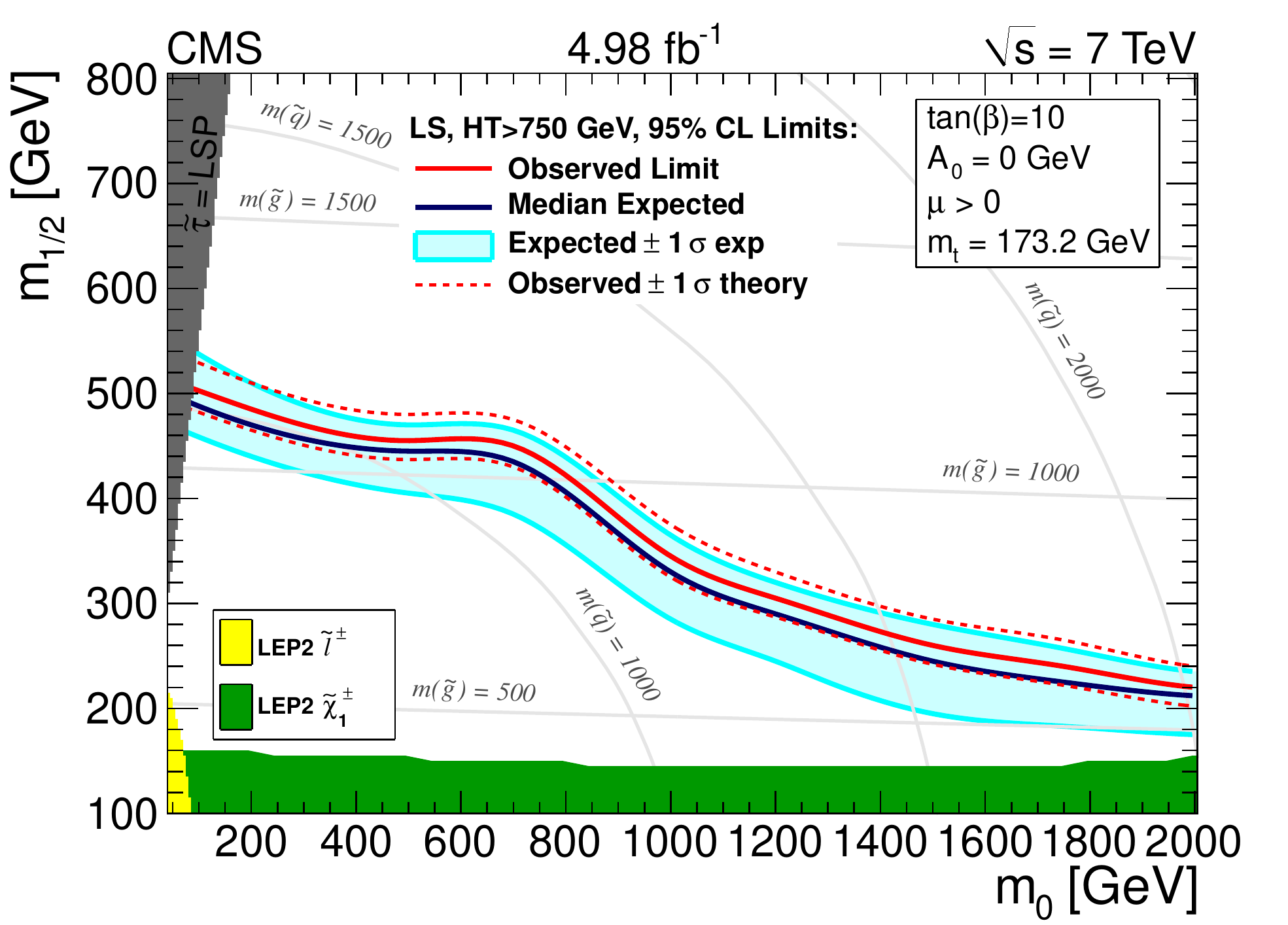}
\end{center}
\caption{LS method: exclusion region in CMSSM parameter space for the $H_\mathrm{T} > 750$\GeV selection.}\label{fig:CMSSM_Exclusion_LS750}
\end{figure}

\begin{figure}[thb]
\begin{center}
\includegraphics[width=\cmsFigWidth]{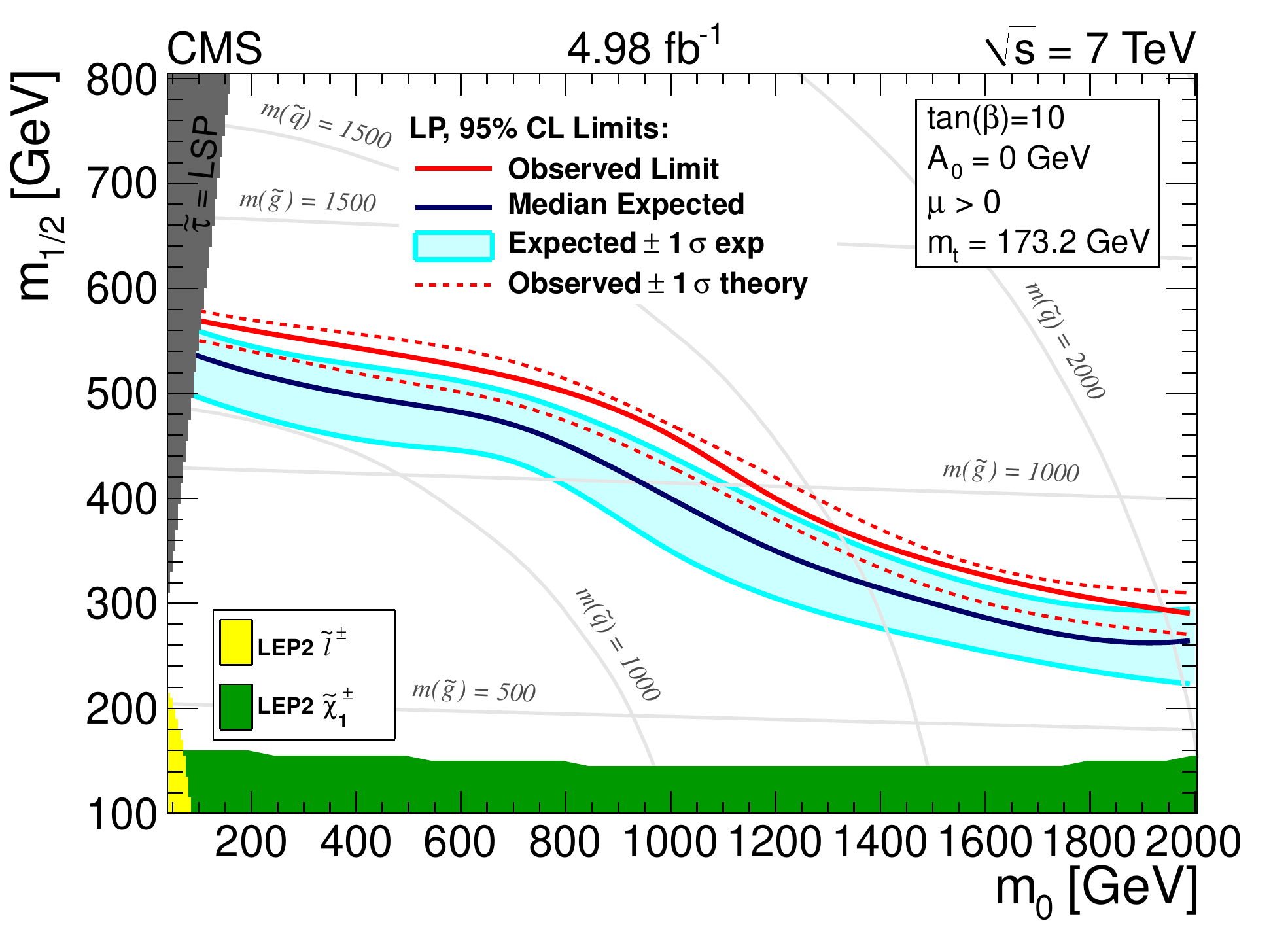}
\caption{\Lp\ method: exclusion region in CMSSM parameter space for all $H_{\mathrm T}$ bins combined.}
\label{fig:CMSSM_Exclusion_Lp}
\end{center}
\end{figure}

\begin{figure}[tbp]
\begin{center}
\includegraphics[width=\cmsFigWidth]{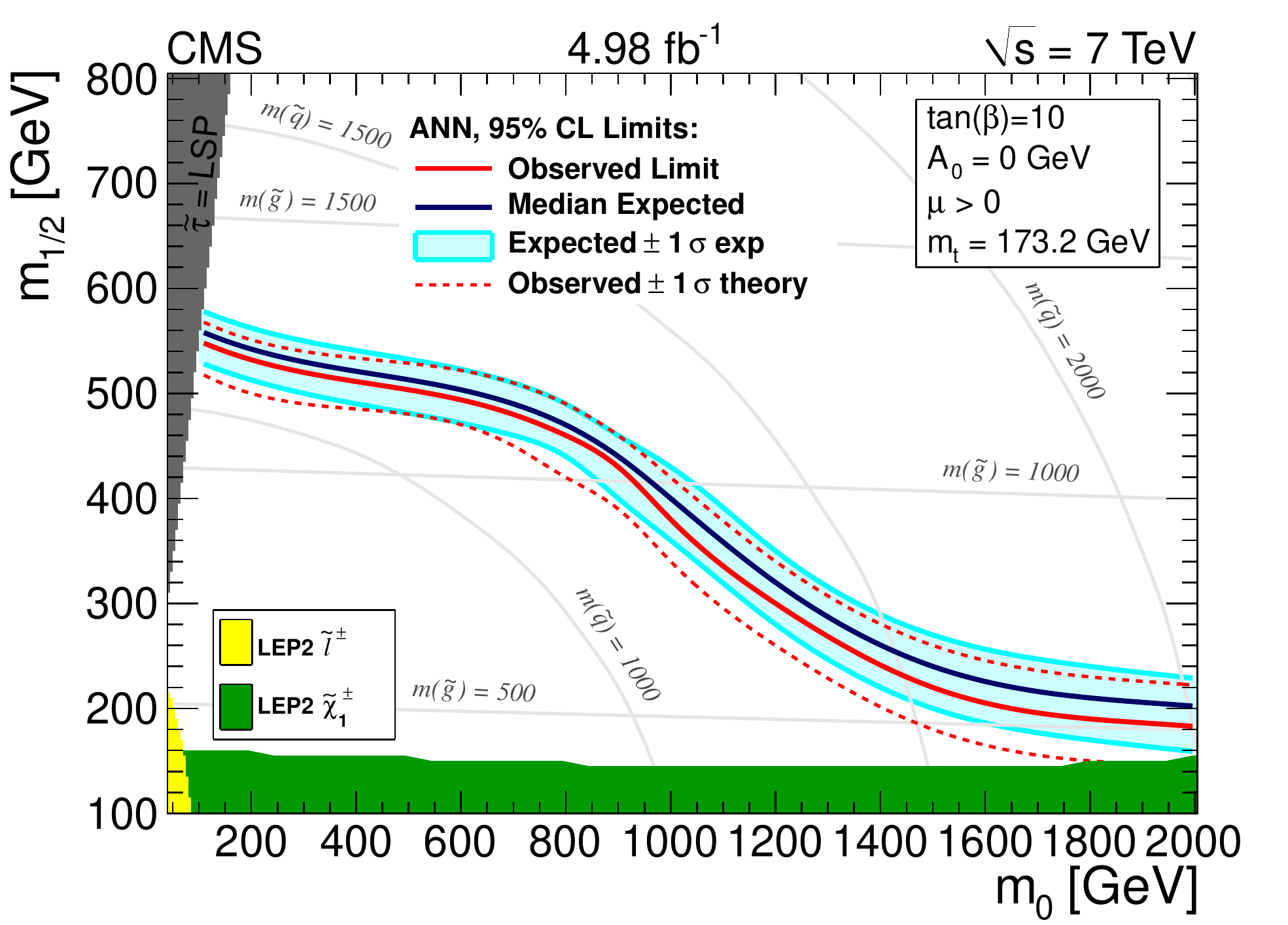}
\end{center}
\caption{ANN method: exclusion region in CMSSM parameter space.}\label{fig:CMSSM_Exclusion_ANN}
\end{figure}

\begin{figure}[tbp]
\begin{center}
\includegraphics[width=\cmsFigWidth]{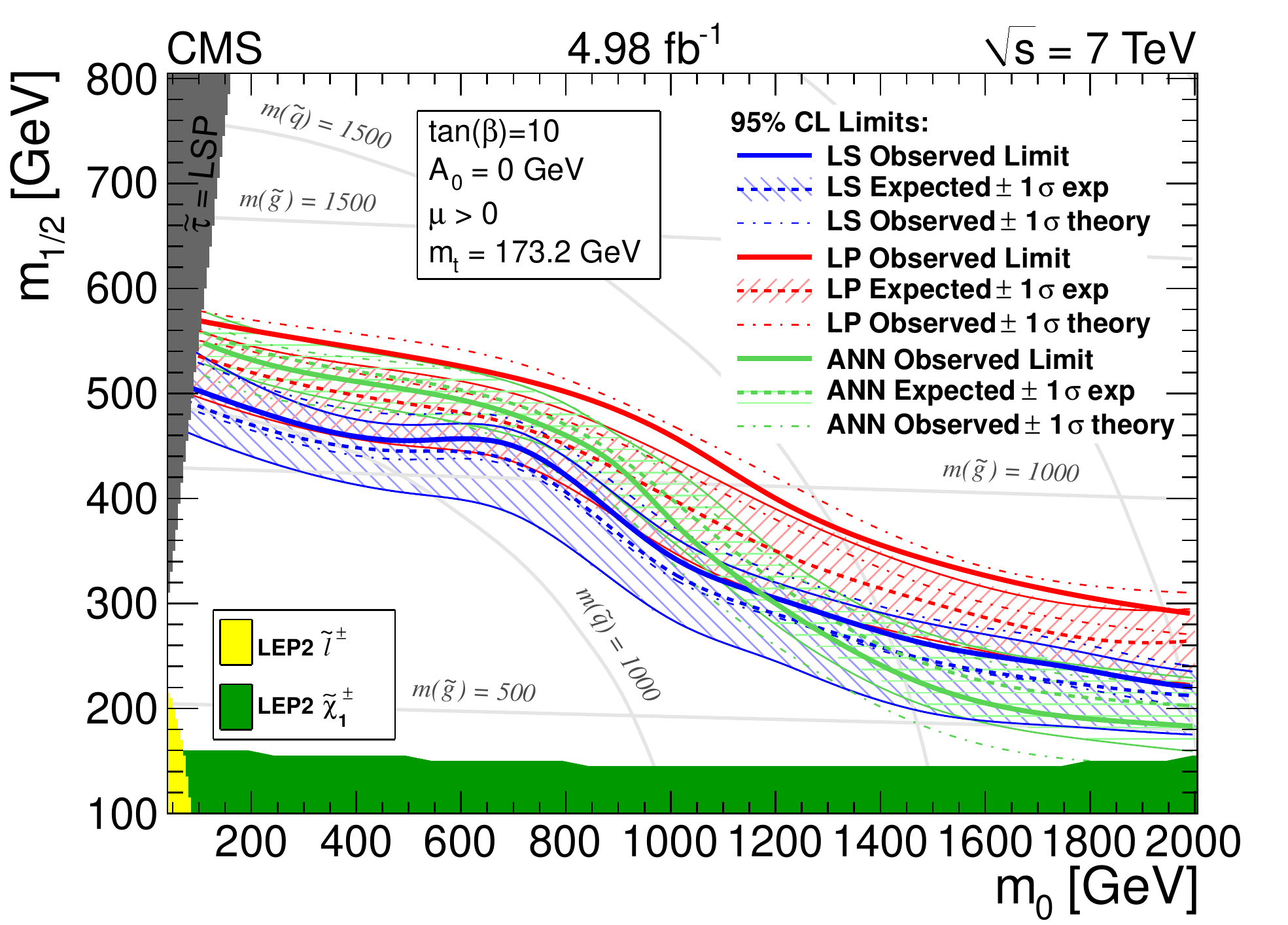}
\end{center}
\caption{Exclusion region for the LS, \Lp, and ANN methods in CMSSM parameter space. Results from
the low- and high-$\ETslash$ signal regions are combined.}\label{fig:CMSSM_allMethods}
\end{figure}

Figures~\ref{fig:CMSSM_Exclusion_LS750}, ~\ref{fig:CMSSM_Exclusion_Lp}, and ~\ref{fig:CMSSM_Exclusion_ANN}
show the CMSSM exclusion region~\cite{Matchev:2012vf} for
the three background estimation methods, evaluated in the
$m_{1/2}$ vs.~$m_0$ plane, with the values of the remaining CMSSM
parameters fixed at $\tan\beta=10$, $A_0=0$\GeV, and $\mu>0$.
Figure~\ref{fig:CMSSM_allMethods} displays all of the results together.
The excluded regions are below the plotted curves, corresponding to
SUSY particle masses below certain values.
For reference, the plots display curves of constant gluino and
squark masses. The lines of constant gluino mass
are approximately horizontal with $m(\PSg)\approx 2.5\, m_{1/2}$.
Lines of constant squark mass are
strongly curved in the $m_{1/2}$ vs.~$m_{0}$ plane.
At low $m_{0}$, the analyses exclude gluinos with masses up to about 1.3\TeV, but
the sensitivity falls with increasing $m_0$.
To determine the one standard deviation ($\sigma$) theoretical uncertainty on the observed limit, the
signal yields are recomputed after changing each of the process-dependent
SUSY production cross sections at each model point by ${\pm}1\sigma$ of their uncertainty
arising from the parton distribution functions and renormalization and factorization scales~\cite{Kramer:2012bx}.

\subsection{Constraints on simplified model parameter space}
The second approach to interpretation is based on the use of
simplified models~\cite{Alwall:2008va,Alves:2011wf}, which
provide a more generic description of new physics signatures. Such models
do not include a full SUSY particle spectrum, but instead include only the
states needed to describe a particular set of decay chains of interest. Rather than excluding a
model, the procedure is to calculate cross section upper limits on a given topological signature.
(Such cross section limits can, however, be converted into limits on particle masses
within the assumptions of the particular model.)
Because simplified models do not describe a full SUSY spectrum, the number of free
parameters is small. Furthermore, the parameters are simply the masses of the
SUSY particles, in contrast to the grand-unified-theory-scale parameters used in the CMSSM. An
advantage of simplified models is that, as a consequence, certain
relationships between particle masses that arise with the CMSSM no longer hold,
and the spectra can be much more generic.

\begin{figure}[tbp]
\begin{center}
\includegraphics[width=0.45\textwidth]{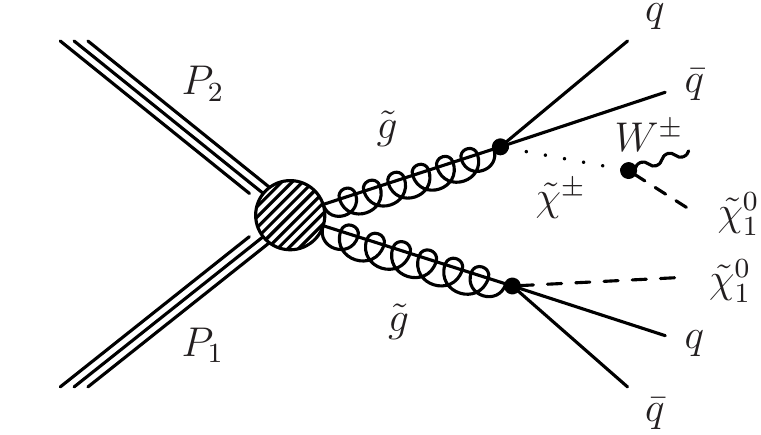}
\end{center}
\caption{Diagram for production and decay in the T3w simplified model.}\label{fig:SMS_T3w}
\end{figure}

\begin{figure}[tbp]
\begin{center}
\includegraphics[width=\cmsFigWidth]{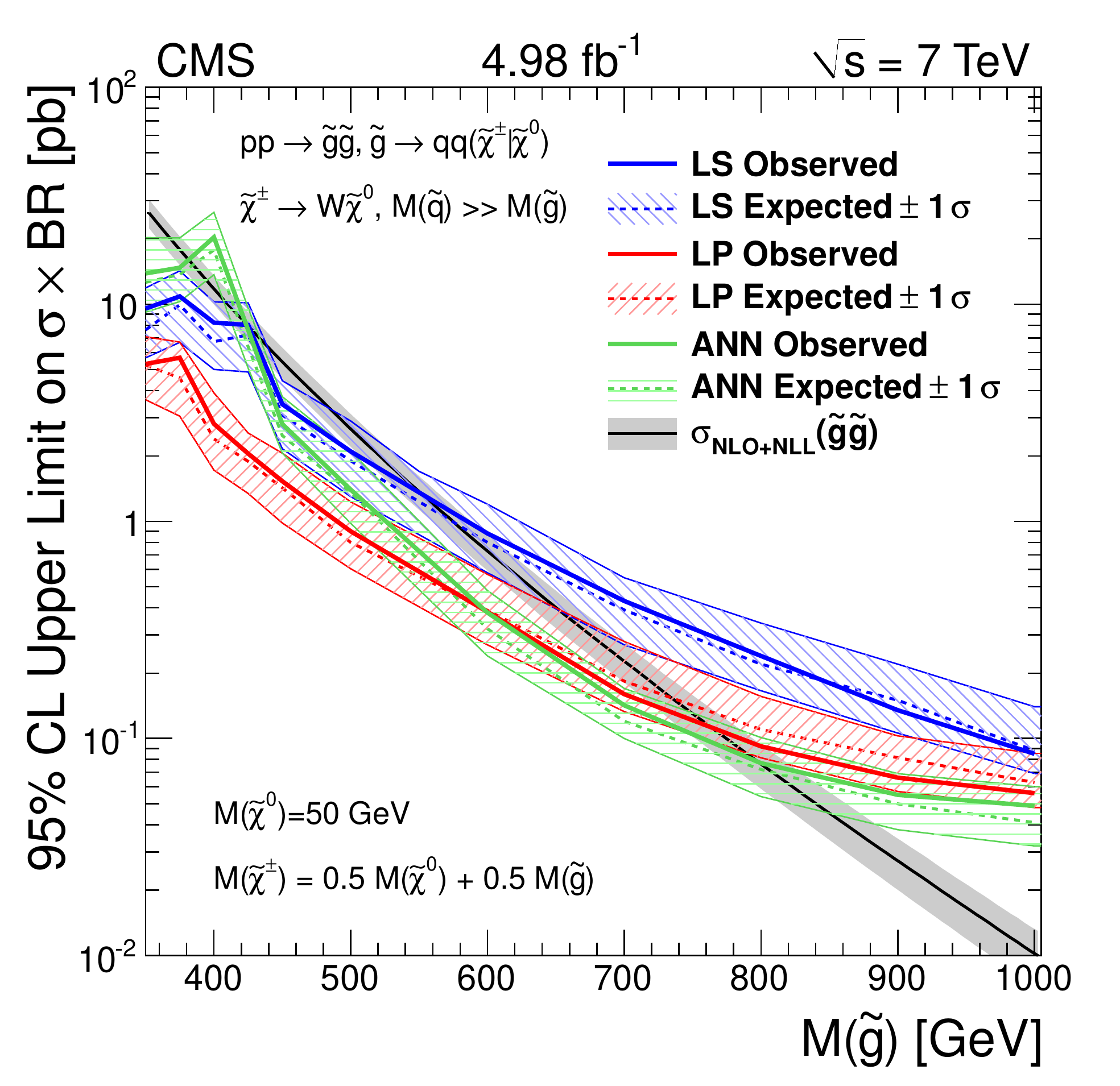}
\end{center}
\caption{Excluded cross sections for the LS, \Lp, and ANN methods for the T3w simplified model.}\label{fig:SMS_T3w_allMethods}
\end{figure}

We consider the ``Topology 3 weakino" (T3w) simplified model, which involves the production of
two gluinos and their decay via the mechanism shown in Fig.~\ref{fig:SMS_T3w}.
One gluino is forced to decay into two quark jets plus the LSP ($\PSGcz$) via the three-body
decay $\PSg\to \Pq\bar \Pq\PSGcz,$ while the other gluino decays via
$\PSg\to \Pq \bar \Pq^{\prime}\PSGc^\pm$, followed by $\PSGc^\pm\to\PW^{\pm}\PSGcz$.
The $\PW^{\pm}$ boson can then decay leptonically.
The T3w model is specified by masses of the gluino, the LSP ($\PSGcz$),
and an intermediate chargino ($\PSGc^\pm$). We calculate
cross section limits as a function of $M(\PSg)$,
assuming a fixed value for the LSP mass  $M(\PSGcz)=50$\GeV
and setting the chargino mass according to
$M(\PSGc^\pm)=0.5( M(\PSGcz) + M(\PSg) )$.
The nominal production cross section for the gluino pair production mechanism
is given in Ref.~\cite{Kramer:2012bx}.
Figure~\ref{fig:SMS_T3w_allMethods} shows
the cross sections excluded by each method for this model.
The limits fluctuate significantly at low $M(\PSg)$ because of the
low signal efficiency in this region.

\subsection{Alternate model exclusions}
The data can be interpreted using a third approach, which is applicable
to models that do not fall within the scope of either the CMSSM or the
simplified model discussed in this section. A model builder can investigate the
sensitivity of any one of the three methods presented in this paper
to a given signal hypothesis by applying the event selection requirements
listed in Table~\ref{tab:Preselection}, together with the final requirements
that define the signal regions. We provide a simple efficiency model
for the most important observables used in the event selections.
The efficiency model can then be applied to a basic (\PYTHIA)
simulation of the signal process.

The efficiency model is based on parametrizations of the efficiencies for the event selection requirements with respect
to the main reconstruction objects and quantities, such as $\Ht$, $\ETslash$, and lepton $\pt$.
The efficiency of the analysis for a given model can be estimated by applying these individual
reconstruction efficiencies, which are given as a function of the most important parameter (such as lepton $\pt$),
to the corresponding kinematic distributions in the model. This procedure would then yield an estimate for the number
of signal events from the model.
Finally, the sensitivity of the analysis to the model can be obtained by comparing the yield of signal events
obtained in this manner with the background yields given in this paper. Kinematic
correlations (which can be model dependent) are not taken into account, but this approach nonetheless provides a
first approximation to the sensitivity.

The efficiencies for each analysis object are described using ``turn-on'' curves, which are simply
error functions,
\begin{equation}
\epsilon(x) = \epsilon_\text{plateau} \frac{1}{2} \left[\erf(\frac{x-x_\text{thresh}}{\sigma}) + 1 \right],
\end{equation}
where $x$ represents the variable most relevant for the reconstruction of the particular object. The
error function is parametrized in terms of the plateau efficiency, $\epsilon_\text{plateau}$;
the turn-on threshold, $x_\text{thresh}$; and the characteristic width
of the turn-on region, $\sigma$. These parameters are
obtained by fitting simulated event samples as a function of the true (generated) value.

The selection efficiency associated with the lepton reconstruction, identification, and isolation requirements
is estimated as a function of lepton $\pt$ by considering muons and electrons (including those
from $\tau$ decay) generated in the {\PYTHIA}-simulated hard-scattering process.
The lepton isolation requirement has a large effect on the efficiency, which consequently
depends on the number of jets in the event. To reduce the model dependence arising from this effect,
two categories of leptons are considered. First, we assign zero efficiency to
leptons that are within $\Delta R<0.4$ of a quark or gluon with
$\pt>40$\GeV in the hard-scattering process. The efficiency for the remaining leptons
is described by a turn-on curve whose parameters are
listed in Table~\ref{tab:lepton_eff_params}. The efficiencies are specified
for both the lepton selection and for the lepton veto.

\begin{table}[tbh!]
\topcaption{Efficiency-model parameters for lepton efficiencies as a function of $x\equiv\pt$.
The leptons are required to lie within the fiducial region and must satisfy the $\pt$ thresholds
specified in Table~\ref{tab:Preselection}.}
\label{tab:lepton_eff_params}
\centering
\begin{tabular}{lccc}
\hline
Lepton     & $\epsilon_\text{plateau}$  & $x_\text{thresh}$[\GeVns{}] & $\sigma$~[\GeVns{}]  \\
\hline
Muon (signal)      & 0.86 & 2.7  & 65 \\
Muon (veto)        & 0.90 & $-17$  & 75 \\
Electron (signal)  & 0.74 &  20  & 61 \\
Electron (veto)    & 0.83 &  2.3 & 54 \\
\hline
\end{tabular}
\end{table}

The number of jets and the resulting \Ht value for each event  are computed using information available
at the generator level. The same clustering algorithm used to reconstruct jets in the data is
applied to the generator-level particles. The resulting generator-level jets are required to satisfy
$\Delta R>0.3$ with respect to the leptons described above. The $\ETslash$ variable is
estimated at the generator level from the transverse momenta of neutrinos and any new weakly interacting
particles, such as the $\PSGcz$. The parametrizations of the efficiency turn-on curves for the
\Ht and $\ETslash$ requirements are listed in Tables~\ref{ht_eff_params} and~\ref{met_eff_params},
respectively. For the requirements used with the LS method, the information given in these tables
generally reproduces the efficiency from full simulation to within about 15\%.

In the $\Lp$\ method, the variables $\Lp$\ and $\stlep$ are functions of
lepton $\pt$ and $\ETslash$. The modeling of lepton $\pt$ is described above.
To emulate $\ETslash$, one needs to apply both a scale shift and smearing to the
generated $\ETslash$ value. The $\ETslash$ scale factor is $\ETslash(\text{reco})/\ETslash(\text{gen})=0.94$.
The value of $\sigma(\ETslash(\text{reco})/\ETslash(\text{gen}))$ is about 0.2 at $\ETslash=100$\GeV. It falls linearly
to about 0.06 at $\ETslash=400$\GeV, and it remains at 0.06 for $\ETslash>400$\GeV.

In the ANN method, the preselection requirements on $\Ht$ and $\ETslash$ are 400 and 100\GeV,
respectively.  The signal regions are specified by $350<\ETslash<500$\GeV and
$\ETslash>500$\GeV together with  $z_\text{ANN} > 0.4$, where $z_\text{ANN} $ is a
function~\cite{ANNfunc} of $n_{\mathrm{jets}}$, $\Ht$, $\Delta\phi(\mathrm{j}_1,\mathrm{j}_2)$,
and \mT. The efficiency turn-on curve for $z_\text{ANN}>0.4$ is approximated by the
parameter values $\epsilon_\text{plateau}=0.98$, $x_\text{thresh}=0.41$, and $\sigma=0.1$.

With these additional procedures, the emulation of the efficiencies for the \Lp\ and ANN methods is
found to be accurate to within ${\sim}15\%$, as for the LS method.

\begin{table}[tbh!]
\topcaption{Efficiency-model parameters for $x\equiv \HT$.}
\label{ht_eff_params}
\centering
\begin{tabular}{lccc}
\hline
Threshold  & $\epsilon_\text{plateau}$ & $x_\text{thresh}$[\GeVns{}] &  $\sigma$ [\GeVns{}]  \\
\hline
$\HT \ge 400$\GeV  & 1.00  & 396   & 65  \\
$\HT \ge 500$\GeV  & 1.00 & 502 & 66 \\
$\HT \ge 750$\GeV  & 1.00 & 760 & 68 \\
$\HT \ge 1000$\GeV & 1.00 & 1013 & 80  \\
\hline
\end{tabular}
\end{table}

\begin{table}[tbh!]
\topcaption{Efficiency-model parameters for $x\equiv\ETslash$.}
\label{met_eff_params}
\centering
\begin{tabular}{lccc}
\hline
Threshold & $\epsilon_\text{plateau}$  & $x_\text{thresh}\,[\GeVns{}]$ & $\sigma\,[\GeVns{}]$  \\
\hline
$\ETslash \ge 100$\GeV  & 1.00  & 103 & 41  \\
$\ETslash \ge 250$\GeV  & 0.99  & 266 & 41 \\
$\ETslash \ge 350$\GeV  & 0.98  & 375 & 45 \\
$\ETslash \ge 450$\GeV  & 0.97  & 485 & 48 \\
$\ETslash \ge 500$\GeV  & 0.94  & 537 & 44  \\
$\ETslash \ge 550$\GeV  & 0.96  & 597 & 59 \\
\hline
\end{tabular}
\end{table}

\section{Summary}
\label{sec:Conclusions}

Using a sample of proton-proton collisions at $\sqrt{s}=7$\TeV corresponding
to an integrated luminosity of 4.98\fbinv,
we have performed a search for an excess of events with
a single, isolated high-\pt lepton, at least three jets, and large missing transverse momentum.
To provide a robust and redundant determination of the SM backgrounds, three methods are used,
each of which relies primarily on control samples in the data.

The Lepton Spectrum (LS) method exploits the relationship between two
key observables, the lepton \pt\ distribution and the $\ETslash$ distribution.
In the dominant SM background processes, which have a single, isolated lepton,
this connection arises from the fact that the lepton and neutrino are produced together
in the two-body decay of the \PW\  boson, regardless of whether the \PW\ is produced
in \cPqt\cPaqt\ or \PW+jets events. In many SUSY models, however, the $\ETslash$ is associated
with the production of two neutralinos, which decouples $\ETslash$ from the lepton
\pt\ spectrum. Smaller backgrounds arising from \cPqt\cPaqt\
dilepton events, from $\tau\to\ell$ decays in \cPqt\cPaqt\ or \PW+jets events, and from QCD multijet
processes are also estimated using control samples in the data. In the sample investigated
with this method, at least four jets are required, which helps to suppress the background
from \PW+jets events.
Nine signal regions are considered, specified by three thresholds on \HT and three bins of \ETslash.
The observed yields in each region are consistent with the background estimates based on control
samples in the data.

The Lepton Projection (\Lp) method exploits information on the W-boson polarization
in \ttbar and W+jets events. The dimensionless \Lp\ variable itself is sensitive to the
helicity angle of the lepton from \PW\ decay, but it also provides discrimination between
signal and background through the ratio of the lepton \pt and the \ETslash values, which is small in SUSY-like events.
The \stlep\ variable maps out a diagonal line in the plane of lepton \pt vs.~\ETslash and reflects the \PW\ transverse momentum for the boosted \PW\
boson.
The \Lp\ distributions are studied in bins of \stlep, and
\HT, and at least three jets are required.
In each signal region, the data are in agreement with expectations from the SM.

The artificial neural network (ANN) method provides a means to obtain the \ETslash distribution of background events
in data by constructing a neural network variable $z_\text{ANN}$, which has a very small
correlation with $\ETslash$. This variable also provides strong discrimination between
signal and background events, so that the background regions do not suffer from large
signal contamination in the models considered. A key element of the
$z_\text{ANN}$ variable is the transverse mass of the lepton-\ETslash system, but
additional variables, such as the number of observed jets, play a role as well.
In the ANN analysis, no excess of events is observed in the
signal regions with respect to the SM background prediction.

Because these methods probe extreme kinematic regions of the background phase space, 
the use of redundant approaches provides confidence in the results. Although the LS and \Lp 
methods both make use of information on the $W$-boson polarization in the background, they are 
based on different kinematic variables and have different signal regions. The LS method breaks the background 
into several pieces (single lepton, $\tau\to\ell$, dilepton, and QCD) and 
provides a direct background prediction for the $\ETslash$ distribution. In contrast, the \Lp method 
defines a powerful kinematic variable that is used to obtain a global background prediction by 
extrapolating an overall background shape from a control region into the signal region.  
The ANN method similarly uses a global approach to estimating the background. The neural-net 
variable incorporates information used in neither of the other two methods.  

The results from each method are interpreted in the context of both the CMSSM
and a so-called simplified model, T3w, which has a minimal SUSY particle spectrum.
The CMSSM limits exclude gluino masses up to approximately $1.3$\TeV in the part of the
parameter space in which $m_0<800$\GeV, but the bound gradually weakens for larger values of
$m_0$. For the T3w simplified model, we obtain cross section upper limits as
a function of gluino mass. Finally, we provide an approximate model of
our signal efficiency that can be used in conjunction with a simple \PYTHIA simulation
to determine whether other models can be probed by these data.
\section*{Acknowledgements}
{\tolerance=800\label{sec:Acknowledgements}

\hyphenation{Bundes-ministerium Forschungs-gemeinschaft Forschungs-zentren} We congratulate our colleagues in the CERN accelerator departments for the excellent performance of the LHC and thank the technical and administrative staffs at CERN and at other CMS institutes for their contributions to the success of the CMS effort. In addition, we gratefully acknowledge the computing centres and personnel of the Worldwide LHC Computing Grid for delivering so effectively the computing infrastructure essential to our analyses. Finally, we acknowledge the enduring support for the construction and operation of the LHC and the CMS detector provided by the following funding agencies: the Austrian Federal Ministry of Science and Research; the Belgian Fonds de la Recherche Scientifique, and Fonds voor Wetenschappelijk Onderzoek; the Brazilian Funding Agencies (CNPq, CAPES, FAPERJ, and FAPESP); the Bulgarian Ministry of Education, Youth and Science; CERN; the Chinese Academy of Sciences, Ministry of Science and Technology, and National Natural Science Foundation of China; the Colombian Funding Agency (COLCIENCIAS); the Croatian Ministry of Science, Education and Sport; the Research Promotion Foundation, Cyprus; the Ministry of Education and Research, Recurrent financing contract SF0690030s09 and European Regional Development Fund, Estonia; the Academy of Finland, Finnish Ministry of Education and Culture, and Helsinki Institute of Physics; the Institut National de Physique Nucl\'eaire et de Physique des Particules~/~CNRS, and Commissariat \`a l'\'Energie Atomique et aux \'Energies Alternatives~/~CEA, France; the Bundesministerium f\"ur Bildung und Forschung, Deutsche Forschungsgemeinschaft, and Helmholtz-Gemeinschaft Deutscher Forschungszentren, Germany; the General Secretariat for Research and Technology, Greece; the National Scientific Research Foundation, and National Office for Research and Technology, Hungary; the Department of Atomic Energy and the Department of Science and Technology, India; the Institute for Studies in Theoretical Physics and Mathematics, Iran; the Science Foundation, Ireland; the Istituto Nazionale di Fisica Nucleare, Italy; the Korean Ministry of Education, Science and Technology and the World Class University program of NRF, Republic of Korea; the Lithuanian Academy of Sciences; the Mexican Funding Agencies (CINVESTAV, CONACYT, SEP, and UASLP-FAI); the Ministry of Science and Innovation, New Zealand; the Pakistan Atomic Energy Commission; the Ministry of Science and Higher Education and the National Science Centre, Poland; the Funda\c{c}\~ao para a Ci\^encia e a Tecnologia, Portugal; JINR (Armenia, Belarus, Georgia, Ukraine, Uzbekistan); the Ministry of Education and Science of the Russian Federation, the Federal Agency of Atomic Energy of the Russian Federation, Russian Academy of Sciences, and the Russian Foundation for Basic Research; the Ministry of Science and Technological Development of Serbia; the Secretar\'{\i}a de Estado de Investigaci\'on, Desarrollo e Innovaci\'on and Programa Consolider-Ingenio 2010, Spain; the Swiss Funding Agencies (ETH Board, ETH Zurich, PSI, SNF, UniZH, Canton Zurich, and SER); the National Science Council, Taipei; the Thailand Center of Excellence in Physics, the Institute for the Promotion of Teaching Science and Technology of Thailand and the National Science and Technology Development Agency of Thailand; the Scientific and Technical Research Council of Turkey, and Turkish Atomic Energy Authority; the Science and Technology Facilities Council, UK; the US Department of Energy, and the US National Science Foundation.

Individuals have received support from the Marie-Curie programme and the European Research Council (European Union); the Leventis Foundation; the A. P. Sloan Foundation; the Alexander von Humboldt Foundation; the Belgian Federal Science Policy Office; the Fonds pour la Formation \`a la Recherche dans l'Industrie et dans l'Agriculture (FRIA-Belgium); the Agentschap voor Innovatie door Wetenschap en Technologie (IWT-Belgium); the Ministry of Education, Youth and Sports (MEYS) of Czech Republic; the Council of Science and Industrial Research, India; the Compagnia di San Paolo (Torino); and the HOMING PLUS programme of Foundation for Polish Science, cofinanced from European Union, Regional Development Fund.
\par}

\bibliography{auto_generated}   

\providecommand{\href}[2]{#2}\begingroup\raggedright\begin{thebibliography}{10}%
\makeatletter
\providecommand{\hrefCMSnoop }[0]{\@secondoftwo}%
\makeatother
\providecommand{\doi}{\texttt{doi:}\begingroup \urlstyle{tt}\Url}

\bibitem{Chatrchyan:2011qs}
\hrefCMSnoop {} {{ CMS} Collaboration, ``{Search for supersymmetry in pp
  collisions at $\sqrt{s}=7$ TeV in events with a single lepton, jets, and
  missing transverse momentum}'',} \textit{ JHEP} \textbf{ 08} (2011) 156,
  \href{http://dx.doi.org/10.1007/JHEP08(2011)156}{\doi{10.1007/JHEP08(2011)156}},
\href{http://www.arXiv.org/abs/1107.1870}{\texttt{ arXiv:1107.1870}}.

\bibitem{Martin:1997ns}
\hrefCMSnoop {} {S.~P. Martin, ``{A Supersymmetry primer}'',} (1997).
\href{http://www.arXiv.org/abs/hep-ph/9709356}{\texttt{ arXiv:hep-ph/9709356}}.

\bibitem{Wess:1974tw}
\hrefCMSnoop {} {J.~Wess and B.~Zumino, ``{Supergauge Transformations in
  Four-Dimensions}'',} \textit{ Nucl. Phys. B} \textbf{ 70} (1974) 39,
\href{http://dx.doi.org/10.1016/0550-3213(74)90355-1}{\doi{10.1016/0550-3213(74)90355-1}}.

\bibitem{Nilles:1983ge}
\hrefCMSnoop {} {H.~P. Nilles, ``{Supersymmetry, Supergravity and Particle
  Physics}'',} \textit{ Phys. Rept.} \textbf{ 110} (1984) 1,
\href{http://dx.doi.org/10.1016/0370-1573(84)90008-5}{\doi{10.1016/0370-1573(84)90008-5}}.

\bibitem{Haber:1984rc}
\hrefCMSnoop {} {H.~E. Haber and G.~L. Kane, ``{The Search for Supersymmetry:
  Probing Physics Beyond the Standard Model}'',} \textit{ Phys. Rept.} \textbf{
  117} (1985) 75,
\href{http://dx.doi.org/10.1016/0370-1573(85)90051-1}{\doi{10.1016/0370-1573(85)90051-1}}.

\bibitem{Barbieri:1982eh}
\hrefCMSnoop {} {R.~Barbieri, S.~Ferrara, and C.~A. Savoy, ``{Gauge Models with
  Spontaneously Broken Local Supersymmetry}'',} \textit{ Phys. Lett. B}
  \textbf{ 119} (1982) 343,
\href{http://dx.doi.org/10.1016/0370-2693(82)90685-2}{\doi{10.1016/0370-2693(82)90685-2}}.

\bibitem{Dawson:1983fw}
\hrefCMSnoop {} {S.~Dawson, E.~Eichten, and C.~Quigg, ``{Search for
  Supersymmetric Particles in Hadron - Hadron Collisions}'',} \textit{ Phys.
  Rev. D} \textbf{ 31} (1985) 1581,
\href{http://dx.doi.org/10.1103/PhysRevD.31.1581}{\doi{10.1103/PhysRevD.31.1581}}.

\bibitem{Farrar:1978xj}
\hrefCMSnoop {} {G.~R. Farrar and P.~Fayet, ``Phenomenology of the production,
  decay, and detection of new hadronic states associated with supersymmetry'',}
  \textit{ Phys. Lett. B} \textbf{ 76} (1978) 575,
\href{http://dx.doi.org/10.1016/0370-2693(78)90858-4}{\doi{10.1016/0370-2693(78)90858-4}}.

\bibitem{Feng:2010gw}
\hrefCMSnoop {} {J.~L. Feng, ``{Dark Matter Candidates from Particle Physics
  and Methods of Detection}'',} \textit{ Ann. Rev. Astron. Astrophys.} \textbf{
  48} (2010) 495,
  \href{http://dx.doi.org/10.1146/annurev-astro-082708-101659}{\doi{10.1146/annurev-astro-082708-101659}},
\href{http://www.arXiv.org/abs/1003.0904}{\texttt{ arXiv:1003.0904}}.

\bibitem{Kane:1993td}
G.~L. Kane\hrefCMSnoop {} { {et~al.}, ``{Study of constrained minimal
  supersymmetry}'',} \textit{ Phys. Rev. D} \textbf{ 49} (1994) 6173,
  \href{http://dx.doi.org/10.1103/PhysRevD.49.6173}{\doi{10.1103/PhysRevD.49.6173}},
\href{http://www.arXiv.org/abs/hep-ph/9312272}{\texttt{ arXiv:hep-ph/9312272}}.

\bibitem{Chamseddine:1982jx}
\hrefCMSnoop {} {A.~H. Chamseddine, R.~L. Arnowitt, and P.~Nath, ``{Locally
  Supersymmetric Grand Unification}'',} \textit{ Phys. Rev. Lett.} \textbf{ 49}
  (1982) 970,
\href{http://dx.doi.org/10.1103/PhysRevLett.49.970}{\doi{10.1103/PhysRevLett.49.970}}.

\bibitem{ArkaniHamed:2007fw}
N.~Arkani-Hamed\hrefCMSnoop {} { {et~al.}, ``{MARMOSET}: The Path from {LHC}
  Data to the New Standard Model via On-Shell Effective Theories'',} (2007).
\href{http://www.arXiv.org/abs/hep-ph/0703088}{\texttt{ arXiv:hep-ph/0703088}}.

\bibitem{Alwall:2008ag}
\hrefCMSnoop {} {J.~Alwall, P.~Schuster, and N.~Toro, ``{Simplified Models for
  a First Characterization of New Physics at the LHC}'',} \textit{ Phys. Rev.
  D} \textbf{ 79} (2009) 075020,
  \href{http://dx.doi.org/10.1103/PhysRevD.79.075020}{\doi{10.1103/PhysRevD.79.075020}},
\href{http://www.arXiv.org/abs/0810.3921}{\texttt{ arXiv:0810.3921}}.

\bibitem{Alwall:2008va}
J.~Alwall\hrefCMSnoop {} { {et~al.}, ``{Model-Independent Jets plus Missing
  Energy Searches}'',} \textit{ Phys. Rev. D} \textbf{ 79} (2009) 015005,
  \href{http://dx.doi.org/10.1103/PhysRevD.79.015005}{\doi{10.1103/PhysRevD.79.015005}},
\href{http://www.arXiv.org/abs/0809.3264}{\texttt{ arXiv:0809.3264}}.

\bibitem{Alves:2011wf}
\hrefCMSnoop {} {{ LHC New Physics Working Group} Collaboration, ``{Simplified
  Models for LHC New Physics Searches}'',} \textit{ J. Phys. G} \textbf{ 39}
  (2012) 105005,
  \href{http://dx.doi.org/10.1088/0954-3899/39/10/105005}{\doi{10.1088/0954-3899/39/10/105005}},
\href{http://www.arXiv.org/abs/1105.2838}{\texttt{ arXiv:1105.2838}}.

\bibitem{Chatrchyan:2012pc}
\hrefCMSnoop {} {{CMS Collaboration}, ``Search for supersymmetry in final
  states with a single lepton, b-quark jets, and missing transverse energy in
  proton-proton collisions at $\sqrt{s} = 7$ {TeV}'',} (2012).
  \href{http://www.arXiv.org/abs/1211.3143}{\texttt{ arXiv:1211.3143}}.
Submitted to Phys. Rev. D.

\bibitem{Aad:2011hh}
\hrefCMSnoop {} {{ ATLAS} Collaboration, ``{Search for supersymmetry using
  final states with one lepton, jets, and missing transverse momentum with the
  ATLAS detector in $\sqrt{s} = 7$ TeV pp collisions}'',} \textit{ Phys. Rev.
  Lett.} \textbf{ 106} (2011) 131802,
  \href{http://dx.doi.org/10.1103/PhysRevLett.106.131802}{\doi{10.1103/PhysRevLett.106.131802}},
\href{http://www.arXiv.org/abs/1102.2357}{\texttt{ arXiv:1102.2357}}.

\bibitem{ATLAS:2011ad}
\hrefCMSnoop {} {{ ATLAS} Collaboration, ``{Search for supersymmetry in final
  states with jets, missing transverse momentum and one isolated lepton in
  $\sqrt{s} = 7$ TeV pp collisions using 1 fb$^{-1}$ of ATLAS data}'',}
  \textit{ Phys. Rev. D} \textbf{ 85} (2012) 012006,
  \href{http://dx.doi.org/10.1103/PhysRevD.85.012006}{\doi{10.1103/PhysRevD.85.012006}},
\href{http://www.arXiv.org/abs/1109.6606}{\texttt{ arXiv:1109.6606}}.

\bibitem{Aad:2012ms}
\hrefCMSnoop {} {{ ATLAS} Collaboration, ``{Further search for supersymmetry at
  $\sqrt{s}=7$ TeV in final states with jets, missing transverse momentum and
  isolated leptons with the ATLAS detector}'',} \textit{ Phys. Rev. D} \textbf{
  86} (2012) 092002,
  \href{http://dx.doi.org/10.1103/PhysRevD.86.092002}{\doi{10.1103/PhysRevD.86.092002}},
\href{http://www.arXiv.org/abs/1208.4688}{\texttt{ arXiv:1208.4688}}.

\bibitem{Chatrchyan:2011bz}
\hrefCMSnoop {} {{ CMS} Collaboration, ``{Search for Physics Beyond the
  Standard Model in Opposite-Sign Dilepton Events at $\sqrt{s} = 7$ TeV}'',}
  \textit{ JHEP} \textbf{ 06} (2011) 026,
  \href{http://dx.doi.org/10.1007/JHEP06(2011)026}{\doi{10.1007/JHEP06(2011)026}},
\href{http://www.arXiv.org/abs/1103.1348}{\texttt{ arXiv:1103.1348}}.

\bibitem{Wpol-PRL}
\hrefCMSnoop {} {{ CMS} Collaboration, ``{Measurement of the Polarization of W
  Bosons with Large Transverse Momenta in W+Jets Events at the LHC}'',}
  \textit{ Phys. Rev. Lett.} \textbf{ 107} (2011) 021802,
  \href{http://dx.doi.org/10.1103/PhysRevLett.107.021802}{\doi{10.1103/PhysRevLett.107.021802}},
\href{http://www.arXiv.org/abs/1104.3829}{\texttt{ arXiv:1104.3829}}.

\bibitem{ref:CMS}
\hrefCMSnoop {} {{ CMS} Collaboration, ``{The CMS experiment at the CERN
  LHC}'',} \textit{ JINST} \textbf{ 3} (2008) S08004,
\href{http://dx.doi.org/10.1088/1748-0221/3/08/S08004}{\doi{10.1088/1748-0221/3/08/S08004}}.

\bibitem{ref:PAS-PFT-09-001}
\href {http://cdsweb.cern.ch/record/1194487} {{ CMS} Collaboration,
  ``Particle--Flow Event Reconstruction in {CMS} and Performance for Jets,
  Taus, and {\MET}'',} CMS Physics Analysis Summary CMS-PAS-PFT-09-001, (2009).

\bibitem{ref:PAS-PFT-10-002}
\href {http://cdsweb.cern.ch/record/1279341} {{ CMS} Collaboration,
  ``Commissioning of the Particle-Flow Reconstruction in Minimum-Bias and Jet
  Events from {\Pp\Pp} Collisions at 7 {TeV}'',} CMS Physics Analysis Summary
  CMS-PAS-PFT-10-002, (2010).

\bibitem{GEANT4}
\hrefCMSnoop {} {{ GEANT4} Collaboration, ``{GEANT4}---a simulation toolkit'',}
  \textit{ Nucl. Instrum. Meth. A} \textbf{ 506} (2003) 250,
\href{http://dx.doi.org/10.1016/S0168-9002(03)01368-8}{\doi{10.1016/S0168-9002(03)01368-8}}.

\bibitem{pythia}
\hrefCMSnoop {} {T.~Sj{\"o}strand, S.~Mrenna, and P.~Z. Skands, ``{PYTHIA} 6.4
  physics and manual'',} \textit{ JHEP} \textbf{ 05} (2006) 026,
  \href{http://dx.doi.org/10.1088/1126-6708/2006/05/026}{\doi{10.1088/1126-6708/2006/05/026}},
\href{http://www.arXiv.org/abs/hep-ph/0603175}{\texttt{ arXiv:hep-ph/0603175}}.

\bibitem{ref:TuneZ2}
\hrefCMSnoop {} {R.~Field, ``{Early LHC Underlying Event Data --- Findings and
  Surprises}'',} (2010).
\href{http://www.arXiv.org/abs/1010.3558}{\texttt{ arXiv:1010.3558}}.

\bibitem{madgraph}
J.~Alwall\hrefCMSnoop {} { {et~al.}, ``{MadGraph 5}: going beyond'',} \textit{
  JHEP} \textbf{ 06} (2011) 128,
  \href{http://dx.doi.org/10.1007/JHEP06(2011)128}{\doi{10.1007/JHEP06(2011)128}},
\href{http://www.arXiv.org/abs/1106.0522}{\texttt{ arXiv:1106.0522}}.

\bibitem{powheg}
\hrefCMSnoop {} {S.~Frixione, P.~Nason, and C.~Oleari, ``{Matching NLO QCD
  computations with Parton Shower simulations: the POWHEG method}'',} \textit{
  JHEP} \textbf{ 11} (2007) 070,
  \href{http://dx.doi.org/10.1088/1126-6708/2007/11/070}{\doi{10.1088/1126-6708/2007/11/070}},
\href{http://www.arXiv.org/abs/0709.2092}{\texttt{ arXiv:0709.2092}}.

\bibitem{PTDR2}
\hrefCMSnoop {} {{ CMS} Collaboration, ``{CMS technical design report, volume
  II: Physics performance}'',} \textit{ J. Phys. G} \textbf{ 34} (2007) 995,
\href{http://dx.doi.org/10.1088/0954-3899/34/6/S01}{\doi{10.1088/0954-3899/34/6/S01}}.

\bibitem{Abdullin:2011zz}
\hrefCMSnoop {} {{ CMS} Collaboration, ``{The fast simulation of the CMS
  detector at LHC}'',} \textit{ J. Phys. Conf. Ser.} \textbf{ 331} (2011)
  032049,
\href{http://dx.doi.org/10.1088/1742-6596/331/3/032049}{\doi{10.1088/1742-6596/331/3/032049}}.

\bibitem{ref:Czarnecki}
\hrefCMSnoop {} {A.~Czarnecki, J.~G. Korner, and J.~H. Piclum, ``{Helicity
  fractions of W bosons from top quark decays at NNLO in QCD}'',} \textit{
  Phys. Rev. D} \textbf{ 81} (2010) 111503,
  \href{http://dx.doi.org/10.1103/PhysRevD.81.111503}{\doi{10.1103/PhysRevD.81.111503}},
\href{http://www.arXiv.org/abs/1005.2625}{\texttt{ arXiv:1005.2625}}.

\bibitem{Aad:2012ky}
\hrefCMSnoop {} {{ ATLAS} Collaboration, ``{Measurement of the W boson
  polarization in top quark decays with the ATLAS detector}'',} \textit{ JHEP}
  \textbf{ 06} (2012) 088,
  \href{http://dx.doi.org/10.1007/JHEP06(2012)088}{\doi{10.1007/JHEP06(2012)088}},
\href{http://www.arXiv.org/abs/1205.2484}{\texttt{ arXiv:1205.2484}}.

\bibitem{ATLAS:2012au}
\hrefCMSnoop {} {{ ATLAS} Collaboration, ``{Measurement of the polarisation of
  W bosons produced with large transverse momentum in pp collisions at
  $\sqrt{s} = 7$ TeV with the ATLAS experiment}'',} \textit{ Eur. Phys. J. C}
  \textbf{ 72} (2012) 2001,
  \href{http://dx.doi.org/10.1140/epjc/s10052-012-2001-6}{\doi{10.1140/epjc/s10052-012-2001-6}},
\href{http://www.arXiv.org/abs/1203.2165}{\texttt{ arXiv:1203.2165}}.

\bibitem{Mangano:2002ea}
M.~L. Mangano\hrefCMSnoop {} { {et~al.}, ``{ALPGEN, a generator for hard
  multiparton processes in hadronic collisions}'',} \textit{ JHEP} \textbf{ 07}
  (2003) 001,
  \href{http://dx.doi.org/10.1088/1126-6708/2003/07/001}{\doi{10.1088/1126-6708/2003/07/001}},
\href{http://www.arXiv.org/abs/hep-ph/0206293}{\texttt{ arXiv:hep-ph/0206293}}.

\bibitem{ref:Blackhat}
Z.~Bern\hrefCMSnoop {} { {et~al.}, ``{Left-Handed W Bosons at the LHC}'',}
  \textit{ Phys. Rev. D} \textbf{ 84} (2011) 034008,
  \href{http://dx.doi.org/10.1103/PhysRevD.84.034008}{\doi{10.1103/PhysRevD.84.034008}},
\href{http://www.arXiv.org/abs/1103.5445}{\texttt{ arXiv:1103.5445}}.

\bibitem{ref:antikt}
\hrefCMSnoop {} {M.~Cacciari, G.~P. Salam, and G.~Soyez, ``{The Anti-k(t) jet
  clustering algorithm}'',} \textit{ JHEP} \textbf{ 04} (2008) 063,
  \href{http://dx.doi.org/10.1088/1126-6708/2008/04/063}{\doi{10.1088/1126-6708/2008/04/063}},
\href{http://www.arXiv.org/abs/0802.1189}{\texttt{ arXiv:0802.1189}}.

\bibitem{JES}
\hrefCMSnoop {} {{ CMS} Collaboration, ``{Determination of Jet Energy
  Calibration and Transverse Momentum Resolution in CMS}'',} \textit{ JINST}
  \textbf{ 6} (2011) P11002,
  \href{http://dx.doi.org/10.1088/1748-0221/6/11/P11002}{\doi{10.1088/1748-0221/6/11/P11002}},
\href{http://www.arXiv.org/abs/1107.4277}{\texttt{ arXiv:1107.4277}}.

\bibitem{Chatrchyan:2012xi}
\hrefCMSnoop {} {{ CMS} Collaboration, ``Performance of {CMS} muon
  reconstruction in pp collision events at {$\sqrt{s} = 7$\TeV}'',} \textit{
  JINST} \textbf{ 7} (2012) P10002,
  \href{http://dx.doi.org/10.1088/1748-0221/7/10/P10002}{\doi{10.1088/1748-0221/7/10/P10002}},
\href{http://www.arXiv.org/abs/1206.4071}{\texttt{ arXiv:1206.4071}}.

\bibitem{ref:PAS-EGM-10-004}
\href {http://cdsweb.cern.ch/record/1299116} {{ CMS} Collaboration, ``Electron
  Reconstruction and Identification at $\sqrt{s} = 7$ {TeV}'',} CMS Physics
  Analysis Summary CMS-PAS-EGM-10-004, (2010).

\bibitem{ref:Pavlunin}
\hrefCMSnoop {} {V.~Pavlunin, ``{Modeling missing transverse energy in V+jets
  at CERN LHC}'',} \textit{ Phys. Rev. D} \textbf{ 81} (2010) 035005,
  \href{http://dx.doi.org/10.1103/PhysRevD.81.035005}{\doi{10.1103/PhysRevD.81.035005}},
\href{http://www.arXiv.org/abs/0906.5016}{\texttt{ arXiv:0906.5016}}.

\bibitem{TMVA_FIXME}
P.~Speckmayer\hrefCMSnoop {} { {et~al.}, ``{The toolkit for multivariate data
  analysis, TMVA 4}'',} \textit{ J. Phys. Conf. Ser.} \textbf{ 219} (2010)
  032057,
\href{http://dx.doi.org/10.1088/1742-6596/219/3/032057}{\doi{10.1088/1742-6596/219/3/032057}}.

\bibitem{tagandprobe}
\href {http://cdsweb.cern.ch/record/1194482} {{ CMS} Collaboration, ``Measuring
  Electron Efficiencies at {CMS} with Early Data'',} CMS Physics Analysis
  Summary CMS-PAS-EGM-07-001, (2008).

\bibitem{Luminosity}
\href {http://cdsweb.cern.ch/record/1434360} {{ CMS} Collaboration, ``Absolute
  Calibration of the Luminosity Measurement at {CMS}: {W}inter 2012 Update'',}
  CMS Physics Analysis Summary CMS-PAS-SMP-12-008, (2012).

\bibitem{Junk:1999kv}
\hrefCMSnoop {} {T.~Junk, ``{Confidence level computation for combining
  searches with small statistics}'',} \textit{ Nucl. Instrum. Meth. A} \textbf{
  434} (1999) 435,
  \href{http://dx.doi.org/10.1016/S0168-9002(99)00498-2}{\doi{10.1016/S0168-9002(99)00498-2}},
\href{http://www.arXiv.org/abs/hep-ex/9902006}{\texttt{ arXiv:hep-ex/9902006}}.

\bibitem{Kramer:2012bx}
M.~Kr{\"a}mer\hrefCMSnoop {} { {et~al.}, ``{Supersymmetry production cross
  sections in pp collisions at $\sqrt{s} = 7$ TeV}'',} (2012).
\href{http://www.arXiv.org/abs/1206.2892}{\texttt{ arXiv:1206.2892}}.

\bibitem{Matchev:2012vf}
\hrefCMSnoop {} {K.~Matchev and R.~Remington, ``{Updated templates for the
  interpretation of LHC results on supersymmetry in the context of mSUGRA}'',}
  (2012).
\href{http://www.arXiv.org/abs/1202.6580}{\texttt{ arXiv:1202.6580}}.

\bibitem{ANNfunc}
\hrefCMSnoop {} {}See supplementary material at [URL will be inserted by
  journal] for a C++ function that evaluates the artificial neural network
  based on the values of the four input variables.

\end{thebibliography}\endgroup

\cleardoublepage \appendix\section{The CMS Collaboration \label{app:collab}}\begin{sloppypar}\hyphenpenalty=5000\widowpenalty=500\clubpenalty=5000\textbf{Yerevan Physics Institute,  Yerevan,  Armenia}\\*[0pt]
S.~Chatrchyan, V.~Khachatryan, A.M.~Sirunyan, A.~Tumasyan
\vskip\cmsinstskip
\textbf{Institut f\"{u}r Hochenergiephysik der OeAW,  Wien,  Austria}\\*[0pt]
W.~Adam, E.~Aguilo, T.~Bergauer, M.~Dragicevic, J.~Er\"{o}, C.~Fabjan\cmsAuthorMark{1}, M.~Friedl, R.~Fr\"{u}hwirth\cmsAuthorMark{1}, V.M.~Ghete, J.~Hammer, N.~H\"{o}rmann, J.~Hrubec, M.~Jeitler\cmsAuthorMark{1}, W.~Kiesenhofer, V.~Kn\"{u}nz, M.~Krammer\cmsAuthorMark{1}, I.~Kr\"{a}tschmer, D.~Liko, I.~Mikulec, M.~Pernicka$^{\textrm{\dag}}$, B.~Rahbaran, C.~Rohringer, H.~Rohringer, R.~Sch\"{o}fbeck, J.~Strauss, A.~Taurok, W.~Waltenberger, G.~Walzel, E.~Widl, C.-E.~Wulz\cmsAuthorMark{1}
% ...
\vskip\cmsinstskip
\dag:~Deceased\\
1:~~Also at Vienna University of Technology, Vienna, Austria\\
% ...

\end{sloppypar}
\end{document}